\documentclass[12pt]{article}
\usepackage{newinutile}
\usepackage{amsmath,amssymb}
\usepackage{mathrsfs}
\usepackage{epsfig,graphics,color}
\usepackage{calc}
\usepackage{supertabular,longtable}
\usepackage{cite}
\usepackage{footnote}

\newtheorem{guess}{Analogy}

\newcommand{\rr}[1]{{\normalfont\textrm{#1}}}

\newcommand{\bb}[1]{{\mathbb{#1}}}

\newcommand{\definisco}[1]{{\em{#1}}}

\newlength{\pecettawidth}
\begin{document}
\title{Residence time estimates for asymmetric simple exclusion dynamics
on strips}

\author{Emilio N.M.\ Cirillo}
\affiliation{Dipartimento di Scienze di Base e Applicate per 
             l'Ingegneria, Sapienza Universit\`a di Roma, 
             via A.\ Scarpa 16, I--00161, Roma, Italy.}
\email{emilio.cirillo@uniroma1.it}
\thanks{ENMC thanks Kerry Landman (Melbourne), Roberto Fern\'andez (Utrecht), 
and Lorenzo Bertini (Roma) for very stimulating discussions and 
useful references.}

\author{Oleh Krehel}
\affiliation{Institute of Complex Molecular Systems
             and Faculty of Chemical Engineering,
             Eindhoven University of Technology,
             P.O.\ Box 513, 5600 MB Eindhoven, The Netherlands}
\email{o.krehel@tue.nl}

\author{Adrian Muntean}
\affiliation{Department of Mathematics and Computer Science, (CASA) Centre for Analysis, Scientific computing and Applications, Institute for Complex Molecular Systems 
             Eindhoven University of Technology,
             P.O.\ Box 513, 5600 MB Eindhoven, The Netherlands}
\email{a.muntean@tue.nl}

\author{Rutger van Santen}
\affiliation{Institute of Complex Molecular Systems
             and Faculty of Chemical Engineering,
             Eindhoven University of Technology,
             P.O.\ Box 513, 5600 MB Eindhoven, The Netherlands}
\email{R.A.v.Santen@tue.nl}

\author{Aditya Sengar}
\affiliation{Indian Institute of Technology Delhi, India}
\email{adityasengariitd@gmail.com}

\begin{abstract}
The target of our study is to approximate numerically and, in
some particular physically relevant cases, also analytically, the residence
time of particles undergoing an asymmetric simple exclusion dynamics on a vertical 
strip.
The source of asymmetry is twofold: (i) the choice of boundary conditions
(different reservoir levels) and (ii) the strong anisotropy from a nonlinear
drift with prescribed directionality.  We focus on the effect of the choice of
anisotropy in the flux on the asymptotic behavior  of the residence time with
respect to the length of the strip. The topic is relevant for situations
occurring in pedestrian flows or biological transport in crowded environments,
where lateral displacements of the particles  occur predominantly affecting
therefore in an essentially way the efficiency of the overall transport
mechanism.
\end{abstract}






\maketitle

\section{Introduction}
\label{s:int}
\par\noindent
The efficiency of transport of active matter in microscopic systems is an 
issue of paramount importance in a number of fields of science including 
biology, chemistry, and logistics. 
Looking particularly at drug--delivery design scenarios  \cite{Kosmidis}, 
ion moving in molecular cytosol \cite{ABC14,ABCM14,ABCM11},
percolation of aggressive acids through reactive porous media \cite{Desirable}, 
the traffic of pedestrians in regions with drastically reduced visibility 
(e.g., in the dark or in the smoke) \cite{CM12,CM13,MCKB} (see also the 
problem of traffic of cars on single--lane highways \cite{Speer}),
we see that the efficiency of a medical treatment, 
the properties of ionic currents thorough cellular membranes,
the durability of a highly permeable material,
or the success of the evacuation of a crowd of humans, strongly depends on
the time spent by the individual particle (colloid, ion, acid molecule, or 
human being) in the constraining  geometry (body, molecule, fabric, 
or corridor). 

In this framework, we focus our attention on the study of the simplest 
2D scenario that mimics alike dynamics. 
The {\em Gedankenexperiment} we make is the following: we 
imagine a vertical strip whose top and bottom entrances are
in touch with infinite particle reservoirs at constant
densities. Assume particles are driven downwards
by the boundary densities difference and/or an
external constant and uniform field (electrical, gravitational, 
generally--accepted crowd opinion, ...).
Let the \emph{residence time} be the typical time a particle
entering the strip at stationarity from the
top boundary needs to exit through the bottom one.
In this framework, under the assumption that
particles in the strip interact only via hard--core exclusion,
we study the {\em ballistic--like} versus the {\em diffusion--like} 
dependence of the residence time on the
external driving force (main source of anisotropy in the system), 
on the length of the strip,
on the horizontal diffusion, and, finally, on the
choice of the boundary condition at the bottom.

We recover the structure of the fluxes as well as the residence times  
proven mathematically by Derrida and co--authors in \cite{DEHP} for the 
asymmetric simple exclusion model on the line; see also \cite{Gorissen} 
for a more recent approach. 
In chemistry single file diffusion has been demonstrated for zeolite 
catalysts \cite{HK} to dramatically reduce the
rate of a reaction. This happens in particular when zeolitic microporous
systems are used with linear micropores with dimensions that are similar as
the size of the molecules that are converted. Since they cannot pass single
file inhibition occurs (see, for instance, \cite{GGS}).

Additionally, we  discover new effects 
that are purely due to the choice of the 2D geometry and 
which are therefore 
absent in a 1D lattice. The most prominent, within the precision 
of our numerical simulations, is the non--monotonic 
behavior in changes in the horizontal displacement 
probability in the bouncing back regime reported in 
Section~\ref{s:caso020}. 
Under certain conditions, particles start accumulating near the bottom exit of the strip. This crowding leads to a bouncing-back effect in which particles trying to escape are reflected in the bulk.   
We observe that, in such a case, 
increasing the frequency of horizontal movements 
help particles to overcome obstacles and to find their way to the 
exit. 

To investigate this model, we employ several working techniques 
including Taylor series truncations for the derivation of the mean--field 
equations, ODE analysis of the stationary case, estimates involving the 
structure of the stationary measure for birth and death processes on a line, 
as well as Monte Carlo simulations to exploit the resources offered by the 
various parameter regimes. In 1D, this model has been 
widely studied both by the mathematical and physical 
communities, see e.g. \cite{Rez,BM,DEHP,Lebowitz,Bertini,Gorissen}.
In 2D, the situation is very much unexplored 
especially in the asymmetric case. We deal with this precise 
problem and we give a rather complete description of the 
phenomenon. Our results, which are based on a thorough study of two
simplified models and extensive Monte Carlo simulations, open 
new mathematical problems concerning the typical time a 
particle need to cross a region in hard--core repulsion regime. 

More precisely, 
in the paper we develop two analytic approaches to predict 
the mean residence time: a macroscopic Mean Field theory and a 
semi--microscopic approach in which the particle motion is 
imagined as that of a single particle against a prescribed 
background density profile borrowed from the macroscopic Mean Field theory. 
These two different predictions are very similar to each other 
but for the regime described above in which the non--monotonic effect 
is found. In this regime the horizontal displacement 
probability tunes the system behavior from the macroscopic prediction
to the semi--microscopic one, with the two limits recovered, respectively, 
in the zero and one horizontal probability displacement cases. 

The importance of the exclusion rule on the time dependence of 
the typical distance covered by a particle is not new in the 
scientific literature. 
Due to the exclusion rule, 
the asymmetric process on a square lattice that we discuss here 
can be considered to occur on a percolating lattice. 
The symmetric exclusion case  has been widely explored for diffusion, 
as for instance the ``ant in the labirinth'' by de Gennes \cite{AS}.
The distance travelled by the ant is proportional to the square root of the 
time (random walk diffusion) as long as the site occupancy is low, 
but, when the critical $0.5928$ site occupation is approached,
this changes to a time dependence close to cubic root of time. 
Beyond this critical site occupancy the order in time rapidly drops.
The excluded volume problem in several dimensions has not yet found however  
a satisfactory solution \cite{vK}.

A related but similar phenomenon occurs in the symmetric 1D
case. When site occupation increases the distance time relation becomes 
the fourth root of time \cite{fedders}.
The asymmetric problem in 2D, that we are interested in, 
can be more easily addressed as asymmetric exclusion with driven diffusion, 
see \cite{Krug} for a paper and \cite{BE} for a complete review.
In the 1D
total asymmwtric case (particles can move only from the top 
to the bottom), 
using a kinematic wave theory related approach and the method of maximum 
transported current \cite{BE}, it is identified a
three parameter space region as a function of the rates at which 
particles would enter (from the top) or exit (from the bottom) the strip. 
These regions closely relate to the particle density where percolation 
sets in (on a square lattice with stochastic bond formation the percolation 
threshold is 0.5). 
In this paper we will be mainly concerned with the case where the relative 
probability for a particle to enter the strip (from the top) 
is one.
In this case the two situations than can be distinguished are the high 
density phase (exit probability smaller than $1/2$) and the 
maximum current phase (exit probability larger than $1/2$). 
Interestingly the difference relates to an occupation density that 
is larger or less than the critical point of  percolation. 

In the lower density phase the average density is $1/2$ 
and current is maximum, at the density higher than the
percolation threshold it decreases with the exit rate. 
\cite{BE}.
We derive an equation for the Mean Field concentration profile 
which, in our setup, depends only on the vertical spatial 
coordinate. The profile depends on the reservoir boundary 
conditions and on the vertical drift, namely, 
the ratio between down and up jump probabilities divided by their sum, 
that, we recall, we do not considered necessarily equal 
to one (as in the total asymmetric case discussed above). 
In the large drift case, we recover density profiles consistent with 
the phase diagram described above which has been proven 
to be correct in the total asymmetric case \cite{DEHP}.

In this scenario, as already mentioned above, we setup two different 
analytic computations of the residence time: a 
semi--microscopic approach based on the analogy with a 1D
birth and death model with not homogeneous background and a purely 
macroscopic Mean Field approximation. In the last approach the 
residence time is finally computed in terms of the stationary 
current. It is important to stress that, for zero drift, these 
two different approach seem to explain the behavior of the model 
in two different regimes, namely, low and high horizontal 
displacement probability. From our point of view this result is 
of absolutely high interest, indeed, the main issue we arise in this 
paper is that of understanding the influence of lateral 
displacements on the typical time 
needed by a particle to cross a pipe. 
Future investigation will be needed to understand deviations 
from the average particle behavior. In this perspective the 
residence time seems to be a very useful observable and its 
distribution will be the key object of our future study. 

The paper is structured in the following fashion:
we describe the dynamics of our stochastic lattice model in 
Section~\ref{s:modello}.  
Section~\ref{s:mf} is concerned primarily with the derivation
of the mean field equations governing the macroscopic evolution of the 
density.  
In the same section we study the stationary mean field 
behavior, the thermalization time of the system (i.e. the time that the system takes to reach the steady state) as well as the numerical 
testing of the accuracy of the mean field prediction of the stationary 
density profile.  
In Section~\ref{s:bd} we make the direct analogy between our
scenario and a biased birth and death model for which we can calculate 
explicitly the unique invariant measure and the use this object to obtain 
analytical lower and upper bounds on the residence time for three distinct 
physically relevant scenarios, viz. (i) a homogeneous case, (ii) a linear 
case, (iii) a totally asymmetric case.   
The core of the paper is 
Section~\ref{s:res} and 
Section~\ref{s:res-num}. 
Herein we use inspiration from the handling of the biased birth and death 
model to get approximate analytical bounds on the residence 
time for our problem for selected parameter regimes. 
We also setup a purely Mean Field macroscopic
computation yielding a prediction of the residence time 
which is fully discussed in the last two sections.
Finally, we explore 
numerically the residence time for all parameter regimes and compare the 
results with the derived analytical bounds. 

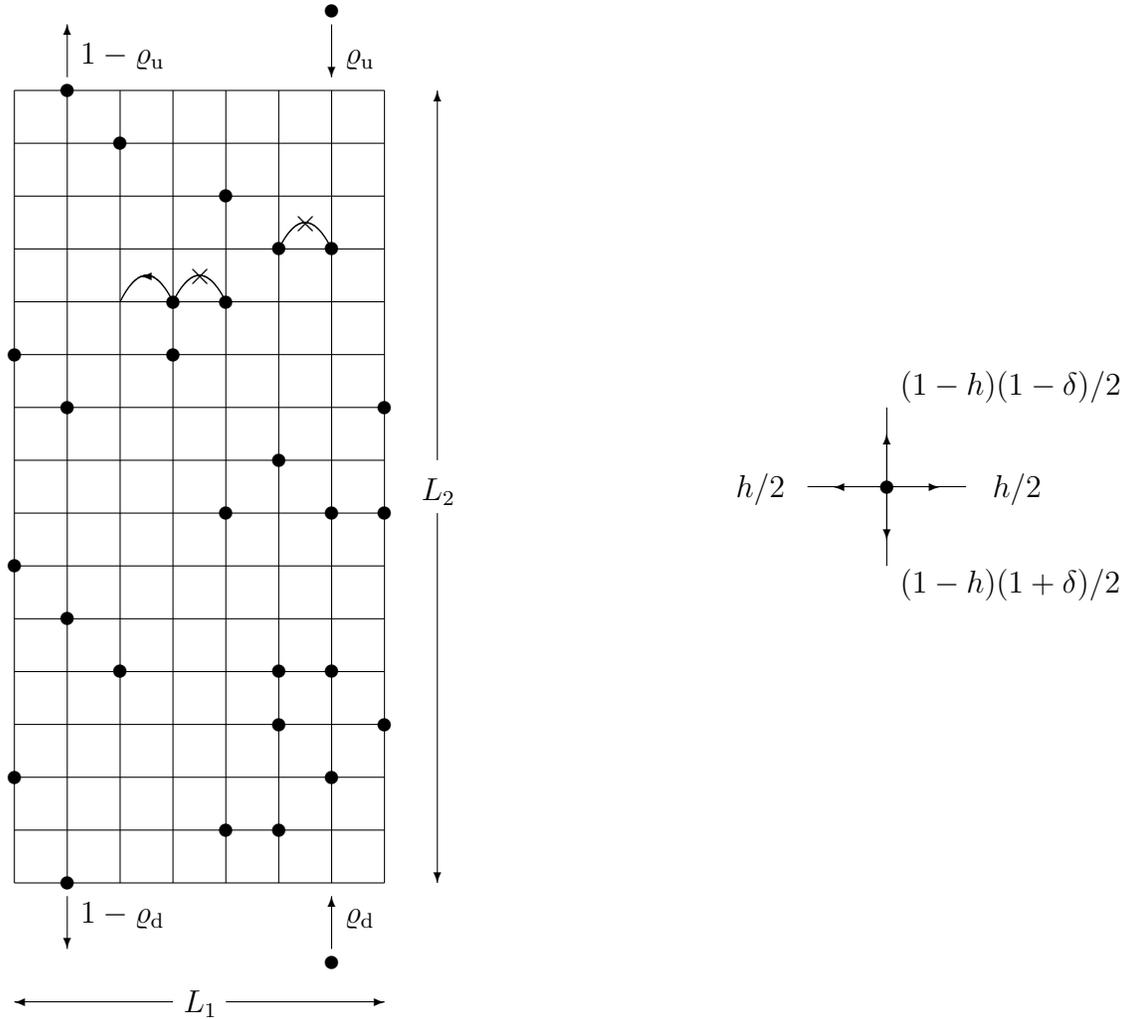
\begin{figure}[h]
\begin{picture}(400,330)(-30,0)
\thinlines
\multiput(0,0)(20,0){8}{\line(0,1){300}}
\multiput(0,0)(0,20){16}{\line(1,0){140}}
\put(60,-45){\vector(-1,0){60}}
\put(80,-45){\vector(1,0){60}}
\put(64,-49){${L_1}$}
\put(160,140){\vector(0,-1){140}}
\put(160,160){\vector(0,1){140}}
\put(154,145){${L_2}$}
\put(0,40){\circle*{5}}
\put(0,120){\circle*{5}}
\put(0,200){\circle*{5}}
\put(20,0){\circle*{5}}
\put(20,100){\circle*{5}}
\put(20,180){\circle*{5}}
\put(20,300){\circle*{5}}
\put(40,80){\circle*{5}}
\put(40,280){\circle*{5}}
\put(60,200){\circle*{5}}
\put(60,220){\circle*{5}}
\put(80,20){\circle*{5}}
\put(80,140){\circle*{5}}
\put(80,220){\circle*{5}}
\put(80,260){\circle*{5}}
\put(100,20){\circle*{5}}
\put(100,60){\circle*{5}}
\put(100,80){\circle*{5}}
\put(100,160){\circle*{5}}
\put(100,240){\circle*{5}}
\put(120,40){\circle*{5}}
\put(120,80){\circle*{5}}
\put(120,140){\circle*{5}}
\put(120,240){\circle*{5}}
\put(140,60){\circle*{5}}
\put(140,140){\circle*{5}}
\put(140,180){\circle*{5}}
\thinlines
\put(20,305){\vector(0,1){20}}
\put(25,310){${1-\varrho_\textrm{u}}$}
\put(120,325){\vector(0,-1){20}}
\put(125,310){${\varrho_\textrm{u}}$}
\put(120,330){\circle*{5}}
\put(20,-5){\vector(0,-1){20}}
\put(25,-15){${1-\varrho_\textrm{d}}$}
\put(120,-25){\vector(0,1){20}}
\put(125,-15){${\varrho_\textrm{d}}$}
\put(120,-30){\circle*{5}}
\qbezier(60,220)(70,240)(80,220)
\put(65.5,226.5){${\times}$}
\qbezier(40,220)(50,240)(60,220)
\put(53,229.5){\vector(-1,0){5}}
\qbezier(100,240)(110,260)(120,240)
\put(105.5,246.5){${\times}$}

\put(330,120){\line(0,1){60}}
\put(300,150){\line(1,0){60}}
\put(330,150){\circle*{5}}
\put(330,150){\vector(0,1){20}}
\put(335,185){$(1-h)(1-\delta)/2$}
\put(330,150){\vector(0,-1){20}}
\put(335,110){$(1-h)(1+\delta)/2$}
\put(330,150){\vector(1,0){20}}
\put(370,146){$h/2$}
\put(330,150){\vector(-1,0){20}}
\put(273,146){$h/2$}
\end{picture}
\vskip 2. cm 
\caption{Schematic representation of the model: the lattice on the 
left and the jumping probabilites on the right.}
\label{f:reticolo} 
\end{figure}

\section{The model}
\label{s:modello}
\par\noindent
Let $L_1,L_2$ be two positive integers. Let $\Lambda\subset\bb{Z}^2$ 
denote the \emph{strip} $\{1,\dots,L_1\}\times\{1,\dots,L_2\}$. 
We say that the coordinate directions $1$ and $2$ of the 
strip are respectively the \definisco{horizontal} and the \definisco{vertical}
direction. We shall accordingly use the words \definisco{top}, 
\definisco{bottom}, \definisco{left}, and \definisco{right}.
On $\Lambda$ we define a discrete time stochastic process controlled by the 
parameters 
\begin{displaymath}
\varrho_\rr{u},\varrho_\rr{d}\in[0,1]
\;\;\;\textrm{ and }\;\;\;
h,u,d\in[0,1]\;\;\textrm{ such that }h+u+d=1
\end{displaymath}
whose meaning will be explained below.

The \emph{configuration} of the system at time $t\in\bb{Z}_+$
is given by the positive integer $n(t)$ denoting  
the number of particles in the system at time $t$ and by the two 
collections of integers
$x_1(1,t),\dots,x_1(n(t),t)\in\{1,\dots,L_1\}$ and 
$x_2(1,t),\dots,x_2(n(t),t)\in\{1,\dots,L_2\}$ denoting, respectively, 
the horizontal and the vertical coordinates of the $n(t)$ particles 
in the strip $\Lambda$ at time $t$. 
The $i$--th particle, with $i=1,\dots,n(t)$, 
is then associated with the site $(x_1(i,t),x_2(i,t))\in\Lambda$ which 
is called \definisco{position} of the particle at time $t$. 
A site associated with a particle a time $t$ will be said to be 
\definisco{occupied} at time $t$, otherwise we shall say that it is 
\definisco{free} or \definisco{empty} at time $t$. 
We let $n(0)=0$. 

At each time $t\ge1$ we first set 
$n(t)=n(t-1)$ and then
repeat the following algorithm $n(t-1)$ times. 
One of the three actions 
\emph{insert a particle through the top boundary},
\emph{insert a particle through the bottom boundary},
and 
\emph{move a particle in the bulk} is performed with 
the corresponding probabilities 
$\varrho_\rr{u}L_1/
(\varrho_\rr{u}L_1+\varrho_\rr{d}L_1+n(t))$, 
$\varrho_\rr{d}L_1/
(\varrho_\rr{u}L_1+\varrho_\rr{d}L_1+n(t))$, 
and
$n(t)/
(\varrho_\rr{u}L_1+\varrho_\rr{d}L_1+n(t))$. 

\emph{Insert a particle through the top boundary.\/}
Chose at random with uniform probability the integer $i\in\{1,\dots,L_1\}$
and, if the site $(1,i)$ is empty, with probability $d$ 
set $n(t)=n(t)+1$ and add a particle to site $(1,i)$.

\emph{Insert a particle through the bottom boundary.\/}
Chose at random with uniform probability the integer $i\in\{1,\dots,L_1\}$
and, if the site $(L_2,i)$ is empty, with probability $u$ 
set $n(t)=n(t)+1$ and add a particle to site $(L_2,i)$.

\emph{Move a particle in the bulk.\/}
Chose at random with uniform probability one of the $n(t)$ particles 
in the bulk.
The chosen particle is moved according to the following rule:
one of the four
neighboring sites of the one occupied by the particle is chosen
at random with probability $h/2$ (left), $u$ (up), $h/2$ (right), and 
$d$ (down). If the chosen site is in the strip (not on 
the boundary) and it is free, the particle is moved there 
leaving empty the site occupied at time $t$. 
If the chosen site is on the boundary of the strip the dynamics is 
defined as follows: 
the left boundary $\{(0,z_2),\;z_2=1,\dots,L_2\}$ and 
the right boundary $\{(L_1+1,z_2),\;z_2=1,\dots,L_2\}$
are \definisco{reflecting} (homogeneous Neumann boundary conditions) 
in the sense that a particle trying to jump there is not moved.
The bottom and the top 
boundary conditions are stochastic in the sense 
that when a particle tries to jump to a site $(z_1,0)$, with 
$z_1=1,\dots,L_1$, such a site has to be considered occupied 
with probability $\varrho_\rr{u}$ and free with probability 
$1-\varrho_\rr{u}$,  
whereas
when a particle tries to jump to a site $(z_1,L_2+1)$, with 
$z_1=1,\dots,L_1$, such a site has to be considered occupied 
with probability $\varrho_\rr{d}$ and free with probability 
$1-\varrho_\rr{d}$. 
If the arrival site is considered free 
the particle trying to jump there is removed 
by the strip $\Lambda$ (it is said to \definisco{exit} the system)
and the number of particles is reduced by one, namely, $n(t)=n(t)-1$.
If the arrival site is occupied the particle is not moved. 

We gave the definition of the model in an algorithmic way, but 
note that the model is a Markov chain $\omega_0,\omega_1,\dots,\omega_t,\dots$ 
on the 
\emph{state} or \emph{configuration space} 
$\Omega:=\{0,1\}^\Lambda$ with transition probability 
that can be deduced by the algorithmic definition. 

\section{Mean Field estimates}
\label{s:mf}
\par\noindent
Let the \emph{occupation number} of the site 
$(z_1,z_2)\in\Lambda$ at time $t$
be equal to $1$ if such a site is occupied by a particle at time $t$ 
and to $0$ otherwise.
Let, also, 
$m_t(z_1,z_2)$ be 
the expectation of the occupation number at site $(z_1,z_2)\in\Lambda$ 
at time $t\ge0$. This quantity is  well defined in a stochastic context. But if one wants a 
more intuitive idea of what such a quantity is, one can think of 
running the dynamics many times independently and then
computing equal time averages with respect to 
these different realizations of the process. 
The mean over those independent realizations of the occupation 
number at site $(z_1,z_2)$ at time $t$ will be 
$m_t(z_1,z_2)$.

Now, we set up a Mean Field 
computation for such a mean occupation time $m_t$. We shall follow
\cite{SLH}, but it is worth noting that, 
in 1D and on the infinite volume $\bb{Z}$, 
the equation we shall obtain 
is proven rigorously to be the macroscopic limit 
of the discrete space random process \cite{dMPS}.

We need to reproduce and slightly generalize the
approach in \cite{SLH} since there
a particular choice for the horizontal probability has been 
considered, whereas we need a more general result.
Let the \emph{drift} be 
\begin{equation}
\label{againmf00-1}
\delta=\frac{d-u}{d+u}
\;\;\;\;.
\end{equation}
This is indeed the physically meaningful definition of drift, since it 
is the ratio between the difference of the probabilities to move 
down and up and the total probability to perform a vertical displacement. 
Note that, since $d+u=1-h$, a simple computation yields 
\begin{equation}
\label{againmf00-2}
d=\frac{1-h}{2}(1+\delta)
\;\;\;\textrm{ and }\;\;\;
u=\frac{1-h}{2}(1-\delta)
\;\;\;\;.
\end{equation}

First of all, note that 
in our dynamics 
the probability that at time $t$ a specific particle is updated 
is of order one, 
since at each time we update at random 
$n(t-1)$ particles, 
where, we 
recall, $n(t)$ is the number of particles in the system at time $t$. 
Thus, 
the Mean Field approximation consists in 
letting
\begin{equation}
\label{newmf000}
\varepsilon=\frac{1}{L_2+1}
\;\;\;\textrm{ and }\;\;\;
\tau=\varepsilon^2
\;\;,
\end{equation}
considering 
the macroscopic variables 
\begin{equation}
\label{newmf010}
z'_1=\varepsilon z_1,\;\;
z'_2=\varepsilon z_2,\;\;
\;\;\;\textrm{ and }\;\;\;
t'=\tau t = \varepsilon^2 t
\;\;,
\end{equation}
and 
writing, for an 
arbitrary point in the bulk, the following 
balance equation:
\begin{displaymath}
\begin{array}{l}
m_{t'+\tau}(z'_1,z'_2)-m_{t'}(z'_1,z'_2)
\\
\phantom{mm}
=(h/2) m_{t'}(z'_1-\varepsilon,z'_2)[1-m_{t'}(z'_1,z'_2)]
 +d m_{t'}(z'_1,z'_2-\varepsilon)[1-m_{t'}(z'_1,z'_2)]
\\
\phantom{mm=}
 +(h/2) m_{t'}(z'_1+\varepsilon,z'_2)[1-m_{t'}(z'_1,z'_2)]
 +u m_{t'}(z'_1,z'_2+\varepsilon)[1-m_{t'}(z'_1,z'_2)]
\\
\phantom{mm=}
 -(h/2) m_{t'}(z'_1,z'_2)[1-m_{t'}(z'_1+\varepsilon,z'_2)]
 -d m_{t'}(z'_1,z'_2)[1-m_{t'}(z'_1,z'_2+\varepsilon)]
\\
\phantom{mm=}
 -(h/2) m_{t'}(z'_1,z'_2)[1-m_{t'}(z'_1-\varepsilon,z'_2)]
 -u m_{t'}(z'_1,z'_2)[1-m_{t'}(z'_1,z'_2-\varepsilon)]
\phantom{mm=}
\\
\phantom{mm}
=(h/2)[m_{t'}(z'_1+\varepsilon,z'_2)-2m_{t'}(z'_1,z'_2)
 +m_{t'}(z'_1-\varepsilon,z'_2)]
\\
\phantom{mm=}
 +((1-h)/2)[m_{t'}(z'_1,z'_2+\varepsilon)
  -2m_{t'}(z'_1,z'_2)+m_{t'}(z'_1,z'_2-\varepsilon)]
\\
\phantom{mm=}
 -\delta((1-h)/2)\{
    [1-m_{t'}(z'_1,z'_2)][m_{t'}(z'_1,z'_2+\varepsilon)
                          -m_{t'}(z'_1,z'_2-\varepsilon)]
\\
\phantom{mm=-\delta((1-h)/2)\{}
    +m_{t'}(z'_1,z'_2)[(1-m_{t'}(z'_1,z'_2+\varepsilon))
                       -(1-m_{t'}(z'_1,z'_2-\varepsilon))]
                  \}
\;\;.
\end{array}
\end{displaymath}
Now, we multiply and divide suitably the space and time units 
$\varepsilon$ and $\tau$ to obtain 
\begin{displaymath}
\begin{array}{l}
[m_{t'+\tau}(z'_1,z'_2)-m_{t'}(z'_1,z'_2)]/\tau
\\
\phantom{mm}
=(h/2)\,[m_{t'}(z'_1+\varepsilon,z'_2)
         -2m_{t'}(z'_1,z'_2)+m_{t'}(z'_1-\varepsilon,z'_2)]/\varepsilon^2
\\
\phantom{mm=}
 +[(1-h)/2]\,[m_{t'}(z'_1,z'_2+\varepsilon)
              -2m_{t'}(z'_1,z'_2)+m_{t'}(z'_1,z'_2-\varepsilon)]
                            /\varepsilon^2
\\
\phantom{mm=}
 -[\delta(1-h)/\varepsilon]\,\{
            [1-m_{t'}(z'_1,z'_2)]
            [m_{t'}(z'_1,z'_2+\varepsilon)-m_{t'}(z'_1,z'_2-\varepsilon)]
\\
\phantom{mm=-[\delta(1-h)/\varepsilon]\{}
                    +m_{t'}(z'_1,z'_2)[(1-m_{t'}(z'_1,z'_2+\varepsilon))
                    -(1-m_{t'}(z'_1,z'_2-\varepsilon))]
                  \}/(2\varepsilon)
\;\;.
\end{array}
\end{displaymath}
Finally, if we assume that in the limit $\varepsilon\to0$, namely, 
$L_2\to\infty$, the drift scales to zero as 
\begin{displaymath}
\delta=\varepsilon\bar\delta
\;\;,
\end{displaymath}
we find the Mean Field limiting equation 
\begin{equation}
\label{againmf000pre}
\frac{\partial m_{t'}}{\partial t'}
=
\frac{1}{2}h
\frac{\partial^2 m_{t'}}{\partial {z'_1}^2}
+
\frac{1}{2}
(1-h)
\frac{\partial^2 m_{t'}}{\partial {z'_2}^2}
-\bar\delta(1-h)
\frac{\partial }{\partial z'_2}[m_{t'}(1-m_{t'})]
\;\;.
\end{equation}

It is worth noting that the above equation is a diffusion--like 
equation with a nonlinear anisotropic flux. 
From the physical point of view the most interesting remark is that 
the diffusion part of the equation is linear. The effect of the drift 
is captured in nonlinear transport term. 
This term vanishes when $\bar\delta=0$, so that linearity is 
approximatively restored at very small $\bar\delta$. It is worth noting that, by the choice of scaling, there is no inbuilt bias towards diffusion- or drift-alone. 

To compare the solution of the Mean Field equation to the numerical 
simulations, we abuse the notation (recall that $t$, $z_1$, and $z_2$ 
denoted above integer numbers) and write 
$t=t'/\tau$, 
$z_1=z'_1/\varepsilon$, 
and 
$z_2=z'_2/\varepsilon$. 
The above limiting equation then reads 
\begin{equation}
\label{againmf000}
\frac{\partial m_t}{\partial t}
=
\frac{1}{2}
h
\frac{\partial^2 m_t}{\partial z_1^2}
+
\frac{1}{2}
(1-h)
\frac{\partial^2 m_t}{\partial z_2^2}
-\delta(1-h)
\frac{\partial }{\partial z_2}[m_t(1-m_t)]
\;\;.
\end{equation}

Since in 
the top and bottom boundary 
densities are constant in space (along the border),
the stationary solutions to \eqref{againmf000} 
do not depend on the space variable $z_1$. We call 
$\varrho(z_2)$ a 
\emph{density profile} 
of the stationary Mean Field equation 
\begin{equation}
\label{againmf040}
\frac{d^2}{dz_2^2}\varrho-b\frac{\rr{d}}{\rr{d}z_2}\varrho(1-\varrho)=0
\;\;\;\textrm{ with }\;\;\;
b=\frac{2\delta(1-h)}{1-h}=2\delta
\end{equation}
with the Dirichlet boundary conditions
\begin{equation}
\label{againmf020}
\varrho(0)=\varrho_\textrm{u}
\;\;\;\textrm{ and }\;\;\;
\varrho(L_2+1)=\varrho_\textrm{d}
\;\;.
\end{equation}

\subsection{Stationary Mean Field behavior}
\label{s:stazMF}
\par\noindent
Finding the stationary profile $\varrho(z_2)$ means 
solving the ordinary differential equation 
\eqref{againmf040} with the Dirichlet boundary 
conditions \eqref{againmf020}.
In the case $\delta=0$ the solution is trivially linear.
We now discuss the case $\delta >0$. 
We integrate equation \eqref{againmf040} once with respect to the space 
variable from $0$ to $z_2$ to get 
\begin{equation}
\label{amf020}
\varrho'=b\varrho(1-\varrho)+c
\;\;\;\textrm{ where }\;\;\;
c=
\varrho'(0)-b\varrho_\textrm{u}(1-\varrho_\textrm{u})
\;\;.
\end{equation}

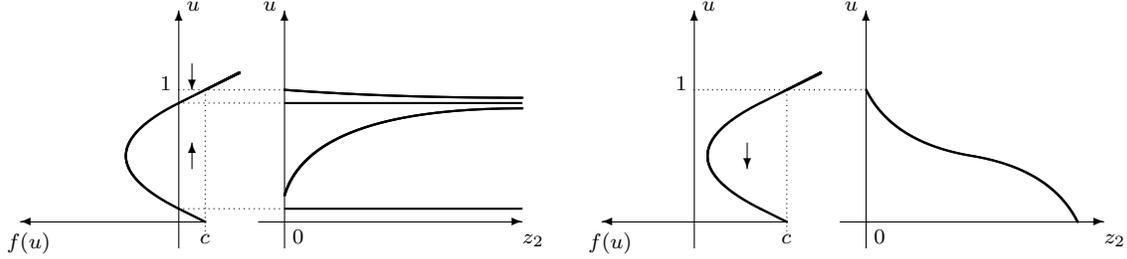
\begin{figure}[ht] 
\begin{picture}(100,80)(55,10) 
\thinlines 
\put(140,0){\vector(0,1){90}} 
\put(150,10){\vector(-1,0){70}} 
\thicklines 
\qbezier(140,15)(100,35)(140,55) 
\qbezier(140,55)(180,75)(150,60) 
\qbezier(140,15)(145,12.5)(150,10) 
\put(75,0){{\scriptsize{$f(u)$}}} 
\put(148,2){{\scriptsize{$c$}}} 
\put(133,60){{\scriptsize{$1$}}} 
\put(143,90){{\scriptsize{$u$}}} 
\thinlines 
\qbezier[21](140,60)(160,60)(180,60)
\qbezier[30](150,5)(150,30)(150,60)
\qbezier[21](140,55)(160,55)(180,55)
\qbezier[21](140,15)(160,15)(180,15)
\put(145,70){\vector(0,-1){10}} 
\put(145,30){\vector(0,1){10}} 
\thinlines 
\put(180,0){\vector(0,1){90}} 
\put(170,10){\vector(1,0){100}} 
\put(183,2){{\scriptsize{$0$}}} 
\put(270,2){{\scriptsize{$z_2$}}} 
\put(172,90){{\scriptsize{$u$}}} 
\thicklines 
\put(180,55){\line(1,0){90}} 
\put(180,15){\line(1,0){90}} 
\qbezier(180,20)(190,53)(270,53)
\qbezier(180,60)(220,57)(270,57)
\thinlines 
\put(335,0){\vector(0,1){90}} 
\put(370,10){\vector(-1,0){70}} 
\thicklines 
\qbezier(360,15)(320,35)(360,55) 
\qbezier(360,55)(400,75)(370,60) 
\qbezier(360,15)(365,12.5)(370,10) 
\put(295,0){{\scriptsize{$f(u)$}}} 
\put(368,2){{\scriptsize{$c$}}} 
\put(328,60){{\scriptsize{$1$}}} 
\put(338,90){{\scriptsize{$u$}}} 
\thinlines 
\qbezier[35](335,60)(365,60)(400,60)
\qbezier[30](370,5)(370,30)(370,60)
\put(355,40){\vector(0,-1){10}} 
\thinlines 
\put(400,0){\vector(0,1){90}} 
\put(390,10){\vector(1,0){100}} 
\put(403,2){{\scriptsize{$0$}}} 
\put(490,2){{\scriptsize{$z_2$}}} 
\put(392,90){{\scriptsize{$u$}}} 
\thicklines 
\qbezier(400,60)(410,40)(440,35)
\qbezier(440,35)(470,30)(480,10)
\end{picture} 
\vskip 0.5 cm 
\caption{Phase diagram corresponding to \eqref{againmf040}.
The case 
$c<0$ and $b/4+c>0$ is depicted on the left. The case $b/4+c<0$ 
is depicted on the right.}
\label{f:mf} 
\end{figure}

The structure of the 
solutions of this equation, namely the phase space trajectories,
can be studied via a simple qualitative 
analysis (see, for instance, \cite[Chapter 1]{Arnold}). 
Let $f(\varrho)=b\varrho(1-\varrho)+c$ be 
the right hand side of \eqref{amf020}. Five different 
situations have to be considered: 
$c>0$, $c=0$, $c<0$ and $b/4+c>0$, $b/4+c=0$, and $b/4+c<0$. 
In Figure~\ref{f:mf} the phase diagram in the extended 
phase space is shown in the two cases 
$c<0$ and $b/4+c>0$, $b/4+c<0$. 

Now, we find the solution of the stationary equation in the cases 
of interest. 
From the picture it is clear that: 
\begin{itemize}
\item[--]
if $1\ge\varrho_\textrm{u}>1/2>\varrho_\textrm{d}\ge0$ the constant $c$ 
has to be such that $b/4+c<0$;
\item[--]
if $1\ge\varrho_\textrm{u}>\varrho_\textrm{d}>1/2$ the constant $c$ 
has to be such that either $b/4+c<0$ or $0>c>-b/4$.
\end{itemize} 

It is important to remark that in the first case the constant 
$c$ has to be necessarily larger that $-b/4$, while in the second 
case there are two different possibilities, so that we will have to decide
which is the correct one. 

\par\noindent
\textit{Case $1\ge\varrho_\textrm{u}>1/2>\varrho_\textrm{d}\ge0$.\/}
Assume $b/4+c<0$, 
by performing a standard computation 
we find the solution 
\begin{equation}
\label{amf050}
\arctan\frac{\varrho(z_2)-1/2}{\sqrt{-c/b-1/4}}
=
\arctan\frac{\varrho_\rr{u}-1/2}{\sqrt{-c/b-1/4}}
-b\sqrt{-\frac{c}{b}-\frac{1}{4}} z_2
\end{equation}
with the constant $c$ given by 
\begin{equation}
\label{amf060}
\arctan\frac{\varrho_\rr{u}-1/2}{\sqrt{-c/b-1/4}}
-
\arctan\frac{\varrho_\rr{d}-1/2}{\sqrt{-c/b-1/4}}
=
b\sqrt{-\frac{c}{b}-\frac{1}{4}} (L_2+1)
\;\;.
\end{equation} 
It is not difficult to verify the following facts: 
as a function of $c\in[-\infty,-b/4]$ the left hand side 
of \eqref{amf060} is a monotonic function 
increasing from $0$ to $-\pi$. Hence, \eqref{amf060}
admits a unique solution in this case. 

We can then conclude that in such a case the stationary Mean Field 
equation has the unique solution given by 
\eqref{amf050} with the constant $c$ as in \eqref{amf060}.

\par\noindent
\textit{Case $1\ge\varrho_\textrm{u}>\varrho_\textrm{d}>1/2$.\/}
In this case,  
the left hand side of
\eqref{amf060} tends to zero for $c\to-b/4$, 
so that \eqref{amf060} has a solution 
provided
\begin{displaymath}
\lim_{c\to-b/4}
\frac{{\displaystyle
       \arctan\frac{\varrho_\rr{u}-1/2}{\sqrt{-c/b-1/4}}
       -
       \arctan\frac{\varrho_\rr{d}-1/2}{\sqrt{-c/b-1/4}}
     }}
     {{\displaystyle
       b\sqrt{-\frac{c}{b}-\frac{1}{4}} (L_2+1)
     }}
>1
\;\;.
\end{displaymath}
By computing the limit above, we get the condition 
\begin{equation}
\label{amf070}
\frac{\varrho_\textrm{u}-\varrho_\textrm{d}}
     {b(L_2+1)(\varrho_\textrm{u}-1/2)(\varrho_\textrm{d}-1/2)}
>1
\;\;.
\end{equation}

We recall, now, that in this case it is also possible to find a solution 
of the Mean Field equation \eqref{againmf040} with $0>c>-b/4$.
By a standard computation we find the solution 
\begin{equation}
\label{amf090}
\varrho(z_2)=\frac{{\displaystyle
                  u_2-u_1\exp\{-bz_2(u_2-u_1)
                    -\log[(\varrho_\textrm{u}-u_1)/(\varrho_\textrm{u}-u_2]\}
                }}
                {{\displaystyle
                  1-\exp\{-bz_2(u_2-u_1)
                    -\log[(\varrho_\textrm{u}-u_1)/(\varrho_\textrm{u}-u_2]\}
                }}
\;\;,
\end{equation}
where 
\begin{equation}
\label{amf100}
u_1=\frac{1-\sqrt{1+4c/b}}{2}
<
u_2=\frac{1+\sqrt{1+4c/b}}{2}
\;\;,
\end{equation}
where 
the constant $c$, hidden in the expressions of $u_1$ and $u_2$, 
can be obtained by requiring $u(L_2+1)=\varrho_\textrm{d}$, 
namely, 
\begin{equation}
\label{amf110}
\varrho_\textrm{d}=\frac{{\displaystyle
                  u_2-u_1\exp\{-b(L_2+1)(u_2-u_1)
                    -\log[(\varrho_\textrm{u}-u_1)/(\varrho_\textrm{u}-u_2]\}
                }}
                {{\displaystyle
                  1-\exp\{-b(L_2+1)(u_2-u_1)
                    -\log[(\varrho_\textrm{u}-u_1)/(\varrho_\textrm{u}-u_2]\}
                }}
\;\;.
\end{equation}

By computing the limit, for $c\to-b/4$ of the ratio on the right 
hand side of \eqref{amf110}, it is possible to show that 
such a limit is either larger or smaller than $\varrho_\textrm{d}$ 
if and only if the equality \eqref{amf070} is satisfied. 
This occurrence is related to the existence of solutions to 
\eqref{amf110}. 
Hence, we have that 
the solution of the stationary Mean Field equation is given by 
\eqref{amf050} provided \eqref{amf070} is satisfied, 
otherwise it is given by \eqref{amf090}.

\begin{figure}[t]
\begin{picture}(200,290)(-10,0)
\put(0,0)
{
\resizebox{14cm}{!}{\rotatebox{0}{\includegraphics{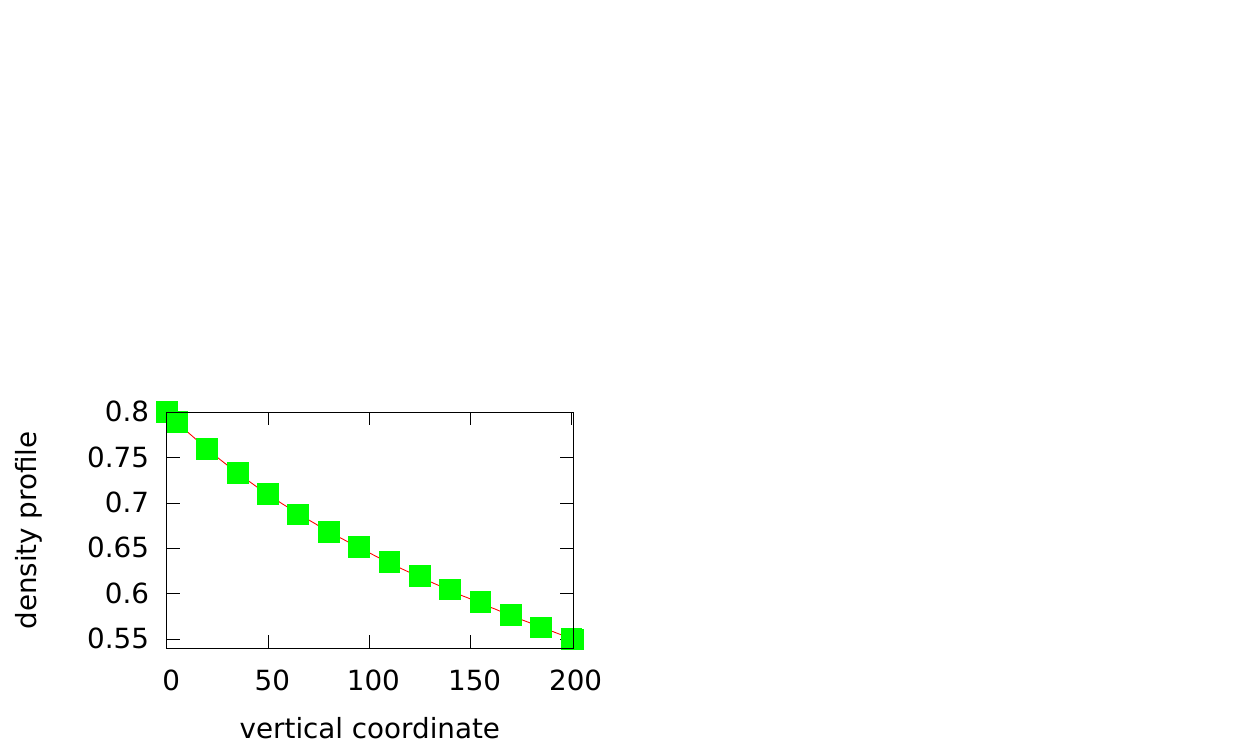}}} 
}
\put(235,0)
{
\resizebox{14cm}{!}{\rotatebox{0}{\includegraphics{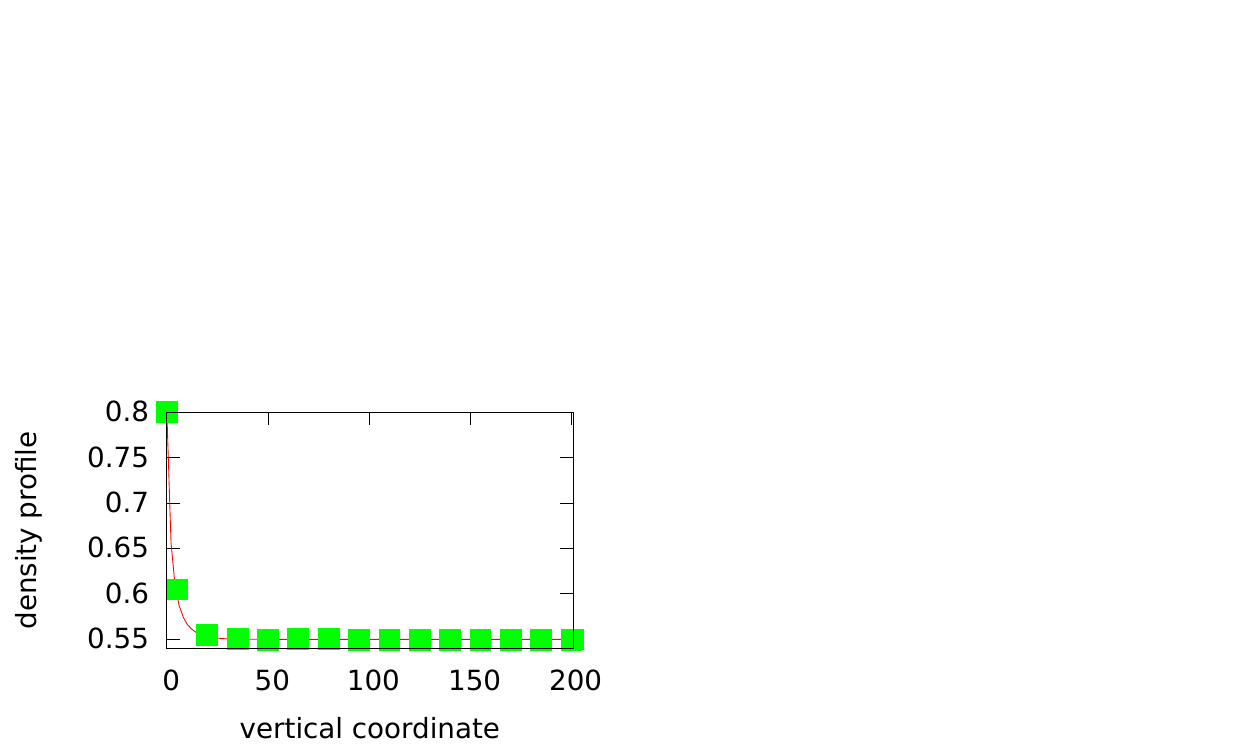}}} 
}
\put(0,150)
{
\resizebox{14cm}{!}{\rotatebox{0}{\includegraphics{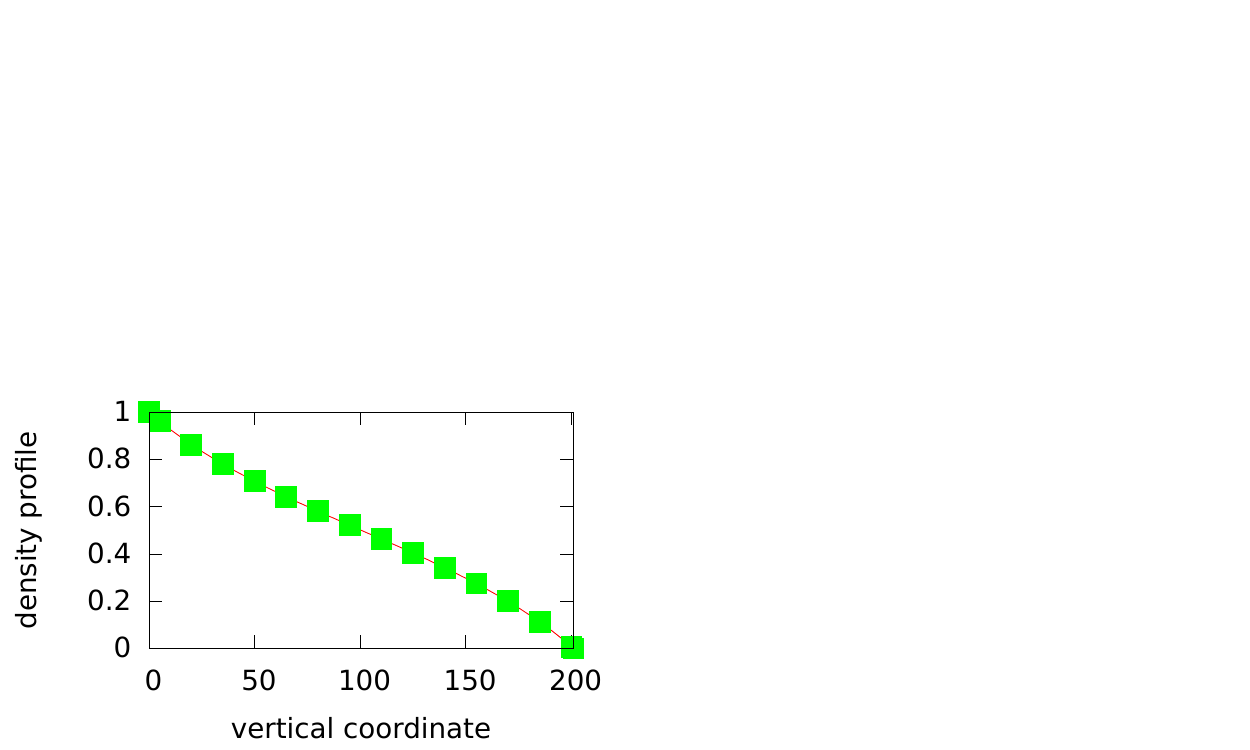}}} 
}
\put(235,150)
{
\resizebox{14cm}{!}{\rotatebox{0}{\includegraphics{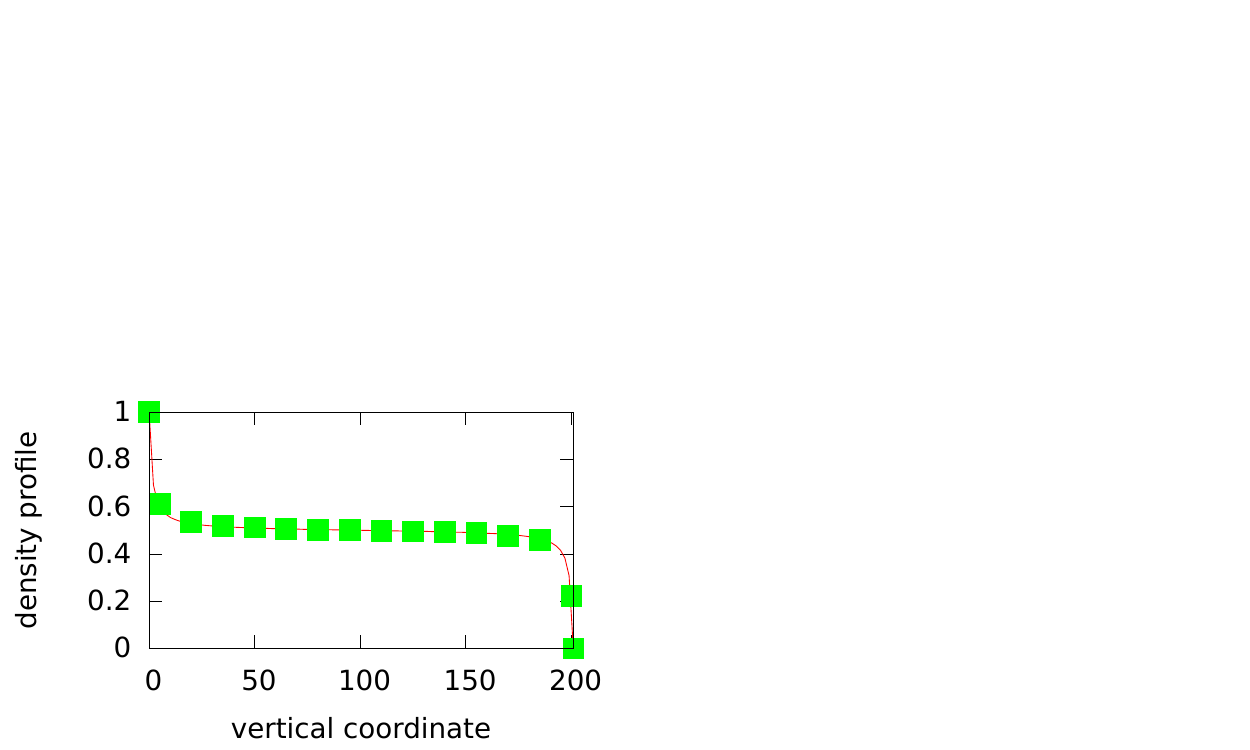}}} 
}
\end{picture}  
\caption{Density profiles. Comparison with numerical results (dots) and 
the Mean Field solution (solid lines). The cases described in 
Section~\ref{s:MFn} are considered:
(i) top left, (ii) top right, (iii) bottom left, and 
(iv) bottom right.
}
\label{f:dens-mf} 
\end{figure} 

\subsection{Numerical verification of the Mean Field prediction}
\label{s:MFn}
\par\noindent
We now test numerically how accurate is the Mean Field prediction of the 
stationary density profile.
To measure experimentally such a profile we proceed as follows. 
We chose two numbers $1\ll t_\textrm{term}\ll t_\textrm{max}$ that 
are called, respectively, \emph{thermalization} and \emph{maximal}
time. 
As an estimator for the density $\varrho(z_2)$ we use 
\begin{equation}
\label{mis030}
\frac{1}{L_1}
\frac{1}{t_\textrm{max}-t_\textrm{term}+1}
\sum_{s=t_\textrm{term}+1}^{t_\textrm{max}} 
\sum_{i=1}^{n(s)}
  \bb{I}_{\{x_2(i,s)=z_2\}}
\;\;,
\end{equation}
where $\bb{I}_{\{\textrm{condition}\}}$ is equal to one if the condition 
is true and to zero otherwise.

The thermalization and maximal time are chosen \emph{ad hoc} so that 
the measure is performed when the system is 
in the stationary state and so that 
the measure is sufficiently stable.

We choose the geometrical parameters $L_1=100$ and $L_2=200$
and the probabilistic one $h=1/2$. Then we vary the remaining 
ones according to the following four cases:
\begin{itemize}
\item[(i)]
$\varrho_\textrm{u}=1$, $\varrho_\textrm{d}=0$, and $\delta=0.008$:
the mean field solution is given by \eqref{amf050} with 
$c=-0.007887$;
\item[(ii)]
$\varrho_\textrm{u}=1$, $\varrho_\textrm{d}=0$, and $\delta=0.8$:
the mean field solution is given by \eqref{amf050} with 
$c=-0.400149$;
\item[(iii)]
$\varrho_\textrm{u}=0.8$, $\varrho_\textrm{d}=0.55$, and $\delta=0.008$:
since the first term of inequality \eqref{amf070} is equal to $5.18242$,
the mean field solution is given by \eqref{amf050} with 
$c=-0.00478572$;
\item[(iv)]
$\varrho_\textrm{u}=0.8$, $\varrho_\textrm{d}=0.55$, and $\delta=0.8$:
since the first term of inequality \eqref{amf070} is equal to $0.0518242$,
the mean field solution is given by \eqref{amf090} with 
$c=-0.396$.
\end{itemize}

The numerical simulations are performed with $t_\textrm{term}=10^5$
and $t_\textrm{max}=5\times10^5$ and the corresponding 
results are depicted in 
Figure~\ref{f:dens-mf}. In all the considered cases the match between the 
numerical data and the Mean Field prediction is strikingly good.

\begin{figure}[t]
\begin{picture}(200,290)(-10,0)
\put(0,0)
{
\resizebox{14cm}{!}{\rotatebox{0}{\includegraphics{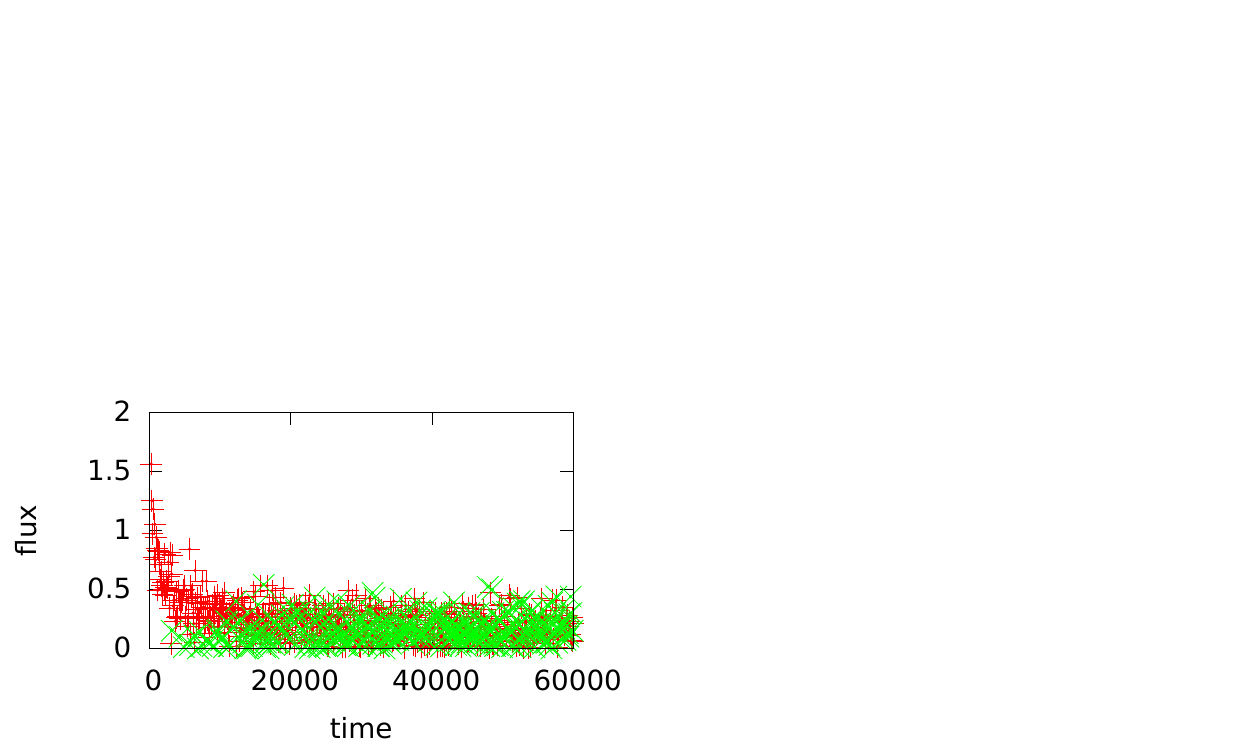}}} 
}
\put(235,0)
{
\resizebox{14cm}{!}{\rotatebox{0}{\includegraphics{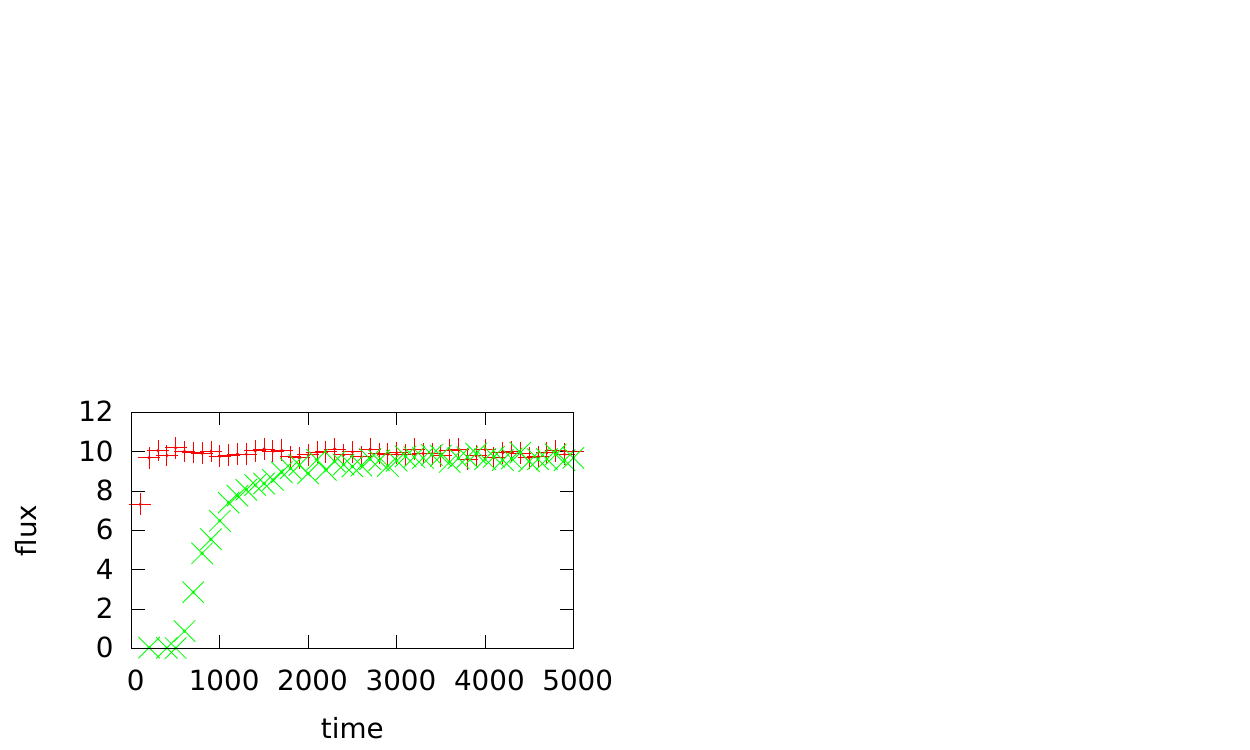}}} 
}
\put(0,150)
{
\resizebox{14cm}{!}{\rotatebox{0}{\includegraphics{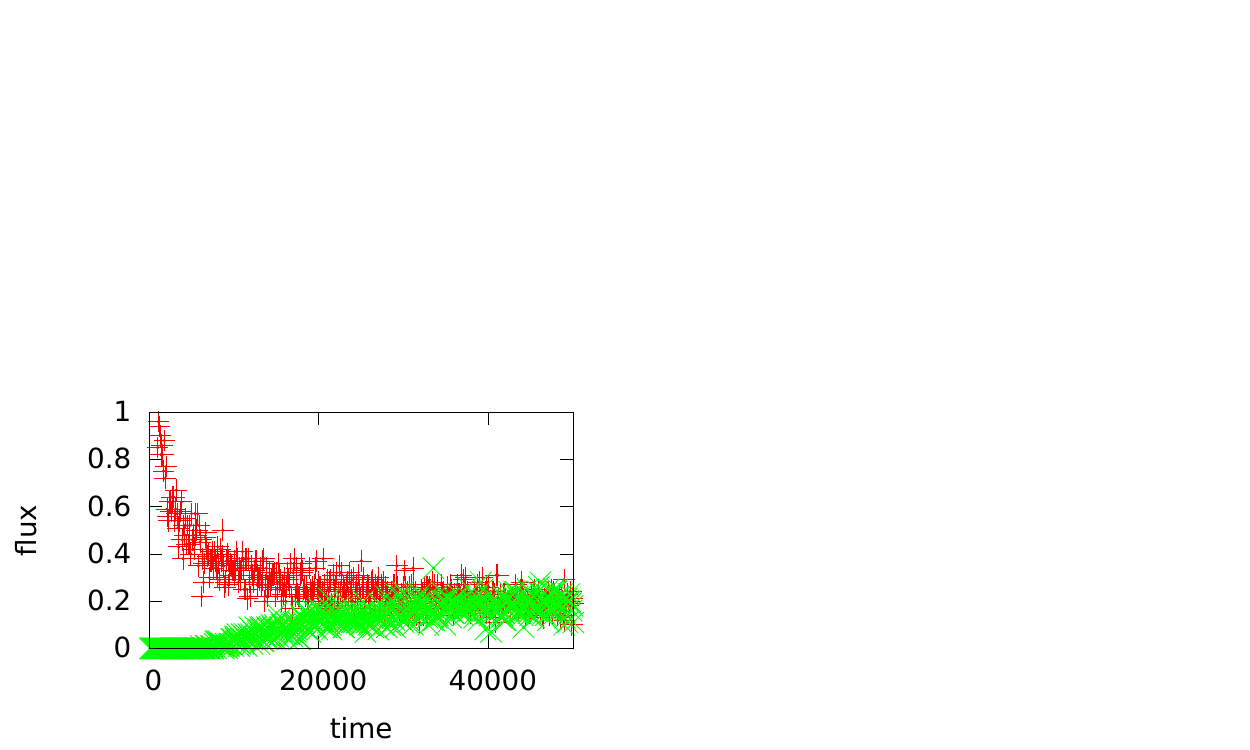}}} 
}
\put(235,150)
{
\resizebox{14cm}{!}{\rotatebox{0}{\includegraphics{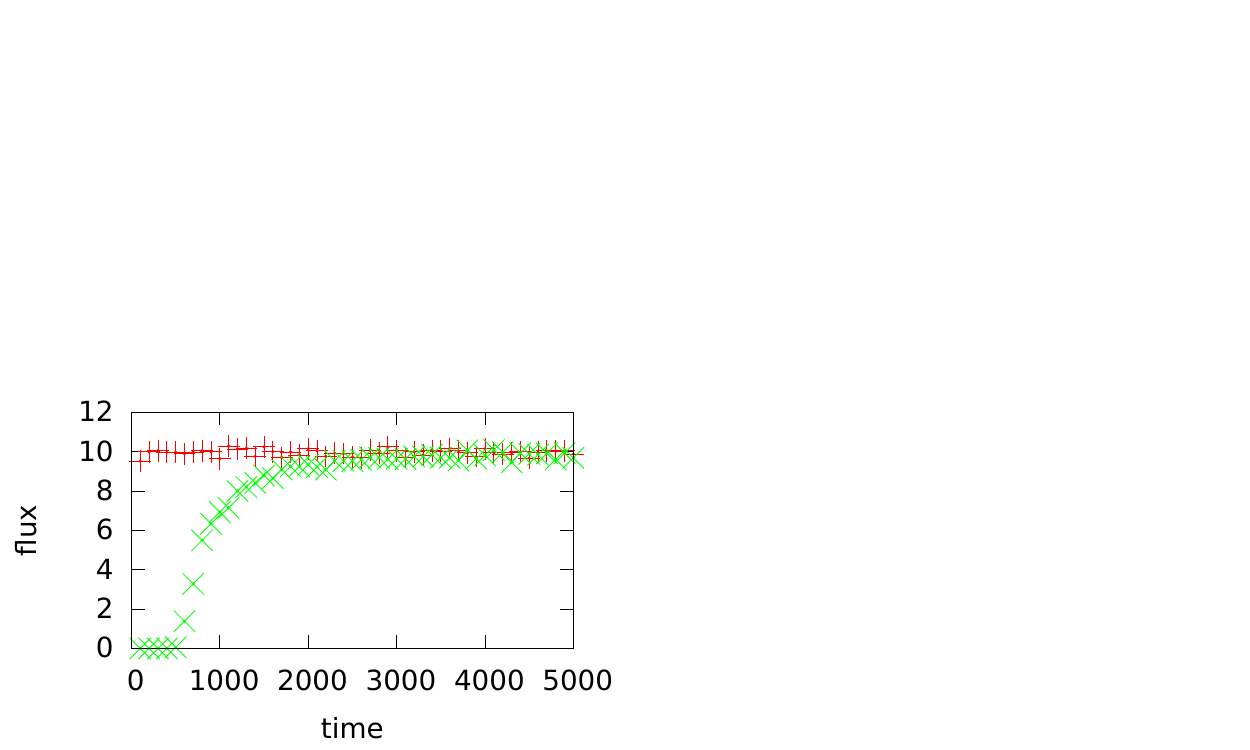}}} 
}
\end{picture}  
\caption{Numerical estimate of the ingoing ($+$) and outgoing ($\times$) 
flux for the four cases described in 
Section~\ref{s:MFn}:
(i) top left, (ii) top right, (iii) bottom left, and 
(iv) bottom right.
}
\label{f:flusso} 
\end{figure} 

\subsection{Thermalization time}
\label{s:term}
\par\noindent
Since the measuring of the density profile has to be performed 
in the stationary states, the 
choice of the thermalization time has to be done with care. 
One possibility to check if the system has reached stationarity is 
to compare the typical number of particles entering the system 
through the top boundary to that of the particles 
exiting from the bottom. 

We proceed as following:
to account for the number of particle crossing the top and 
bottom boundaries, we let 
$I(t)$ be the difference between the number of particles 
that entered at time $t$ through the 
top boundary and that of the particles that exited through the same 
boundary.
On the other hand, 
let
$O(t)$ be the difference between the number of particles 
that exited at time $t$ through the 
bottom boundary and that of the particles that entered through the same 
boundary.
Both $I(t)$ and $O(t)$ are stochastic variables that can be 
positive or negative. We can expect that, if a 
stationary state is observed, in such a state $I(t)$ and $O(t)$ 
will both 
fluctuate around the same value. 

We also define two additional quantities:
given the positive integer $T$, we let 
the \emph{local ingoing and outgoing} fluxes at time $t$ 
be as 
\begin{equation}
\label{mis010}
F^\rr{i}_T(t)=\frac{1}{T}\sum_{s=\max\{0,t-T\}}^t I(s)
\;\;\;\textrm{ and }\;\;\;
F^\rr{o}_T(t)=\frac{1}{T}\sum_{s=\max\{0,t-T\}}^t O(s)
\;\;,
\end{equation}
respectively.

In Figure~\ref{f:flusso} we plot the local ingoing and outgoing fluxes 
measured on the interval $T=100$ for the four cases (i)--(iv) 
considered above. 
The data shown in the figures indicate the choice 
$10^5$ for the thermalization time.

\section{A biased birth and death model}
\label{s:bd}
\par\noindent
The main physical problem we discuss in this paper
is that of the typical time that (at stationarity)
a particle spends
in the strip before finding its way to go out. 
We refer to this time as the 
\emph{residence time} and we 
will 
discuss its definition in detail in the next section.

Since from this point of view 
only vertical displacements are relevant, we can think 
to the ``vertical" history of the particle as to the 
evolution of a non--homogeneous birth and death process for the 
vertical coordinate with a rule depending on the 
stationary density profile. In this section we recall 
general results on this birth and death model and in the next section we 
discuss how to apply such results to our 2D model. 
We mainly follow \cite{BBF} for the general discussion. We 
derive explicit formulas in two specific simple cases, that 
will turn out to be very important from the physical point of view. 
For the sake of completeness we outline these  
computations in Subsections~\ref{s:cost} and \ref{s:lin}. 

Let $L$ be a positive integer, 
$[[0,L]]:=\{0,1,\dots,L\}$, and 
consider a random walk on $[[0,L]]$
with transition probabilities 
$p(x,y)$ with $x,y\in[[0,L]]$. 
We denote by $X_a(t)$, with $t=0,1,\dots$ the position of the 
walker at time $t$ with initial position $a$.
We assume that at each time the walker either does not move 
or moves to one of the neighboring sites, i.e., 
we assume $p(x,y)=0$ for $x,y\in[[0,L]]$ such that $|x-y|\ge2$. 
Moreover, 
we let 
$p_x=p(x,x+1)$ with $x=0,\dots,L-1$ be the probabilities 
to go to the right and 
$q_x=p(x,x-1)$ with $x=1,\dots,L$ be the probabilities 
to go to the left.
We assume $0\le p_{x-1}\le q_x<1$ for 
$x=1,\dots,L$, namely, at each site the walker has 
the chance to go both to the left and to the right and 
the walk is left--biased.
Furthermore, 
we assume that at least one of the ``rest probabilities"
$p(x,x)=1-(p_x+q_x)$ is different from zero so that the 
chain is aperiodic. 
We note that the more general situation in which the 
bias condition is removed can be treated as well, but 
for the sake of clarity, we consider the biased case. 

In the case $0< p_{x-1}\le q_x<1$ for $x=1,\dots,L$, 
since the birth and death process is aperiodic and 
positive recurrent, it has a unique invariant measure that can be 
written as 
\begin{equation}
\label{art040}
\pi(x)=\pi(0)\prod_{i=1}^x\frac{p_{i-1}}{q_i}
\;\;\textrm{ for all }x=1,\dots,L
\;\;,
\end{equation}
see \cite[equation~(4.1)]{BBF}. Note that the proof of the above statement 
is immediate, since one just has to verify that the above measure 
satisfies the detailed balance equation $\pi(x)p(x,x+1)=\pi(x+1)p(x+1,x)$
for all $x=1,\dots,L-1$.

In \cite{BBF} the authors study in detail the properties of the 
first hitting time of the chain to any point of the lattice $[[0,L]]$ 
with any initial condition (initial position of the walker). 
We are interested
only to the hitting time 
\begin{equation}
\label{hitting}
T:=\inf\{t\ge1,\,X_L(t)=0\}
\;\;,
\end{equation}
i.e.,
the random time that 
the walker started at $L$ needs to reach the origin for the first 
time. 
The expectation of such a hitting time 
is given by 
\begin{equation}
\label{art050}
\bb{E}[T]
=
\sum_{i=1}^{L}\frac{1}{q_i\pi(i)}\,\sum_{j=i}^{L}\pi(j)
=
\frac{1}{q_L}
+
\sum_{i=1}^{L-1}\frac{1}{q_i}\,
      \Big(1+\sum_{j=i+1}^{L}\prod_{k=i+1}^j\frac{p_{k-1}}{q_k}
      \Big)
\;\;,
\end{equation}
see \cite[equation~(4.3)]{BBF}.

The first very simple remark is that,
since the velocity of the 
particle is bounded by one, the mean hitting time to $0$ cannot be 
smaller than $L$. More precisely, by using \eqref{art050} 
and recalling that $q_x<1$ for $x=1,\dots,L$
we get the ballistic lower bound 
\begin{equation}
\label{hitlow}
\bb{E}[T]
\ge
L
\;\;.
\end{equation}

A natural question is under which assumptions on the bias there 
exists a ballistic upper bound to the first hitting time to zero. 
We prove this bound in a very simple case, namely, 
when we assume that each bond is left--biased. More precisely, 
we show that if there exists $0<\eta<1$ such that
$p_{x-1}/q_x\le\eta$ for $x=1,\dots,L$, then
\begin{equation}
\label{hitup}
\bb{E}[T]
\le
\frac{L}{\underline{q}(1-\eta)}
\end{equation}
where we have let $\underline{q}=\min\{q_1,\dots,q_L\}$.
Indeed, from \eqref{art050} we have that 
\begin{displaymath}
\bb{E}[T]
\le
\frac{1}{\underline{q}}
+
\sum_{i=1}^{L-1}\frac{1}{\underline{q}}\,
      \Big(1+\sum_{j=i+1}^{L}\prod_{k=i+1}^j\eta
      \Big)
=
\frac{1}{\underline{q}}
+
\frac{1}{\underline{q}}
\sum_{i=1}^{L-1}
      \sum_{k=0}^{L-i}\eta^k
\le
\frac{1}{\underline{q}}
+
\frac{1}{\underline{q}}
\sum_{i=1}^{L-1}
      \sum_{k=0}^\infty\eta^k
\end{displaymath}
where we have omitted few simple steps. The statement \eqref{hitup}
follows recalling that $\eta<1$.

Finally, we remark that using equations \eqref{art040} 
and \eqref{art050} we can 
compute the expected value of the first hitting time $T$. In practice 
this is explicitly feasible only for some particularly easy choices 
of the probabilities $p_x$ and $q_x$. In the next subsections 
we discuss two physically relevant cases. 

\subsection{Homogeneous case}
\label{s:cost}
\par\noindent
Let $0<p\le q<1$ be two real numbers and assume 
$p_x=p$ and $q_x=q$ for all $x=0,\dots,L$. 
Note that this choice satisfies all the basic assumptions on 
the birth and death chain. 

To compute the expected value of the first hitting time $T$ 
it is convenient to set $\lambda=p/q$, so that, from 
\eqref{art040}, we have
$\pi(x)=\pi(0)\lambda^x$.
Equation 
\eqref{art050} yields
\begin{displaymath}
\bb{E}[T]
=
\frac{1}{q}
\sum_{i=1}^{L} \frac{1}{\lambda^i} \,\sum_{j=i}^{L} \lambda^j
=
\frac{1}{q}
\sum_{i=1}^{L} \,\sum_{j=i}^{L} \lambda^{j-i}
=
\frac{1}{q}
\sum_{i=1}^{L} \,\sum_{k=0}^{L-i} \lambda^{k}
=
\frac{1}{q(1-\lambda)}
\sum_{i=1}^{L} (1-\lambda^{L-i+1})
\;\;.
\end{displaymath}
By reordering the sum, we obtain
\begin{displaymath}
\bb{E}[T]
=
\frac{1}{q(1-\lambda)}
\Big(L-\sum_{k=1}^{L} \lambda^{k}\Big)
=
\frac{1}{q(1-\lambda)}
\Big(L+1-\frac{1-\lambda^{L+1}}{1-\lambda}\Big)
=
\frac{L}{q(1-\lambda)}
-
\frac{\lambda(1-\lambda^L)}{q(1-\lambda)^2}
\;\;.
\end{displaymath}
Recalling $\lambda=p/q$, we finally have 
the expression 
\begin{equation}
\label{art110}
\bb{E}[T]
=
\frac{L}{q-p}-\frac{p}{(q-p)^2}\Big[1-\Big(\frac{p}{q}\Big)^L\Big]
\;\;.
\end{equation}

A physical comment is useful: 
if $p/q<1$, the behavior of the mean first hitting time
on $L$ is ballistic. 
On the other hand, we can prove that
\begin{displaymath}
\lim_{p\to q} \bb{E}[T]=\frac{L(L+1)}{2q}
\end{displaymath}
Hence, in the symmetric limiting case the 
diffusive dependence 
of the mean hitting time on the length $L$ is found. 
Note that the totally asymmetric limit $p\to0$ will be considered in 
Section~\ref{s:tot} and the ballistic scaling will be found. 

\subsection{A linear case}
\label{s:lin}
\par\noindent
We now consider a case in which the transition probabilities 
$q_x$ and $p_x$ decrease linearly in the interval $[[0,L]]$.
The physical interest of the peculiar choice we shall do will 
be clear in the next section. 
Let $A>0$ and take
\begin{equation}
\label{lin00}
q_x=2A+A(L-x)
\;\;\;\textrm{ and }\;\;\;
p_x=A(L-x)
\;\;.
\end{equation}

By using \eqref{art040} for the stationary measure,
we find that 
\begin{displaymath}
\pi(x)=
       \pi(0)\,
       \frac{L}{L+1}\,
       \frac{L-1}{L}\,
       \frac{L-2}{L-1}\,\cdots\cdots
       \frac{L-x+1}{L-x+2}
=
 \pi(0)\,
 \frac{L-x+1}{L+1}
\;\;.
\end{displaymath}
By \eqref{art050}, we get 
\begin{displaymath}
\bb{E}[T]
=
\frac{1}{A}
\sum_{i=1}^{L}\frac{1}{L+2-i}\,\frac{1}{L+1-i}
\sum_{j=i}^{L}(L+1-j)
\;\;.
\end{displaymath}
Since, it holds that  
\begin{displaymath}
\sum_{j=i}^{L}(L+1-j)
=
\frac{1}{2}(L+2-i)(L-i+1)
\;\;,
\end{displaymath}
we finally get
\begin{equation}
\label{lin040}
\bb{E}[T]
=
\frac{L}{2A}
\;\;.
\end{equation}
It is worth noting that, 
if $A$ is a constant then the scaling is ballistic. But
if $A$ is small with $L$, then we can possibly expect to have 
a diffusive scaling. 

\subsection{The totally asymmetric case}
\label{s:tot}
\par\noindent
A situation that will be useful in our discussion and that is not 
included in the results discussed above is the case 
$0= p_{x-1}<q_x<1$ for $x=1,\dots,L$, which we refer as 
the \textit{totally asymmetric} case. In such a case, by using the 
same strategy of proof as in \cite{BBF}, it is easy to show that
\begin{equation} 
\label{hitting-tot}
\bb{E}[T]
=
\sum_{i=1}^{L}\frac{1}{q_i}
\;\;.
\end{equation} 

The dependence of the mean hitting time to zero on the length 
of the system is, in this case, trivially ballistic. Indeed, 
from \eqref{hitting-tot} we get the bounds 
\begin{equation}
\label{hittot}
\bb{E}[T]\ge\frac{L}{\bar{q}}
\;\;\textrm{ and }\;\;
\bb{E}[T]\le\frac{L}{\underline{q}}
\;\;,
\end{equation} 
where we have set 
$\underline{q}=\min\{q_1,\dots,q_L\}$ and 
$\bar{q}=\max\{q_1,\dots,q_L\}$.
Note that in the totally asymmetric homogeneous case, 
that is to say, $0<q_x=q<1$ for $x=1,\dots,L$, we get $\bb{E}[T]=L/q$.

\section{Residence time at stationarity}
\label{s:res}
\par\noindent
The main question we pose in 
this paper is to find estimates for 
the typical time spent by a particle 
in the strip before exiting through the bottom boundary. 

To give a precise definition of such a time for the 
simple exclusion model on the strip defined 
in Section~\ref{s:modello} we 
set $\varrho_\rr{u}=1$ so that particles cannot exit the 
strip through the top boundary. 
Once the stationary state is reached, we look at the 
particles that enter from the top boundary and exit from the 
bottom one. 
We count after how many steps of the dynamics the particle exits 
from the bottom boundary and call \textit{residence time} the average 
of such a time over all the particles entered in the system 
after the stationary state is reached. 
Note that, at each step of the dynamics, a number of particles 
equal to the number of particles in the system at the end 
of the preceding time step is tentatively moved.


Despite its evident physical interest, the residence time 
is quite a difficult object to treat mathematically. 
In this section we discuss a microscopic approach in which we follow 
the motion of a single particle in a stationary density profile 
that is treated as a fixed background. In Section~\ref{s:res-mf}
we shall treat the problem macroscopically, by using the Mean Field
theory. 

Recall that, at stationarity, the average density 
profile is given by a function that we have denoted by 
$\varrho(z_2)$. Making a thought experiment,  
imagine that a new particle is injected into the stationary system 
through the top boundary.
To estimate the typical time this particle needs to find its way 
out through the bottom boundary we note that 
only vertical displacements are relevant.
Moreover,  we can think 
of the ``vertical" history of the particle as to the 
evolution of the not homogeneous birth and death process 
defined in Section~\ref{s:bd}
with the peculiar choice of the jump probabilities $p_x$ and $q_x$ 
that will be discussed below.  
We let $L=L_2$ and
imagine that 
the value $x=0$ of the birth and death process represents the particle 
at the row $L_2+1$ of the lattice (bottom boundary), 
the value $x=L_2$ represents the particle 
at the row $1$ of the lattice (the row close to the top boundary), 
and 
the generic value $x$ represents the particle 
at the row $L_2+1-x$ of the lattice.
The only not zero transition probabilities of the birth and death 
process are chosen as  
\begin{equation}
\label{art010}
\left\{
\begin{array}{ll}
{\displaystyle
 q_x
 =\frac{1-h}{2}(1+\delta)[1-\varrho(L_2+1-x+1)]
}
&
\;\;\;\textrm{ for }
x=1,\dots,L_2
\vphantom{\Bigg\{_\}}
\\
{\displaystyle
 p_x
 =\frac{1-h}{2}(1-\delta)[1-\varrho(L_2+1-x-1)]
}
&
\;\;\;\textrm{ for }
x=0,\dots,L_2-1
\\
\end{array}
\right.
\;\;.
\end{equation}
Indeed, 
recalling the expressions \eqref{againmf00-2} for 
the probabilities $u$ and $d$,
the prefactor $(1-h)(1+\delta)/2$ 
in the expression of $q_x$ 
is nothing but the probability $d$ to move the particle in the 
real space downwards; while the second factor is nothing 
but the probability that, at stationarity, the site where 
the particle tries to jump is empty. 
The expression for $p_x$ can be justified similarly. The rest probability $1-(p_x+q_x)$ is nonzero as the particle can stay at the same site or perform a unit step in the horizontal direction. 

We now propose a conjecture for the residence 
time of the exclusion model and we will test it numerically 
in the next section.

\begin{guess}
The residence time\footnote{Since in the real model a particle at the 
fictitious row $L_2+1$ cannot jump back to the real row $L_2$, 
one should set the probability $p(0,1)=0$. This would be a problem 
for the birth and death process, where all the jumping probabilities 
has to be assumed strictly positive, but for our purpose it is 
not necessary, since we are only interested to the first hitting time 
to $0$, so that when the 1D birth and death process 
reaches such a state  it is stopped. Hence all our results will not 
depend on the choice of the probability $p(0,1)$.}
of the model introduced in Section~\ref{s:modello} is equal to 
the mean value of the first hitting time to $0$ for the 
birth and death process started at $L_2$. 
\end{guess}

The main properties of birth and death processes have been 
recalled in Section~\ref{s:bd}. 
Those results and the conjecture based on the above analogy 
suggest that, for $\delta<1$,  
the residence time is given by equation \eqref{art050} where
the stationary measure $\pi$ is given by \eqref{art040} with 
the jump probabilities $p_i$ and $q_i$ defined in \eqref{art010}.
Since the stationary density profile is given with great accuracy by 
the stationary Mean Field equation, we can use 
these equations to give an estimate of the residence time of the model. 
It is worth noting that, since the expression of the 
stationary density profile is rather complicated, see \eqref{amf050}
and \eqref{amf090}, it will not be possible to derive in general explicit 
formulas for the residence time. On the other hand, since 
in \eqref{art050} and \eqref{art040} only finite product and 
sums are involved, we will be able to 
compute estimates for the residence time 
numerically for any values of the parameters of the model.
In the case $\delta=1$ the expression \eqref{hitting-tot} should 
be used. 

Let us now discuss three simple cases in which explicit expressions
of the residence time can be derived explicitly.

\subsection{Totally vertically asymmetric model}
\label{s:td}
\par\noindent
Recall we assumed $\varrho_\rr{u}=1$.
Consider, more, the case $\delta=1$. 
From the profiles in Figure~\ref{f:dens-mf} (graphs on the right) 
it is rather clear 
that in this situation the density profile is with very good approximation 
constant throughout the strip. 

Denote by $\bar\varrho$ such a constant value of the 
density profile. By \eqref{art010}, it follows that 
the birth and death model has the jump probabilities 
\begin{displaymath}
q_x=(1-h)(1-\bar\varrho)
\;\;x=1,\dots,L_2
\;\;\textrm{ and }\;\;
p_x=0
\;\;x=0,\dots,L_2-1
\end{displaymath}
Hence, by the results in Section~\ref{s:tot}, 
we have that the residence time is given by 
\begin{equation}
\label{td010}
R
=
\frac{L_2}{(1-h)(1-\bar\varrho)}
\;\;.
\end{equation}
In particular, 
the large $L_2$ behavior of the residence time given by \eqref{td010} 
is ballistic with a 
slope depending on the parameters $h$ and $\bar\varrho$.

\subsection{Large drift case}
\label{s:ld}
\par\noindent
Recall we assumed $\varrho_\rr{u}=1$.
Consider $\delta$ close to one. 
From the profiles in Figure~\ref{f:dens-mf} (graphs on the right) 
it is rather clear 
that in this situation the density profile is with very good approximation 
equal to a constant, denoted again by $\bar\varrho$. 
By \eqref{art010} it follows that 
the birth and death model 
has the jump probabilities 
\begin{displaymath}
q_x=\frac{1-h}{2}(1+\delta)(1-\bar\varrho)
\;\;x=1,\dots,L_2
\;\;\textrm{ and }\;\;
p_x=\frac{1-h}{2}(1-\delta)(1-\bar\varrho)
\;\;x=0,\dots,L_2-1
\end{displaymath}
Hence, by \eqref{art110}, we have that the residence time is given by 
\begin{equation}
\label{ld010}
R
=
\frac{L_2}{(1-h)(1-\bar\varrho)\delta}
-
\frac{1-\delta}{2(1-h)(1-\bar\varrho)\delta^2}
\Big[
     1-\Big(\frac{1-\delta}{1+\delta}\Big)^{L_2}
\Big]
\;\;.
\end{equation}
The formula \eqref{ld010} suggests that in this case 
the dependence of the residence time on $L_2$ is 
ballistic, but this statement needs more care. Indeed, 
the equation above is based on the assumption that 
the density profile is constant with good approximation 
in this regime and such an assumption is based on the 
results obtained via the Mean Field equation. But we have to recall 
that the Mean Field equation has been (not rigorously) 
derived in the large 
volume limit with the drift parameter scaling to zero  
with $L_2$. For this reason it is not correct, in principle, 
to fix $\delta$ and let $L_2\to\infty$ in the above formula. 

We remark, that in this case we shall as well be able to deduce 
the ballistic scaling of the residence time with $L_2$ 
in Section~\ref{s:scaling} via a different argument. 

\subsection{Zero drift}
\label{s:zd}
\par\noindent
Consider, now, the case $\delta=0$ (zero drift)
again for $\varrho_\rr{u}=1$.
In the absence of drift the stationary density 
profile is linear, hence
\begin{equation}
\label{zd000}
\varrho(z_2)=1-\frac{1-\varrho_\textrm{d}}{L_2+1}z_2
\end{equation}
for $z_2=0,\dots,L_2+1$.
By \eqref{art010}, we get also that 
\begin{displaymath}
q_x=\frac{1-h}{2}\frac{1-\varrho_\textrm{d}}{L_2+1}(L_2+2-x)
\;\;\;\textrm{ and }\;\;\;
p_x=\frac{1-h}{2}\frac{1-\varrho_\textrm{d}}{L_2+1}(L_2-x)
\;\;.
\end{displaymath}
Thus, by using the result \eqref{lin040} from Section~\ref{s:lin}, we find 
the residence time 
\begin{equation}
\label{zd020}
R=
\frac{L_2(L_2+1)}{(1-h)(1-\varrho_\textrm{d})}
\;\;.
\end{equation}
As expected, 
the large volume ($L_2\to\infty)$ behavior of the residence 
time is quadratic in 
the purely diffusive zero drift case.

\subsection{Dependence of the residence time on the length of the strip}
\label{s:scaling}
\par\noindent
In this section we focus on 
the dependence of the residence time 
on the length $L_2$ of the strip. 
We expect that for $\delta>0$, since there is a 
preferred direction in the movement, the particles 
will have a not zero average vertical velocity. As a 
consequence, we expect a ballistic behavior and a 
linear dependence of the residence time on $L_2$. 

We have already shown in Section~\ref{s:td} that this is indeed the 
case for $\delta=1$. Assume, now, that the drift $\delta$ is fixed and 
$0<\delta<1$. From \eqref{art010}, we see
\begin{displaymath}
\frac{p_{x-1}}{q_x}
=
\frac{1-\delta}{1+\delta}
\frac{1-\varrho(L_2+1-x)}{1-\varrho(L_2+1-x+1)}
\end{displaymath}
for any $x=1,\dots,L_2$.
Since, $\varrho_\rr{u}=1>\varrho_\rr{d}\ge0$
we can reasonably assume that the density profile is a decreasing 
function of the vertical spatial coordinate. Thus, 
$\varrho(L_2+1-x)>\varrho(L_2+1-x+1)$ implies that 
there exists $\eta<1$ such that 
$p_{x-1}/q_x\le\eta$.
This remark and \eqref{hitup} ensure that, for any positive 
finite $\delta$, the residence 
time has a ballistic dependence on the length of the strip. 

Finally, for $\delta=0$ we have shown in Section~\ref{s:zd}, cf.\
\eqref{zd020}, that the residence time is quadratic 
in $L_2$ (diffusive scaling). We stress that this 
result is not trivial at all. Indeed, even in the case 
$\delta=0$ the birth and death model that we use to 
investigate the property of the residence time is not 
symmetric. The lack of symmetry is due to 
\begin{displaymath} 
q_x-p_{x-1}
=
\frac{1-h}{2}\,
\frac{1-\varrho_\rr{d}}{L_2+1}
\end{displaymath} 
for any $x=1,\dots,L_2$. The diffusive scaling is a consequence of 
the fact that this difference vanishes as $1/(L_2+1)$. 

\subsection{Single file regime}
\label{s:single}
\par\noindent
The model we study is two-dimensional. Particles move 
in a strip from the top boundary towards the bottom one due 
to the presence of a drift $\delta$ or just because 
of a vertical biased diffusion due to the difference between 
the boundary densities on the top and on the bottom 
end of the strip. The particles are subjected 
also to horizontal displacements whose probability 
is controlled by the parameter $h$. 

In the particular case $h=0$, our model becomes 1D
and particles move, each on its vertical line, as in a single 
file system. In other words in this case our model 
reduces to the 1D simple exclusion 
model with open boundaries. The specification 
``open boundaries" means that the system is finite 
and at the boundaries there is a rule prescribing the 
rate at which particles can either enter or leave the system. 

This model has been widely studied, see for instance \cite{DEHP}
for the seminal paper where the model was solved exactly
in the totally asymmetric case. See, also, 
\cite{BE} for a general review. 
Our results can be compared easily to those in \cite{DEHP} which 
refer to the totally asymmetric case, namely, for our case $\delta=1$. 
In \cite{DEHP} one computes the stationary current 
$J$, namely, the number of particles that for unit of time 
crosses one bond at stationarity. With our notation 
one finds
\begin{equation}
\label{derrida}
J=
\left\{
\begin{array}{ll}
1/4 
& 
\;\;\;\;\;\textrm{ for } \varrho_\rr{d}\le 1/2\\
\varrho_\rr{d}(1-\varrho_\rr{d})
& 
\;\;\;\;\;\textrm{ for } \varrho_\rr{d}> 1/2\\
\end{array}
\right.
\end{equation}
in the case $\varrho_\rr{u}=1$ (see \cite[equations (58) and (60)]{DEHP}). 

Now, since in the totally asymmetric case the stationary density 
throughout the system is equal to a constant $\bar\varrho$, 
we can write 
$J\approx\bar\varrho L_2/R$. Hence, we have that
\begin{displaymath}
R\approx\bar\varrho\,\frac{L_2}{J}
\end{displaymath}
which reduces to our result \eqref{td010} with $h=0$
once we use \eqref{derrida} and notice that 
$\bar\varrho=1/2$ for $\varrho_\rr{d}\le1/2$ and 
$\bar\varrho=\varrho_\rr{d}$ for $\varrho_\rr{d}>1/2$.

Additionally, 
we mention here an interesting result which is valid for the symmetric 
simple exclusion model in 1D on the whole line (no boundary).
In both \cite{fedders} and \cite{DF}, the authors compute 
the variance of the position of a tagged particle and they 
prove that it is proportional to the square root of time, meaning that 
the typical distance spanned by the (symmetric) 
walker is proportional to $t^{1/4}$.
In our problem we do not find any $L_2^4$ scaling for the 
residence time even in the single--file regime. Indeed, 
we think that there is no direct connection between the two problems. 
One important point is that the results in \cite{fedders,DF} 
deal with a simple exclusion model without drift (symmetric)
and without boundaries. In our problem, even in the zero drift 
($\delta=0$) case, a net flux is present due to the boundary conditions. 

\subsection{Mean Field approximation for the residence time}
\label{s:res-mf}
\par\noindent
In this section we approach the residence time computation 
from a macroscopic point of view. In the Mean Field approximation
the evolution of the systems can be described in terms 
of the density profile $m_t(z_1,z_2)$ evolving according 
to the Partial Differential Equation \eqref{againmf000}. 
Such an equation can be tought of as a continuity equation 
for the current two--dimensional vector 
\begin{displaymath}
\vec{J}_t=-\frac{1}{2}h\frac{\partial m_t}{\partial z_1}\vec{e}_1
        +\Big(
           -\frac{1}{2}(1-h)\frac{\partial m_t}{\partial z_2}
           +\delta(1-h)m_t(1-m_t)
         \Big)\vec{e}_2
\end{displaymath}
where $\vec{e}_1$ and $\vec{e}_2$ are the unitary vectors of the 
lattice.

At stationarity, the density profile $\varrho(z_2)$ does not depend on 
the horizontal coordinate, so that the current is 
parallel to the vertical direction and its intensity is given by
\begin{equation}
\label{res-mf000}
J= -\frac{1}{2}(1-h)\frac{\partial \varrho}{\partial z_2}(z_2)
           +\delta(1-h)\varrho(z_2)[1-\varrho(z_2)]
      =-\frac{1}{2}(1-h)\varrho'(0)
\end{equation}
where we have used \eqref{againmf040}, \eqref{amf020}, and the fact that 
$\varrho_\rr{u}=1$.

As we have already done in Section~\ref{s:single}, we can 
relate the stationary current to the residence time. 
Indeed, it is quite reasonable to assume that, at stationarity, 
the average velocity of a particle $v(z_2)$ is such that 
$\varrho(z_2)v(z_2)=J_2$. Then, an easy integration gives
\begin{equation}
\label{res-mf010}
R=
\int_0^{L_2+1}\frac{\varrho(z_2)}{J}\,\rr{d}z_2
=
-\frac{2}{(1-h)\varrho'(0)}
\int_0^{L_2+1}\varrho(z_2)\,\rr{d}z_2
\end{equation}

In general, we cannot use the above equation to write an explicit 
expression of the residence time in terms of the parameter of the model, 
indeed the quantity $\varrho'(0)$ is nothing but the constant 
$c$ in equation \eqref{amf020} (recall $\varrho_\rr{u}=1$) which 
is, in turn, the solution of either the equation 
\eqref{amf060} ot \eqref{amf110}. 

But in the zero drift case, i.e., 
$\delta=0$, the density profile is the linear function 
$\varrho(z_2)=1-(1-\varrho_\rr{d})z_2/(L_2+1)$ for $z_2\in[0,L_2+1]$.
Hence an easy computation yields 
\begin{equation}
\label{res-mf020}
R=\frac{1}{(1-h)}\frac{1+\varrho_\rr{d}}{1-\varrho_\rr{d}}(L_2+1)^2.
\end{equation}
This formula has to be compared to the prediction \eqref{zd020} 
for the residence time in the zero drift case obtained in the 
framework of the birth and death model approximation. 

The mean field residence time expression (\ref{res-mf020}) can be considered the residence time according to the birth and death model corrected by  a term proportional to $\frac{\varrho_\rr{d}}{1-\varrho_\rr{d}}$. When $\varrho_\rr{d}$ is zero both expressions are the same. The correction can be understood is due to the correlated motion  in the one dimensional bouncing back  case. Fedders \cite{fedders} found that the diffusion constant  has to be corrected by a similar term in the stationary case.
We will see that expression  (\ref{res-mf020}) only agrees well the Monte Carlo  simulations  as long as $h=0$. On the other hand, the expression (\ref{zd020}) applies rigorously as long as $\delta=0$.

\section{Numerical estimates of the residence time}
\label{s:res-num}
\par\noindent

In this section we test numerically the validity of the analytic
computations developed in 
Section~\ref{s:res}. We choose the 
parameters of the model as follows: take $\varrho_\rr{u}=1$, 
$L_1=100$, and consider the cases 
\begin{displaymath}
L_2=100,200,\;\;\;
\delta=0,0.2,0.4,0.6,0.8,1,\;\;
\varrho_\rr{d}=0,0.4,0.8,\;\;
h=0,0.4,0.8.
\end{displaymath}

\begin{figure}[h]
\begin{picture}(200,125)(0,10)
\put(0,0)
{
\resizebox{10cm}{!}{\rotatebox{0}{\includegraphics{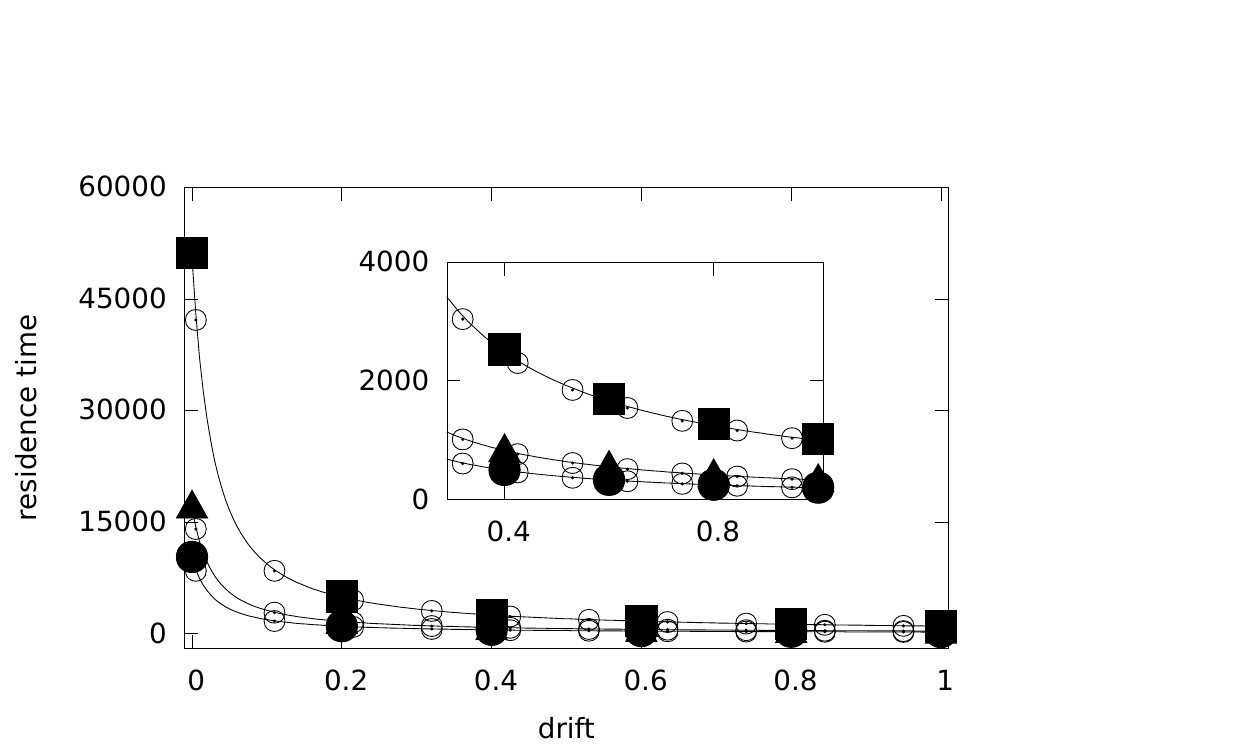}}} 
}
\put(235,0)
{
\resizebox{10cm}{!}{\rotatebox{0}{\includegraphics{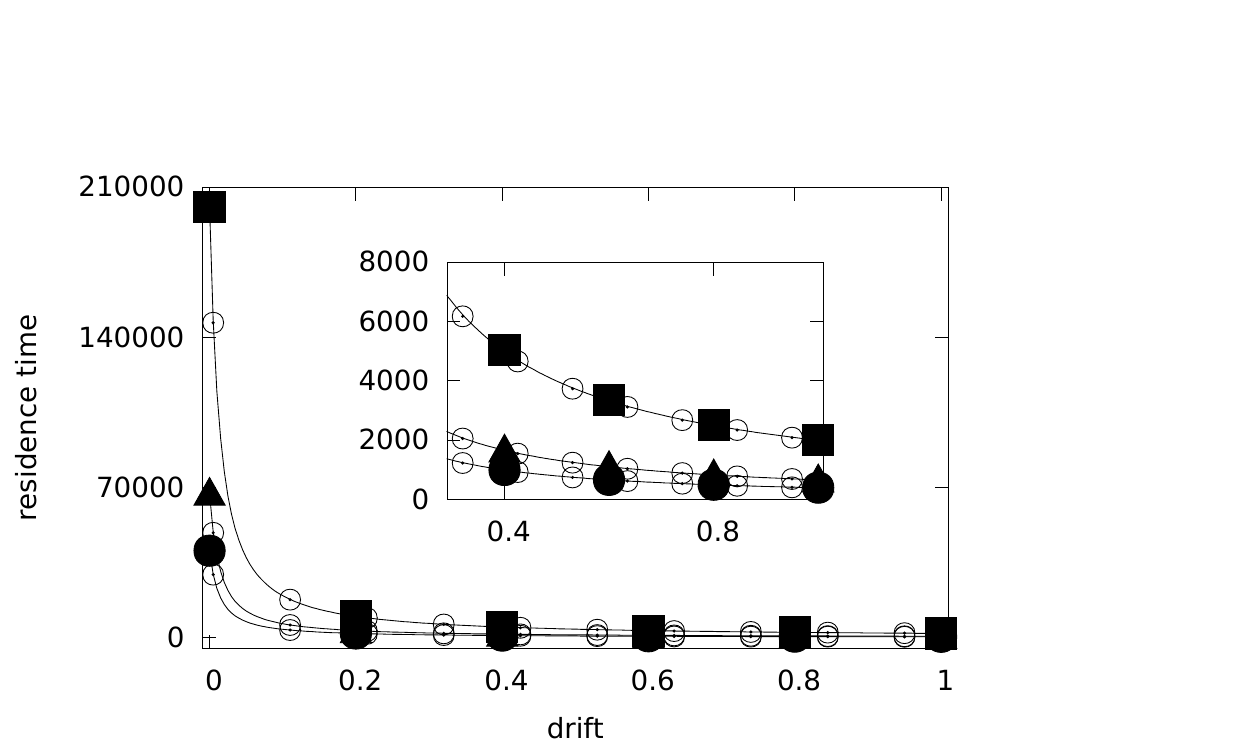}}} 
}
\end{picture}  
\caption{Residence time versus drift in the cases 
$\varrho_\rr{d}=0$ and $L_2=100$ (left) 
and $L_2=200$ (right).
The symbols $\bullet$, $\blacktriangle$, and $\blacksquare$
refer, respectively, to the cases $h=0,0.4,0.8$. 
The inset in the left picture is just a zoom 
of part of the data of the same picture. 
Solid lines are the birth and death theoretical prediction. 
Open circles are the Mean Field theoretical prediction \eqref{res-mf010}. 
}
\label{f:res-dr-100} 
\end{figure} 

\begin{figure}[h]
\begin{picture}(200,125)(0,10)
\put(0,0)
{
\resizebox{10cm}{!}{\rotatebox{0}{\includegraphics{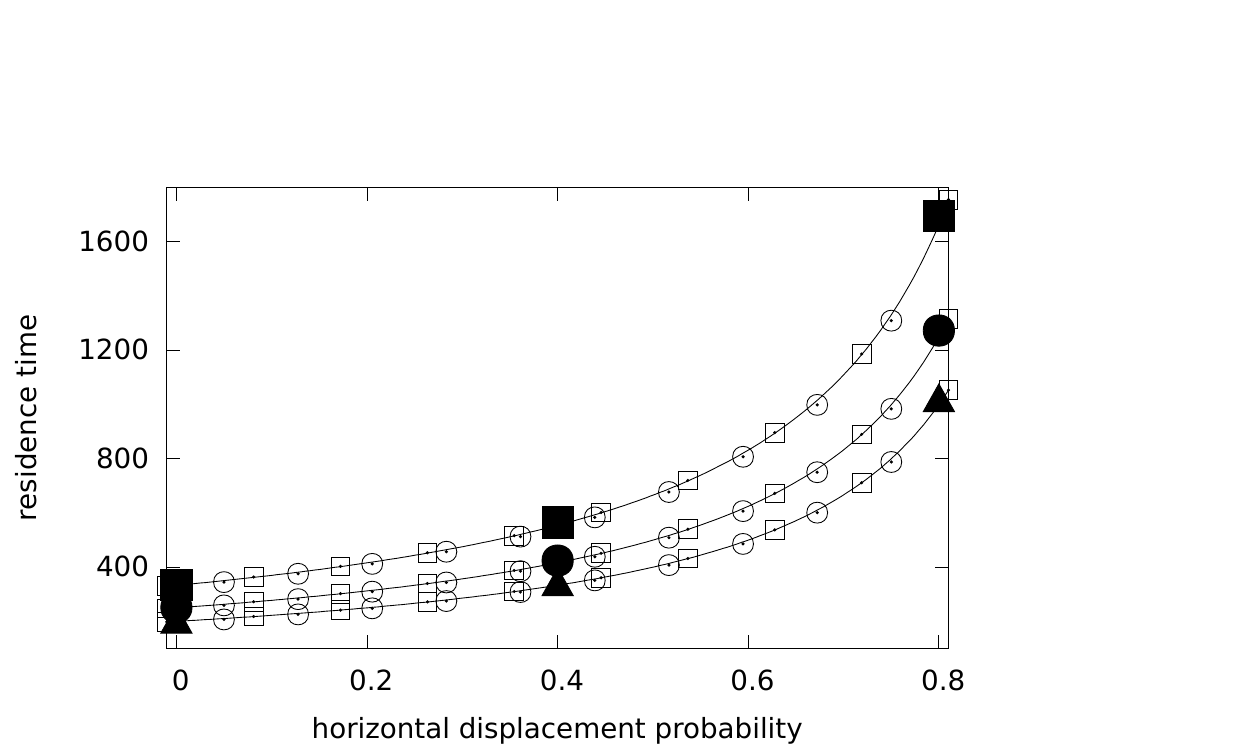}}} 
}
\put(235,0)
{
\resizebox{10cm}{!}{\rotatebox{0}{\includegraphics{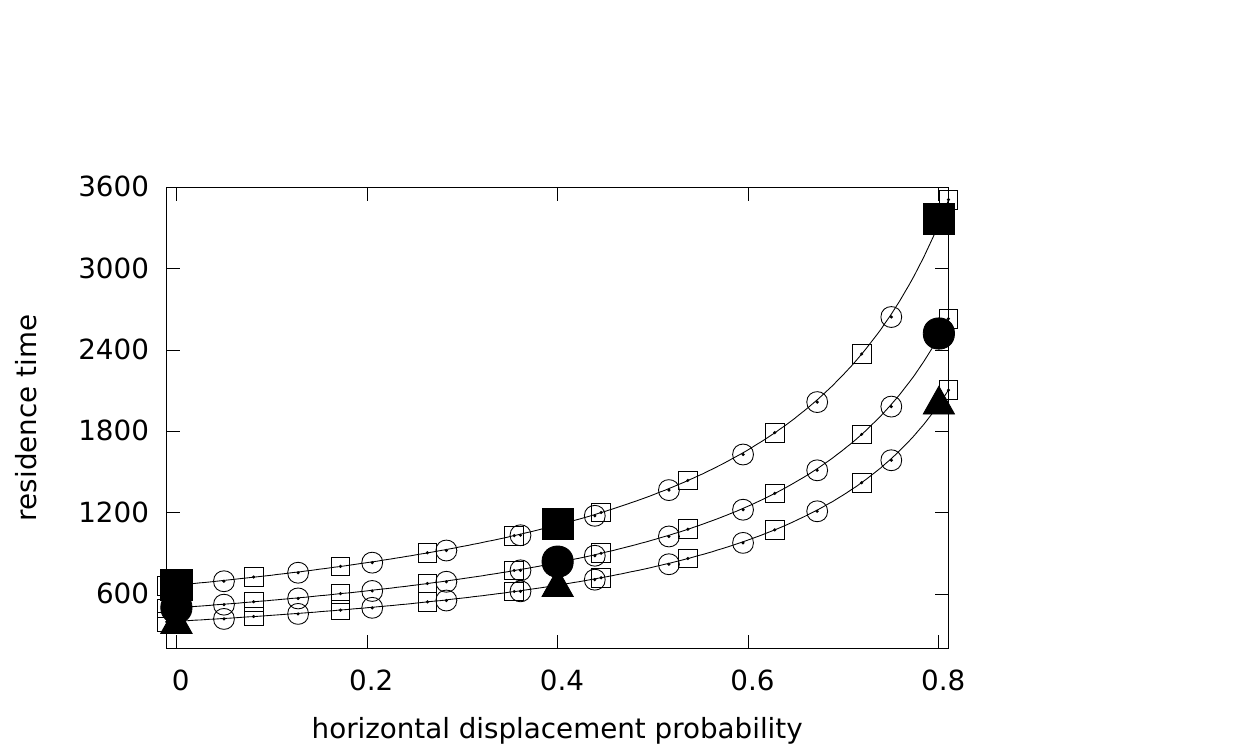}}} 
}
\end{picture}  
\caption{Residence time versus the horizontal displacement probability 
$h$ in the cases $\varrho_\rr{d}=0$ and $L_2=100$ (left) 
and $L_2=200$ (right).
The symbols $\blacksquare$, $\bullet$, and $\blacktriangle$
refer, respectively, to the cases $\delta=0.6,0.8,1$. 
Solid lines are the birth and death theoretical prediction. 
Open squares are the approximated theoretical prediction 
\eqref{ld010} with $\bar\varrho=1/2$, which is 
valid in the large drift regime.
Open circles are the Mean Field theoretical prediction \eqref{res-mf010}. 
}
\label{f:res-ph-100} 
\end{figure} 

\begin{figure}[h]
\begin{picture}(200,125)(0,10)
\put(0,0)
{
\resizebox{10cm}{!}{\rotatebox{0}{\includegraphics{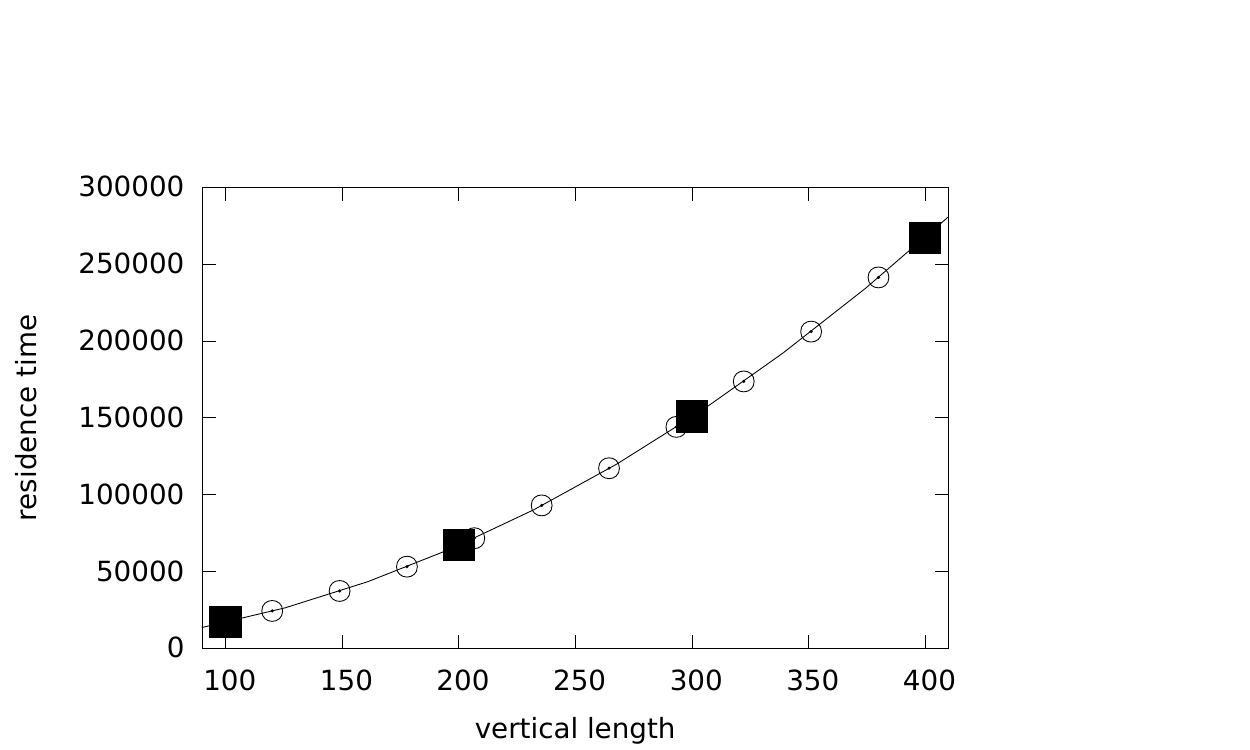}}} 
}
\put(235,0)
{
\resizebox{10cm}{!}{\rotatebox{0}{\includegraphics{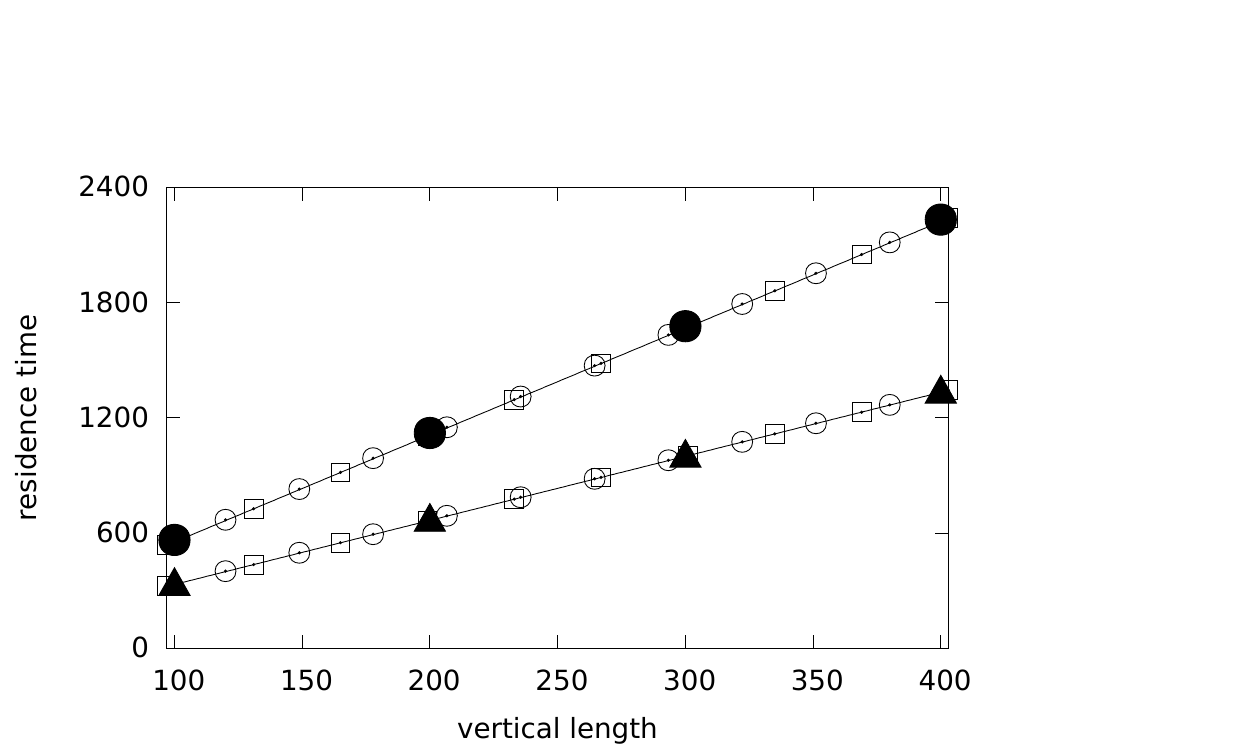}}} 
}
\end{picture}  
\caption{Residence time versus length of the strip $L_2$ 
in the case $\varrho_\rr{d}=0$ and $h=0.4$. 
The symbol $\blacksquare$ refers to the 
case $\delta=0$ (left picture) while 
the symbols $\bullet$ and $\blacktriangle$
refer, respectively, to the cases $\delta=0.6,1$. 
Solid lines are the theoretical prediction based on the birth and death 
model;
in the picture in the left the explicit expression \eqref{zd020} is used, 
Open squares are the approximated theoretical prediction 
\eqref{ld010} with $\bar\varrho=1/2$, which is 
valid in the large drift regime.
Open circles are the Mean Field theoretical prediction \eqref{res-mf010}. 
}
\label{f:res-el} 
\end{figure} 

\begin{figure}[h]
\begin{picture}(200,125)(0,10)
\put(0,0)
{
\resizebox{10cm}{!}{\rotatebox{0}{\includegraphics{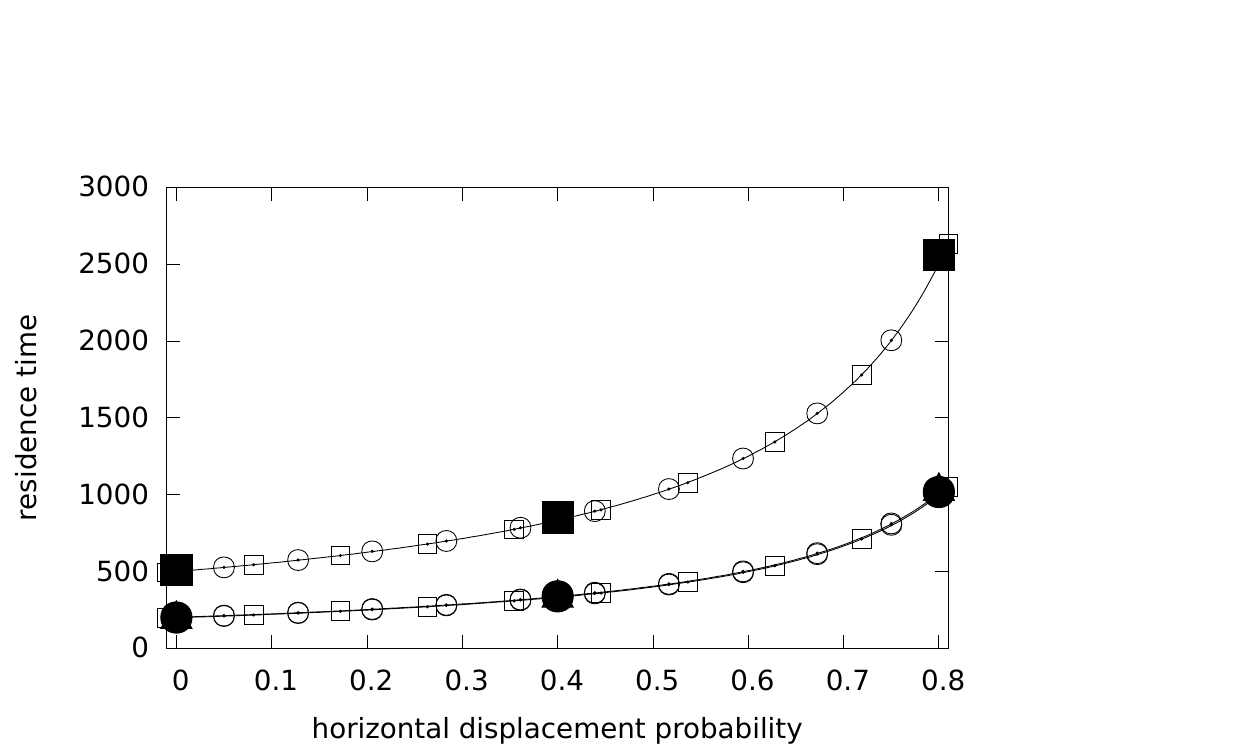}}} 
}
\put(235,0)
{
\resizebox{10cm}{!}{\rotatebox{0}{\includegraphics{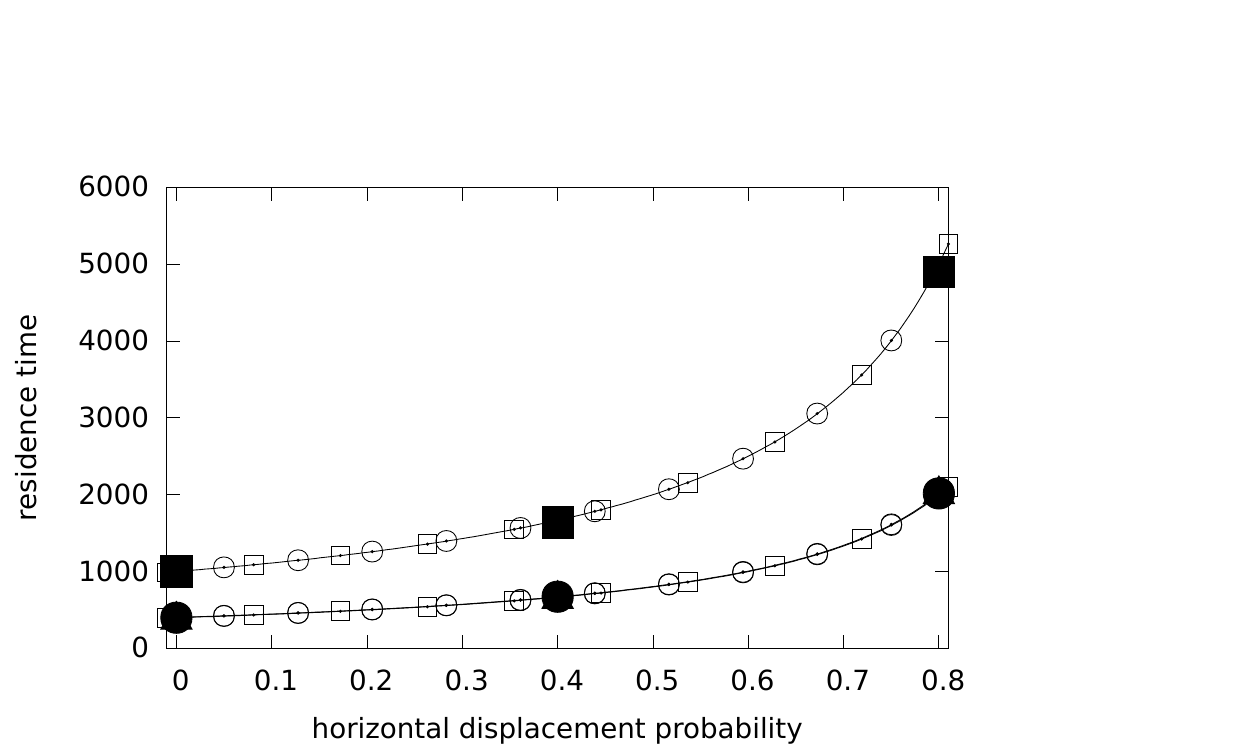}}} 
}
\end{picture}  
\caption{Residence time versus the horizontal displacement probability 
$h$ in the cases $\delta=1$ and $L_2=100$ (left) 
and $L_2=200$ (right).
The symbols $\bullet$, $\blacktriangle$, and $\blacksquare$
refer, respectively, to the cases $\varrho_\rr{d}=0,0.4,0.8$. 
Note that bullets and black triangle are perfectly coinciding, 
this is due to the fact that in this regime the residence time 
does not depend that much on the bottom boundary density, provided 
it is smaller that $1/2$. In such a case, indeed, 
the density profile inside the strip is almost perfectly constant and equal 
to $1/2$. 
Solid lines are the theoretical prediction based on the birth and 
death model.
Open squares are the approximated theoretical prediction 
\eqref{ld010} valid in the large drift regime 
with $\bar\varrho=1/2$ for the cases $\varrho_\rr{d}=0,0.4$
and 
with $\bar\varrho=8/10$ in the case $\varrho_\rr{d}=0.8$.
Indeed, in the first two cases the density profile is almost constantly
equal to $1/2$, while in the last case it is almost constantly equal 
to $8/10$. 
Open circles are the Mean Field theoretical prediction \eqref{res-mf010}. 
}
\label{f:res-ph-100-dr1} 
\end{figure} 

\begin{figure}[h]
\begin{picture}(200,125)(0,10)
\put(0,0)
{
\resizebox{10cm}{!}{\rotatebox{0}{\includegraphics{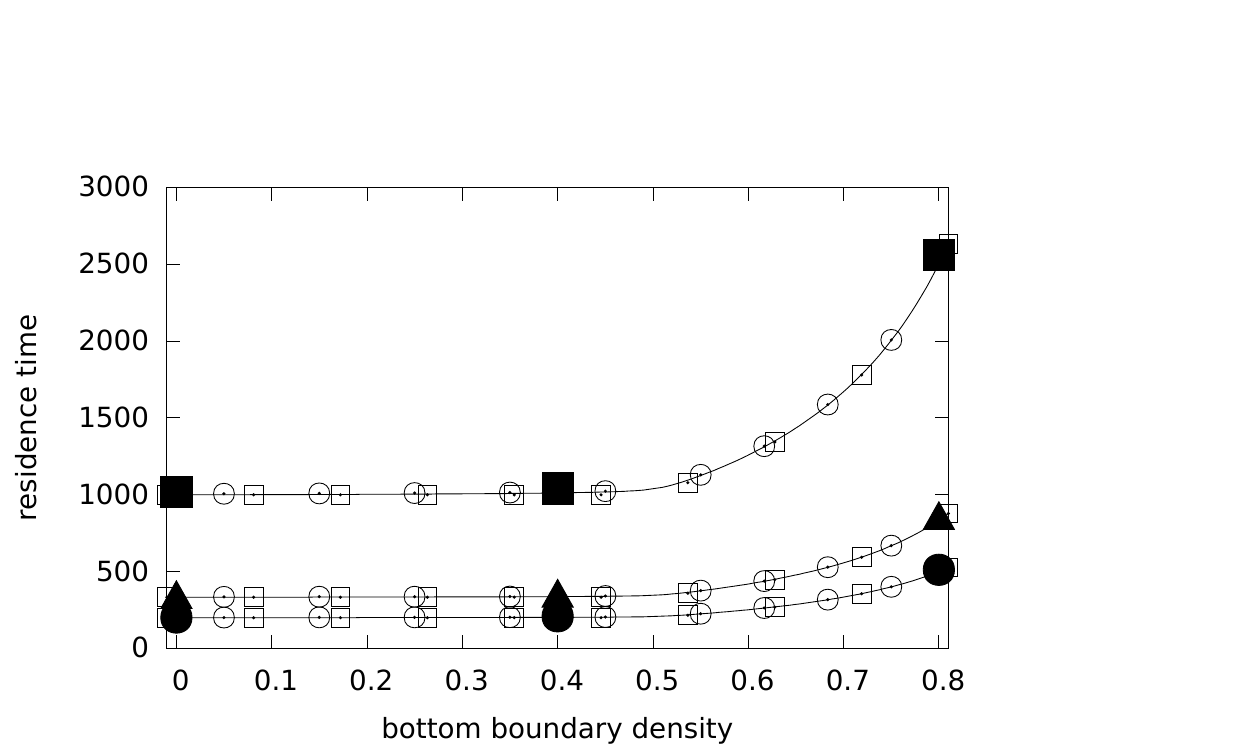}}} 
}
\put(235,0)
{
\resizebox{10cm}{!}{\rotatebox{0}{\includegraphics{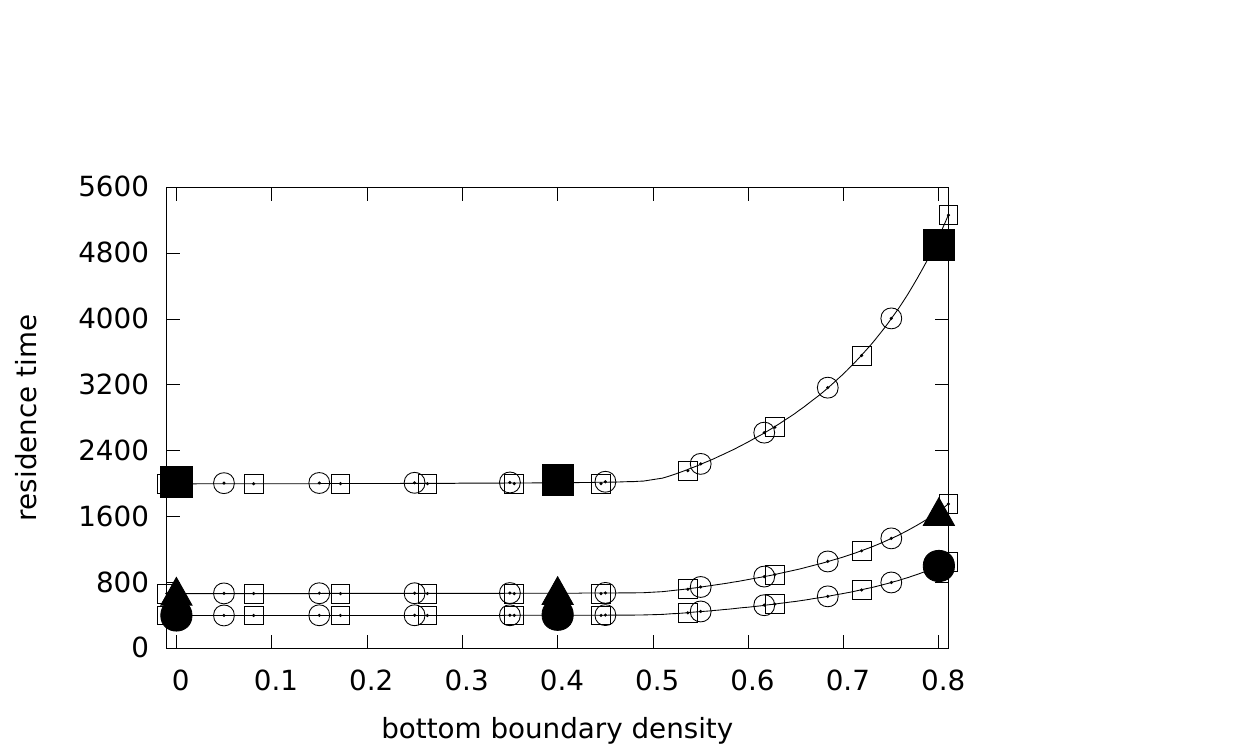}}} 
}
\end{picture}  
\caption{Residence time versus the bottom boundary density
$\varrho_\rr{d}$ 
in the cases $\delta=1$ and 
$L_2=100$ (left) 
and $L_2=200$ (right).
The symbols $\bullet$, $\blacktriangle$, and $\blacksquare$
refer, respectively, to the cases $h=0,0.4,0.8$. 
Solid lines are the theoretical prediction based on the birth and 
death model. 
Open squares are the approximated theoretical prediction 
\eqref{ld010} valid in the large drift regime 
with $\bar\varrho=1/2$ for $\varrho_\rr{d}\le 1/2$
and 
with $\bar\varrho=\varrho_\rr{d}$ for $\varrho_\rr{d}>1/2$.
Indeed, when $\varrho_\rr{d}<1/2$ the density profile is almost constantly
equal to $1/2$, while for $\varrho_\rr{d}>1/2$ 
it is almost constantly equal to $\varrho_\rr{d}$. 
Open circles are the Mean Field theoretical prediction \eqref{res-mf010}. 
}
\label{f:res-rd-100-dr1} 
\end{figure} 

\begin{figure}[h]
\begin{picture}(200,140)(0,10)
\put(80,0)
{
\resizebox{12cm}{!}{\rotatebox{0}{\includegraphics{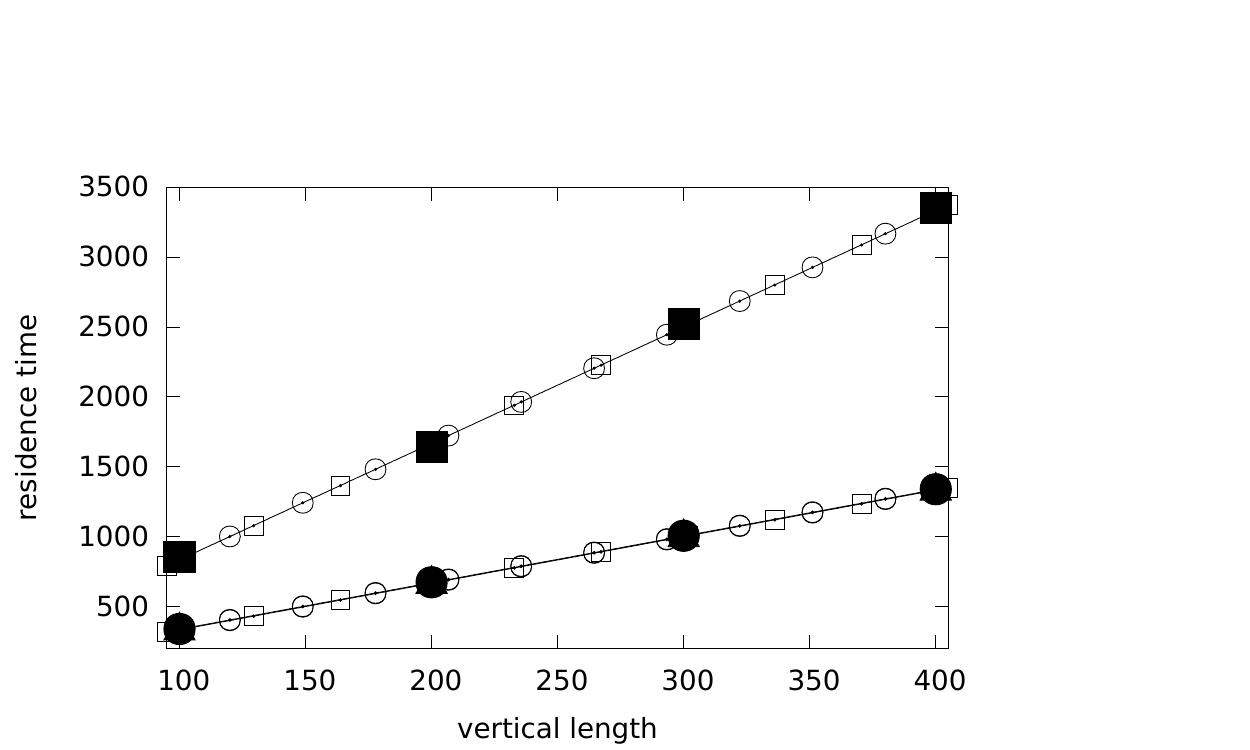}}} 
}
\end{picture}  
\caption{Residence time versus the vertical length of the strip $L_2$
in the case $\delta=1$ 
and $h=0.4$.
The symbols $\bullet$, $\blacktriangle$, and $\blacksquare$
refer, respectively, to the cases $\varrho_\rr{d}=0,0.4,0.8$. 
Note that bullets and black triangle are perfectly coinciding, 
this is due to the fact that in this regime the residence time 
does not depend that much on the bottom boundary density, provided 
it is smaller that $1/2$. In such a case, indeed, 
the density profile inside the strip is almost perfectly constant and equal 
to $1/2$. 
Solid lines are the theoretical prediction based on the birth and death 
model.
Open squares are the approximated theoretical prediction 
\eqref{ld010} valid in the large drift regime 
with $\bar\varrho=1/2$ for the cases $\varrho_\rr{d}=0,0.4$
and 
with $\bar\varrho=8/10$ in the case $\varrho_\rr{d}=0.8$.
Indeed, in the first two cases the density profile is almost constantly
equal to $1/2$, while in the last case it is almost constantly equal 
to $8/10$. 
Open circles are the Mean Field theoretical prediction \eqref{res-mf010}. 
}
\label{f:res-el-dr1} 
\end{figure} 

\begin{figure}[h]
\begin{picture}(200,125)(0,10)
\put(0,0)
{
\resizebox{10cm}{!}{\rotatebox{0}{\includegraphics{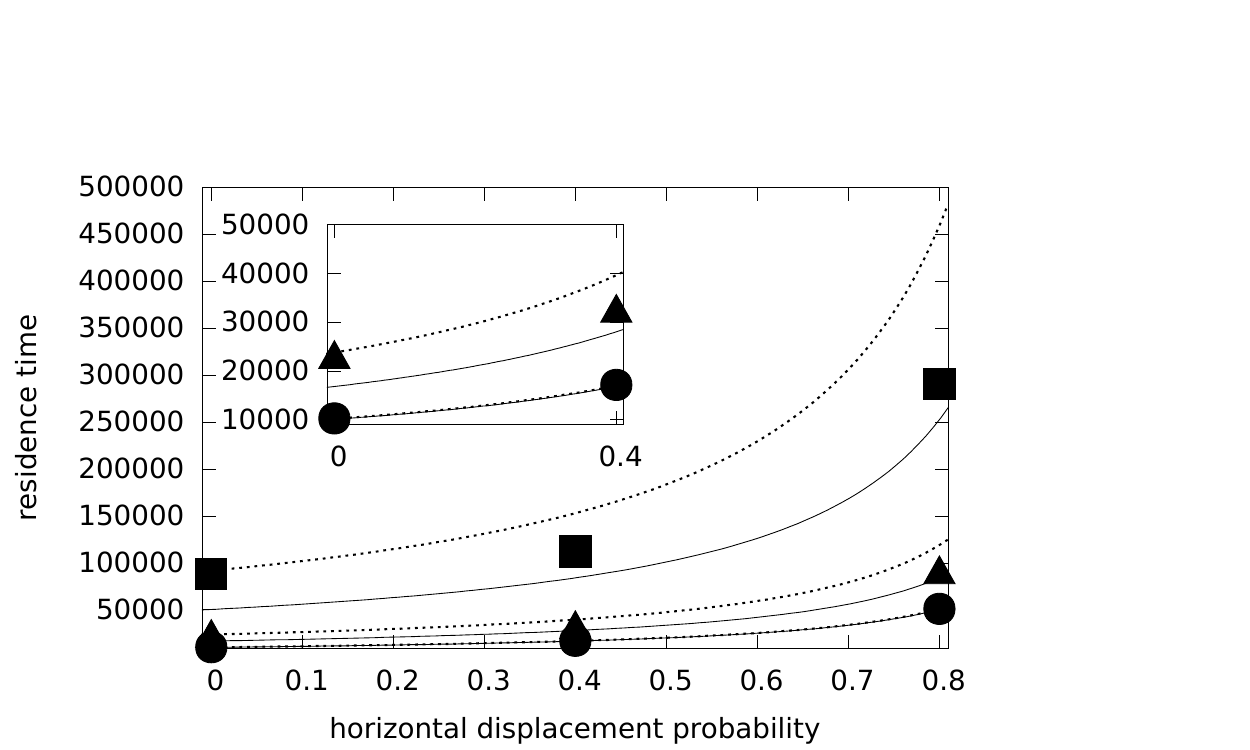}}} 
}
\put(235,0)
{
\resizebox{10cm}{!}{\rotatebox{0}{\includegraphics{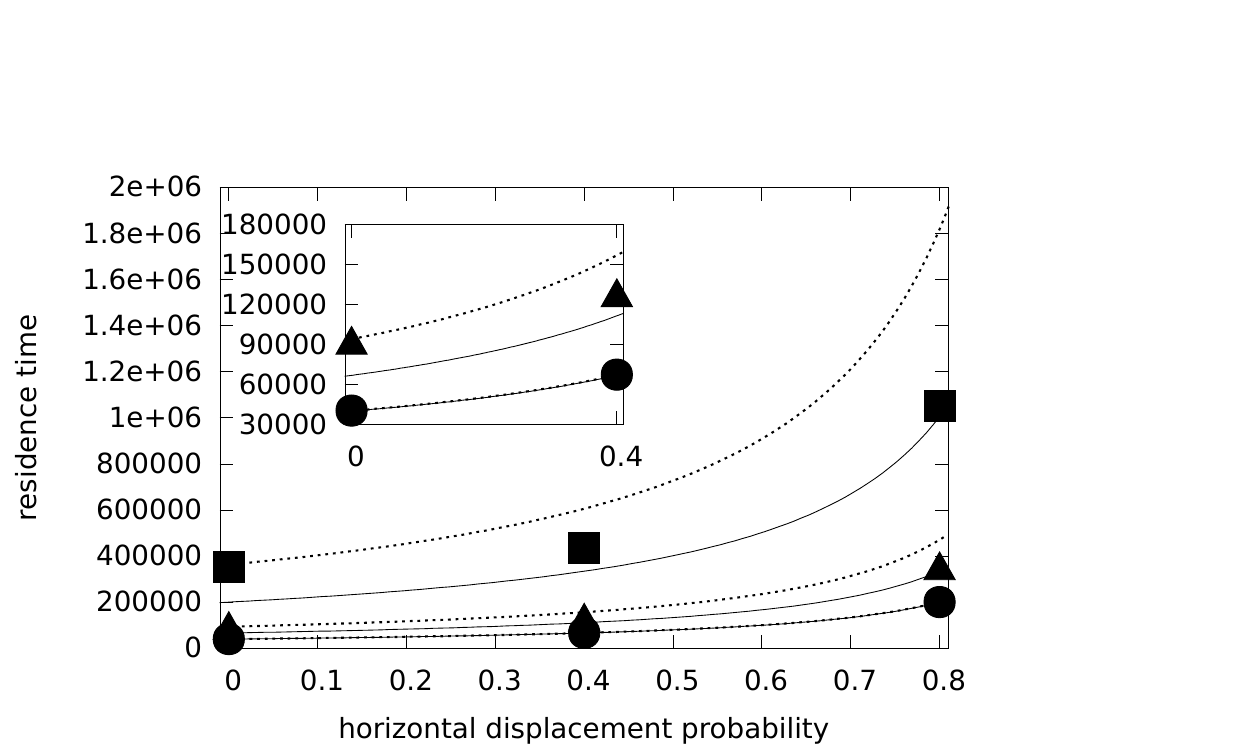}}} 
}
\end{picture}  
\caption{Residence time versus the horizontal displacement probability 
$h$ in the cases $\delta=0$ and $L_2=100$ (left) 
and $L_2=200$ (right).
The symbols $\bullet$, $\blacktriangle$, and $\blacksquare$
refer, respectively, to the cases $\varrho_\rr{d}=0,0.4,0.8$. 
Solid lines are the theoretical prediction \eqref{zd020}
from the bottom to the top for the cases $\varrho_\rr{d}=0,0.4,0.8$. 
Dotted lines are the Mean Field  prediction \eqref{res-mf020}
from the bottom to the top for the cases $\varrho_\rr{d}=0,0.4,0.8$. 
The dotted and the solid lines corresponding to the case 
$\varrho_\rr{d}=0$ cannot be distinguished in the picture.
The insets are just a zoom 
of part of the data of the same picture. 
}
\label{f:res-ph-100-dr0} 
\end{figure} 

\begin{figure}[h]
\begin{picture}(200,125)(0,10)
\put(0,0)
{
\resizebox{10cm}{!}{\rotatebox{0}{\includegraphics{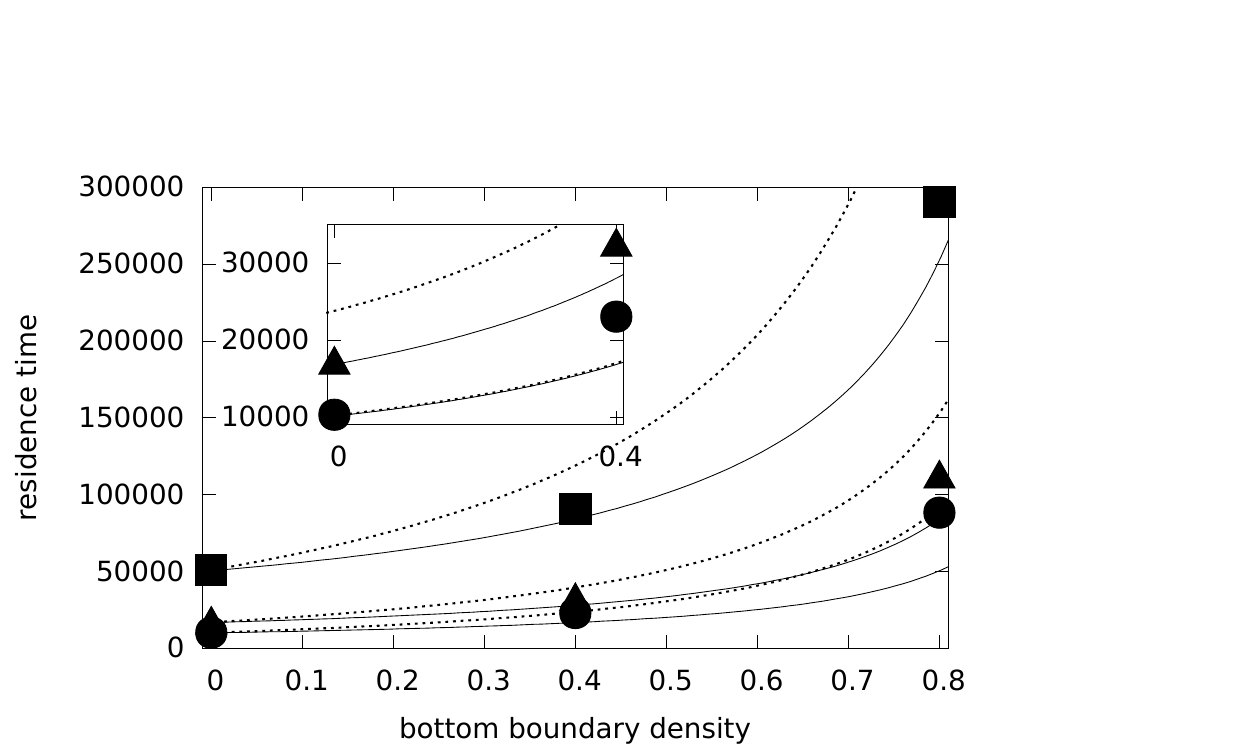}}} 
}
\put(235,0)
{
\resizebox{10cm}{!}{\rotatebox{0}{\includegraphics{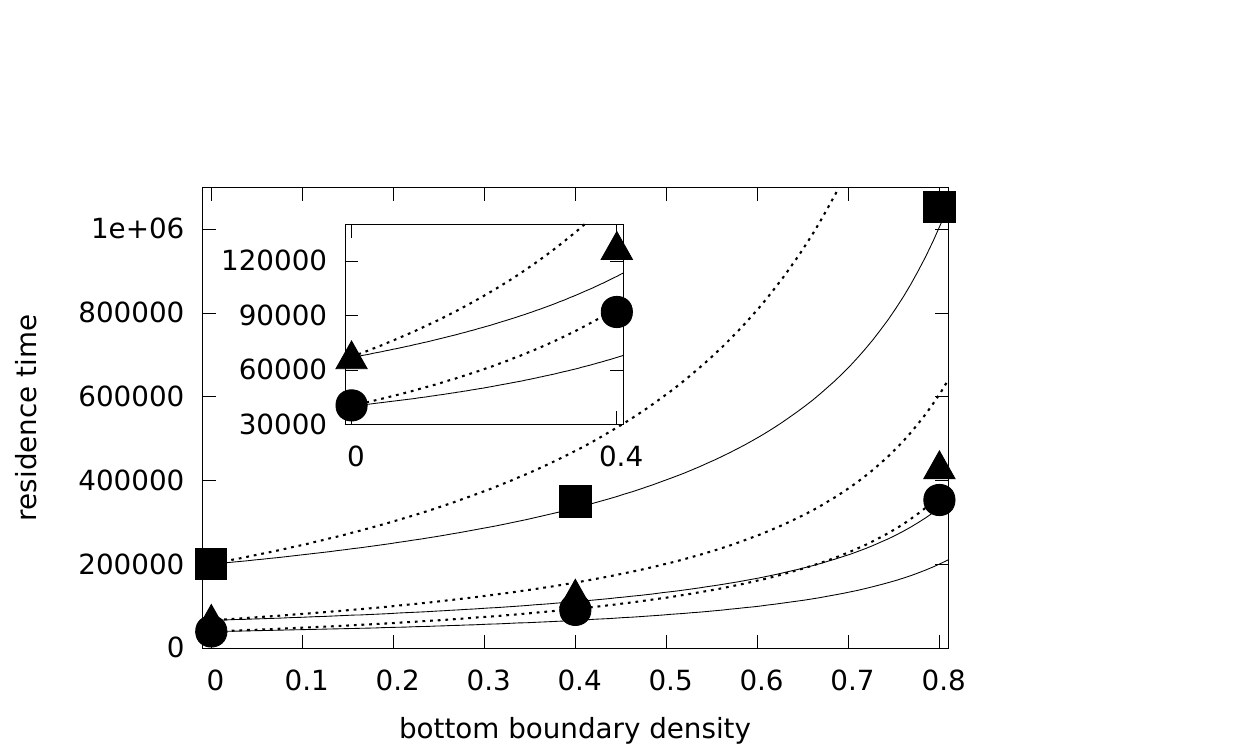}}} 
}
\end{picture}  
\caption{Residence time versus the bottom boundary density
$\varrho_\rr{d}$ 
in the cases $\delta=0$ and 
$L_2=100$ (left) 
and $L_2=200$ (right).
The symbols $\bullet$, $\blacktriangle$, and $\blacksquare$
refer, respectively, to the cases $h=0,0.4,0.8$. 
Solid lines are the theoretical prediction \eqref{zd020}
from the bottom to the top for the cases $h=0,0.4,0.8$. 
Dotted lines are the Mean Field  prediction \eqref{res-mf020}
from the bottom to the top for the cases $h=0,0.4,0.8$. 
The insets are just a zoom 
of part of the data of the same picture. 
}
\label{f:res-rd-100-dr0} 
\end{figure} 

\begin{figure}[h]
\begin{picture}(200,375)(0,10)
\put(0,280)
{
\resizebox{10cm}{!}{\rotatebox{0}{\includegraphics{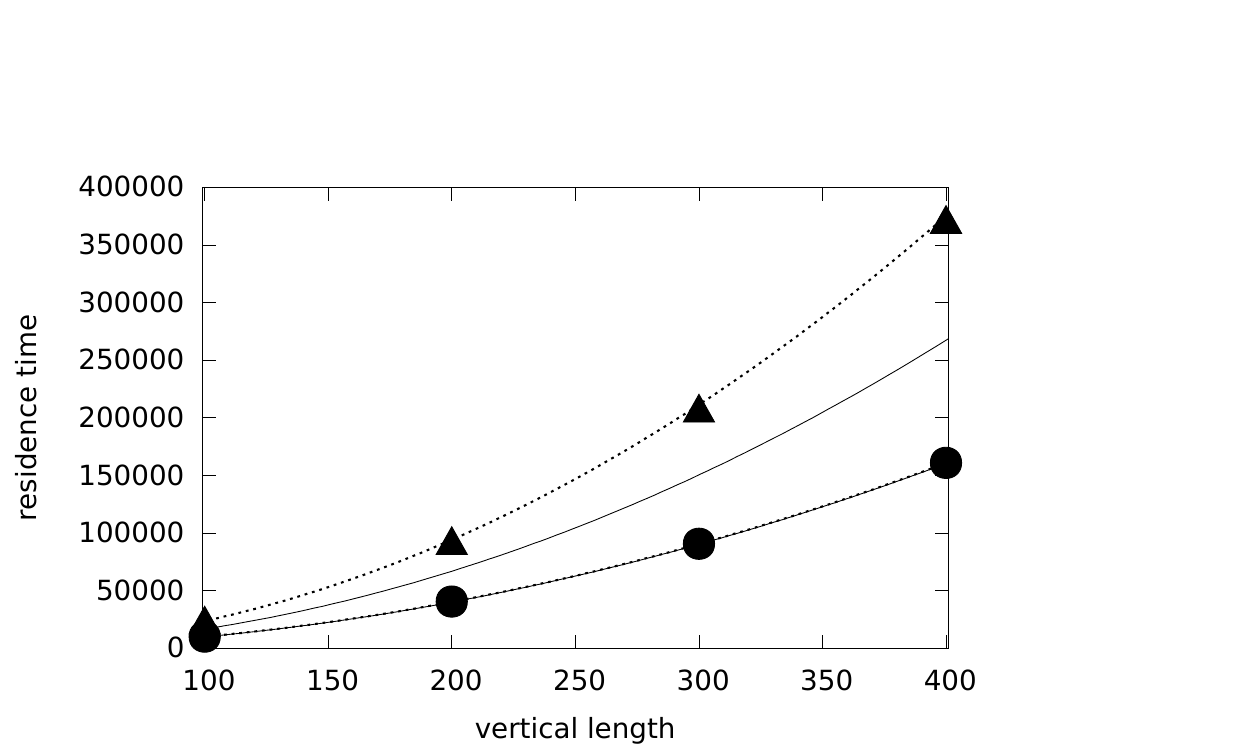}}} 
}
\put(235,280)
{
\resizebox{10cm}{!}{\rotatebox{0}{\includegraphics{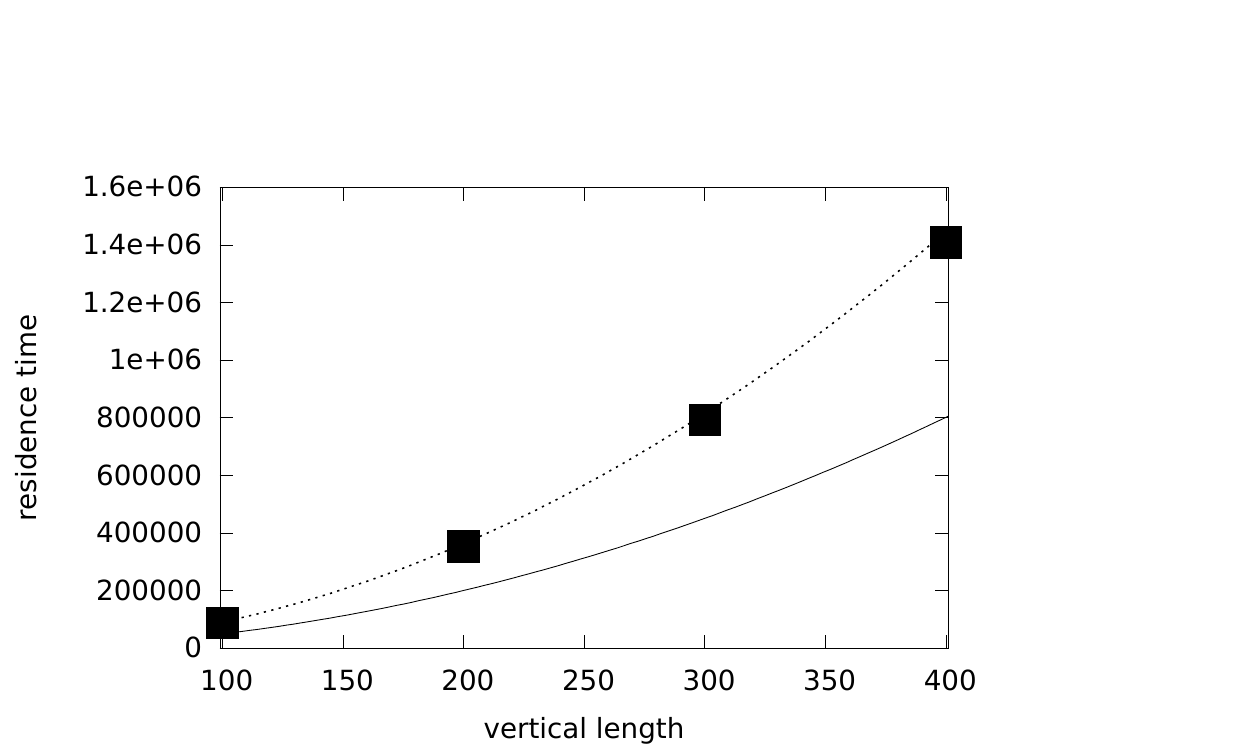}}} 
}
\put(0,140)
{
\resizebox{10cm}{!}{\rotatebox{0}{\includegraphics{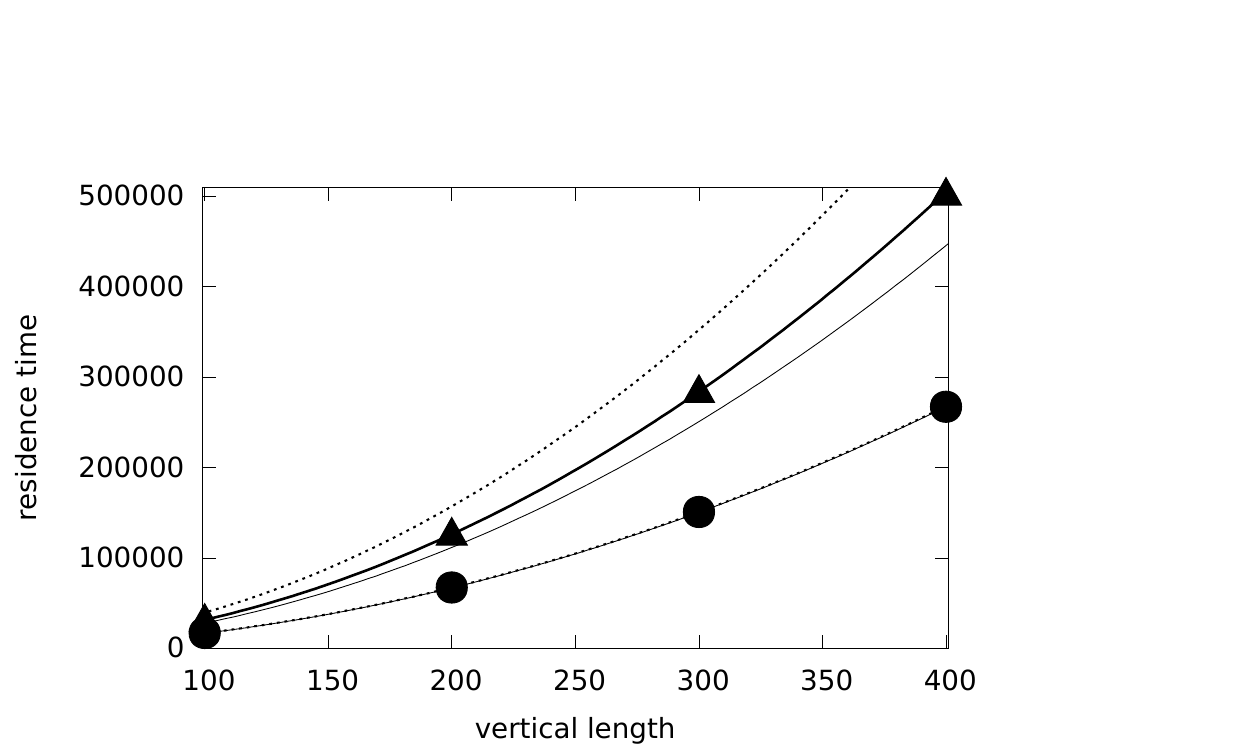}}} 
}
\put(235,140)
{
\resizebox{10cm}{!}{\rotatebox{0}{\includegraphics{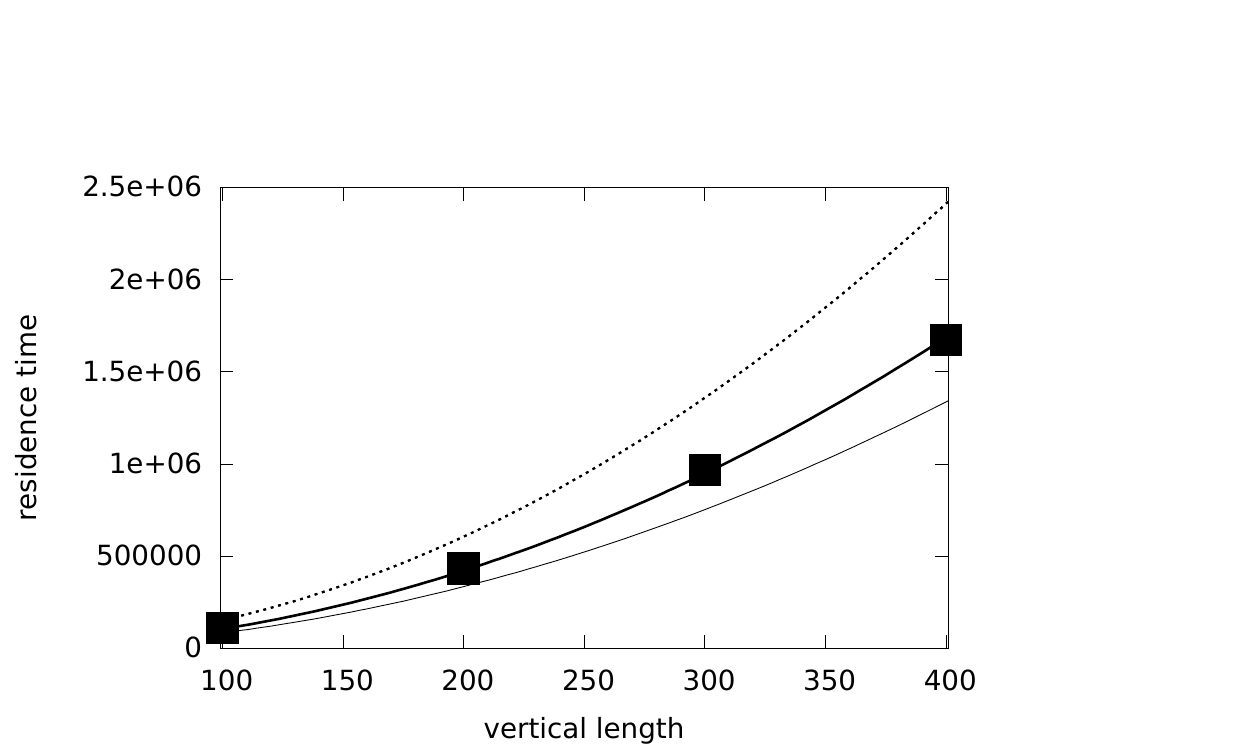}}} 
}
\put(0,0)
{
\resizebox{10cm}{!}{\rotatebox{0}{\includegraphics{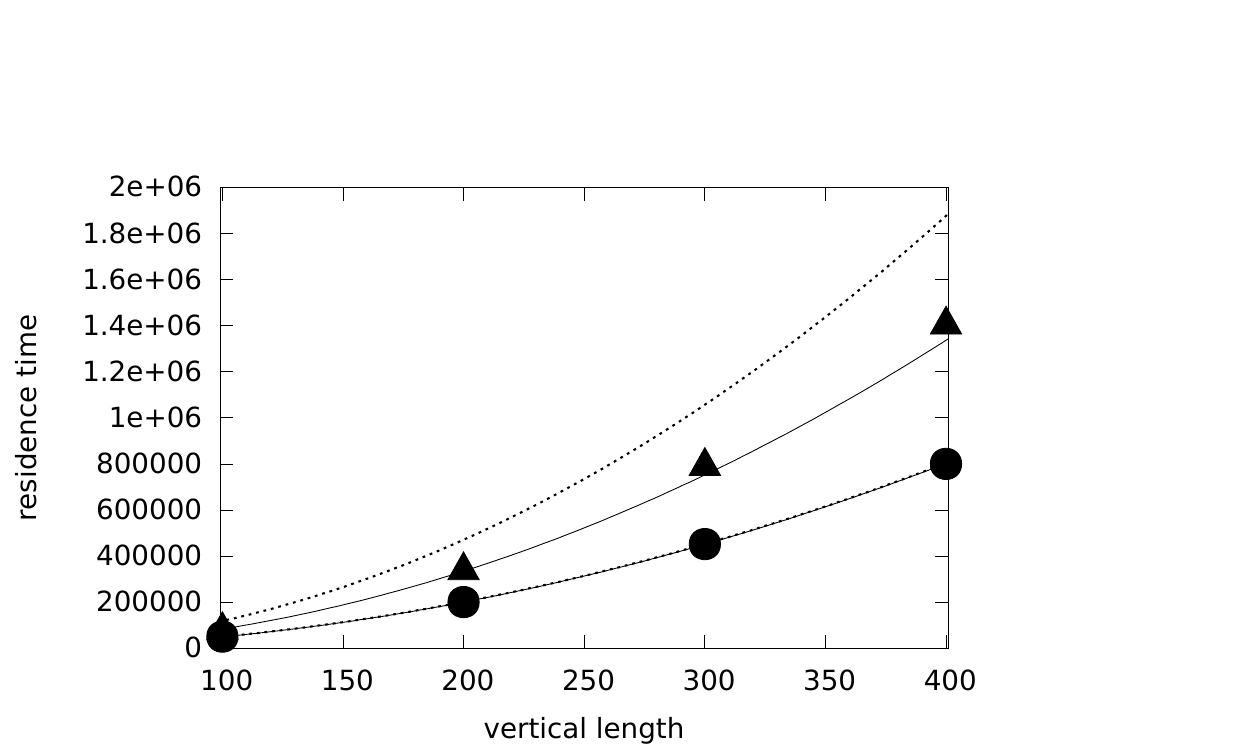}}} 
}
\put(235,0)
{
\resizebox{10cm}{!}{\rotatebox{0}{\includegraphics{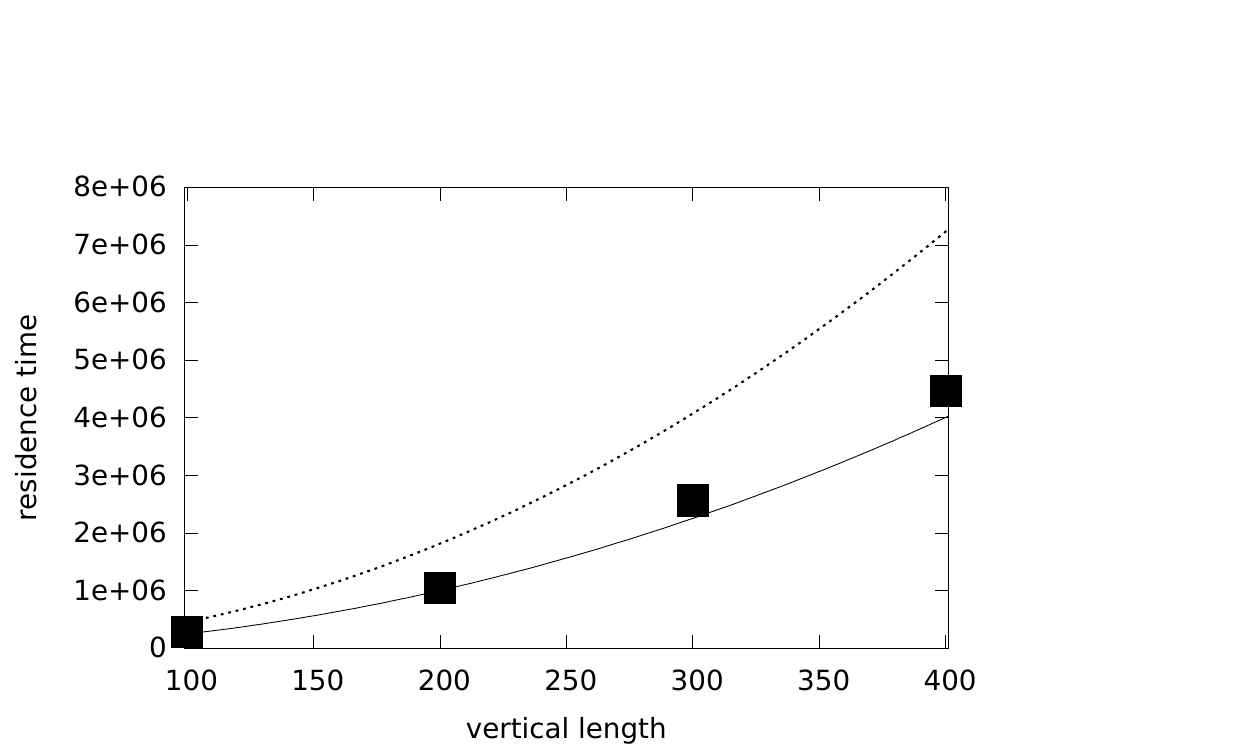}}} 
}
\end{picture}  
\caption{Residence time versus the vertical length of the strip $L_2$
in the case $\delta=0$ 
and $h=0,0.4,0.8$ from the top to the bottom.
The symbols $\bullet$, $\blacktriangle$, and $\blacksquare$
refer, respectively, to the cases $\varrho_\rr{d}=0,0.4,0.8$. 
Solid thin lines are the theoretical prediction \eqref{zd020}.
Dotted lines are the Mean Field prediction \eqref{res-mf020}.
The solid thick lines are quadratic fitting of the experimental data: 
$3.16\times L_2^2$ in the case $h=0.4$ and $\varrho_\rr{d}=0.4$ and 
$10.53\times L_2^2$ in the case $h=0.4$ and $\varrho_\rr{d}=0.8$, 
The dotted and the solid lines corresponding to the cases 
$\varrho_\rr{d}=0$ cannot be distinguished in the picture.
}
\label{f:res-el-dr0} 
\end{figure} 

\begin{figure}[h]
\begin{picture}(200,125)(0,10)
\put(0,0)
{
\resizebox{10cm}{!}{\rotatebox{0}{\includegraphics{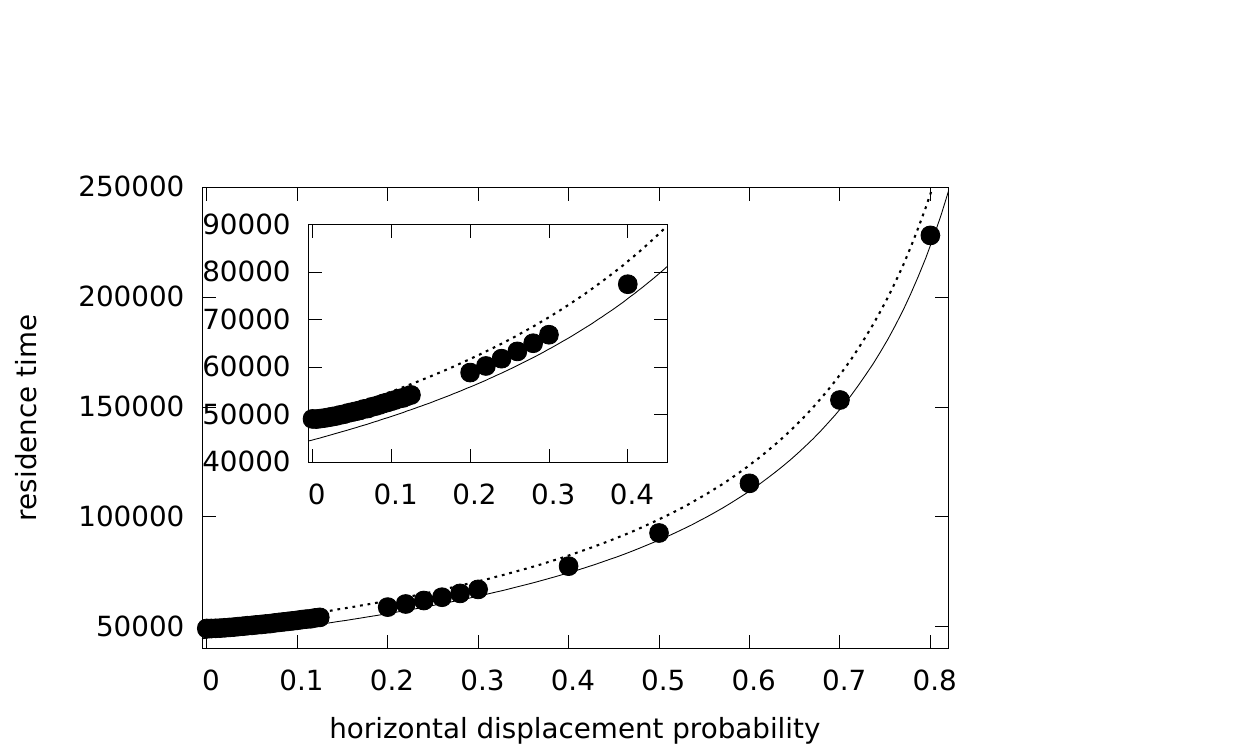}}} 
}
\put(235,0)
{
\resizebox{10cm}{!}{\rotatebox{0}{\includegraphics{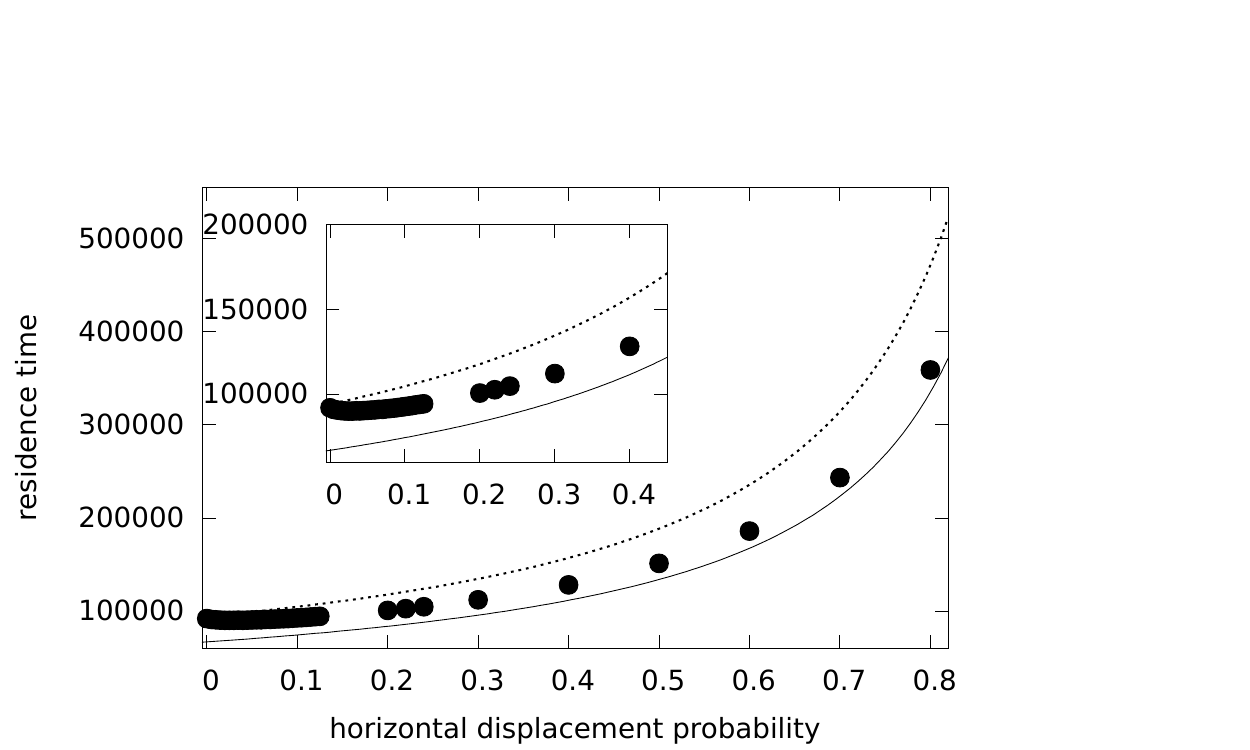}}} 
}
\end{picture}  
\caption{Residence time versus the horizontal displacement 
probability in the case 
$L_2=200$, $\delta=0$, and 
and $\varrho_\rr{d}=0.1$ (left) and $\varrho_\rr{d}=0.4$ (right).
Solid thin lines are the theoretical prediction \eqref{zd020}.
Dotted lines are the Mean Field prediction \eqref{res-mf020}.
}
\label{f:res-ph-nuova} 
\end{figure} 

In all these cases, the residence time has been evaluated 
by averaging over all the particles entered in the system through 
the top boundary after the time $t_\rr{term}$, namely, at 
stationarity, and exited through the bottom one at a time
smaller than $t_\rr{max}$. 
The simulations have been performed with 
$t_\rr{term}=5\times10^5$
and 
$t_\rr{max}=5\times10^6$.

To check the dependence of the residence time on the length 
of the strip we had to consider a few more cases with larger $L_2$, 
viz.\
$L_2=300,400$.
In these cases, due to the length of the strip, we had to use 
$t_\rr{term}=10^6$
and increase $t_\rr{max}$ up to $4\times10^7$ in the more 
delicate situations ($h=0.8$). 
It is important to note that in the cases $L_2=300,400$
and $\varrho_\rr{d}=0.8$, 
the initial configuration of the system has been chosen by 
populating the system by putting particles 
with probability $0.95$. Indeed, by starting the system with all 
the sites empty, the dynamics was trapped in a sort of ``meta--stationary" 
state with density approximatively constant and slightly larger than 
$0.8$. 

In all the pictures the vertical bar representing the 
statistical error is not visible since it is smaller than 
the symbol representing the measured value of the residence time. 

We compare the numerical results with the theoretical predictions based 
both on the 
birth and death analogy and on the Mean Field approach.
In general the birth and death theoretical prediction of the residence time is 
the value given by \eqref{art050} with
the stationary measure $\pi$ given by \eqref{art040} where
the jump probabilities $p_i$ and $q_i$ are defined in \eqref{art010}.
Since we could not derive explicit expressions in terms of the 
model parameters, 
the birth and death theoretical prediction has been computed by performing 
sums and products numerically. 
In the case $\delta=1$ (see, the discussion in 
Section~\ref{s:caso010})
the birth and death theoretical prediction is 
the value given by \eqref{hitting-tot} with
the $q_i$'s defined in \eqref{art010}.
In the case $\delta=0$ (see for instance the discussion in 
Section~\ref{s:caso020})
the theoretical prediction is given explicitly by \eqref{zd020}.

We find that the match between the theoretical predictions 
based both on the birth and death analogy and on the 
Mean Field approach and the numerical data 
is perfect for any choice of the parameters 
of the model provided either $\varrho_\rr{d}=0$
or $\delta=1$ (or both). 
To illustrate this fact we first discuss 
our data at $\varrho_\rr{d}=0$ and let 
$h=0,0.4,0.8$ and $\delta=0,0.2,0.4,0.6,0.8,1$.
Then, we consider $\delta=1$ and let 
$h=0,0.4,0.8$ and $\varrho_\rr{d}=0,0.4,0.8$.

Finally, for the region where the match between the theoretical 
predictions and the numerical results is not good, 
we consider 
the worst case from the point of view of drift, namely, 
we choose $\delta=0$
and let, again, 
$h=0,0.4,0.8$ and $\varrho_\rr{d}=0,0.4,0.8$.
We shall notice that the match between theory and experiment, even 
if qualitatively correct, it becomes progressively worse with increasing $\varrho_\rr{d}$.

\subsection{Case $\varrho_\rr{d}=0$}
\label{s:caso000}
\par\noindent
We discuss first the case $\varrho_\rr{d}=0$ and show that here
the theoretical predictions of the residence time 
based on both the birth and death analogy and the 
Mean Field computation discussed in Section~\ref{s:res} 
agree perfectly with the numerical results. 

In Figure~\ref{f:res-dr-100} it is shown the dependence 
of the residence time on the drift $\delta$ for 
$\varrho_\rr{d}=0$ and for different values 
of the horizontal displacement probability $h$. 
The dependence 
of the residence time on the horizontal displacement 
probability for different values of the drift
is shown in Figure~\ref{f:res-ph-100}.

The fact that the residence time decreases with the drift and increases 
with the horizontal displacement probability is completely 
reasonable. 
The match between the simulation data and the theoretical predictions is 
perfect. 
It is remarkable the fact that even the approximated 
expression \eqref{ld010} (which in the case $\delta=1$ 
reduces to \eqref{td010}) 
gives a perfect estimate of the residence time. This is due to the 
fact that for the values of the parameter that we have chosen 
the stationary density profile throughout the system 
is constantly equal to $1/2$  with very a good approximation. 

In Figure~\ref{f:res-el},
the residence time in the case $\varrho_\rr{d}$ has been plotted as 
a function of the 
length $L_2$ of the strip for different values of the drift and 
for $h=0.4$. 
It is remarkable to note the striking match between theory and 
simulation. In particular the fact that the 
behavior is linear (ballistic) for $\delta>0$ and 
quadratic (diffusive) for $\delta=0$, see the theoretical 
discussion in Section~\ref{s:scaling}, is confirmed by the 
numerical experiment. 

\subsection{Case $\delta=1$}
\label{s:caso010}
\par\noindent
This is 
the case of totally asymmetric simple exclusion rule 
along the vertical direction. Here, 
particles can never jump upwards. 
We show that, for this scenario, 
the theoretical predictions of the residence time 
based on both the birth and death analogy and the 
Mean Field computation discussed in Section~\ref{s:res} 
agree perfectly with the numerical results. 

We show three pictures:
the dependence of the residence time 
on the horizontal displacement probability $h$ (Figure~\ref{f:res-ph-100-dr1}), 
the dependence on the bottom boundary density $\varrho_\rr{d}$
(Figure~\ref{f:res-rd-100-dr1}), 
and, finally, 
the dependence on the length of the strip $L_2$
(Figure~\ref{f:res-el-dr1}).

We do not repeat the discussion in detail. We just mention 
that the linearity of the residence time with respect to 
the length of the strip 
is confirmed by the experimental data and refer the 
reader to the caption of the pictures for more specific 
comments. 

\subsection{Case $\delta=0$}
\label{s:caso020}
\par\noindent
We discuss now the case of a
symmetric simple exclusion rule 
along the vertical direction. 
As we already mentioned at the beginning of this 
section, in this case the match between the theoretical 
prediction and the numerical data is only qualitatively good and 
quantitavely worst and worst 
when $\varrho_\rr{d}$ is increased. 

We also remark that, fixed $\varrho_\rr{d}$, the match with the 
prediction based on the birth and death analogy is better 
at larger values of the horizontal displacement probability $h$. 
On the other hand, 
fixed $\varrho_\rr{d}$, the match with the 
Mean Field prediction is better 
at smaller values of the horizontal displacement probability $h$ and 
perfect at $h=0$.

The physical interpretation of this fact is that at large 
$\varrho_\rr{d}$ particles accumulate 
at the bottom exit and, due to bouncing back of particles, 
fluctuations are not more negligible. For this reason our theoretical 
predictions, which are based only on the stationary shape of the 
density profile, are not anymore always efficient. 
We shall discuss this point also in Section~\ref{s:caso030}.

We show three pictures:
the dependence of the residence time 
on the horizontal displacement probability $h$ (Figure~\ref{f:res-ph-100-dr0}), 
the dependence on the bottom boundary density $\varrho_\rr{d}$
(Figure~\ref{f:res-rd-100-dr0}), 
and, finally, 
the dependence on the length of the strip $L_2$
(Figure~\ref{f:res-el-dr0}).

Note that in this section, since 
$\delta=0$, the theoretical prediction based on the birth and death analogy is
given by the explicit formula \eqref{zd020}, whereas the Mean Field 
prediction is \eqref{res-mf020}.

The relevant comment now, see Figures~\ref{f:res-ph-100-dr0}, 
\ref{f:res-el-dr0}, and \ref{f:res-ph-nuova},
is that the match between the Mean Field prediction and 
the experiment is perfect at $h=0$, whereas it gets worst and 
worst when $h$ is increased. 
On the other hand, the birth and death analogy poorly predicts
the residence time behavior 
$h=0$, whereas it captures the phenomenon better and better when
$h$ is increased. 
This phenomenon can be explained via two mechanisms: bouncing back and 
the possibility to bypass clusters of blocking particles. 
The former suggests that correlations become important in this 
regime so that a model based only on the average stationary 
density profile cannot explain the behavior of the system. 
On the other hand, the latter phenomenon suggests that the effect 
of bouncing back is milder when the horizontal displacement probability is
larger, since particles have a good chance to avoid blocking 
clusters. 

In is interesting to remark that 
the Mean Field expression \eqref{res-mf020} for the case $\delta=0$, 
and $L_2$ large, 
is given by the expression \eqref{zd020} predicted by the 
birth and death analogy plus a term that can be ascribed 
the the two point correlations, see \cite{fedders}.
We could then guess that the correct expression of the residence time 
could be the birth and death prediction plus the two point 
correlation contribution weighted by a function depending on $h$ 
tending to $1$ for $h\to0$ and to $0$ for $h\to1$. 

\subsection{Non--monotonic behavior in the bouncing back regime}
\label{s:caso030}
\par\noindent
As it has been seen in the previous section 
the residence time is typically an increasing function of the 
horizontal displacement probability. 
This is an obvious fact. Indeed, when $h$ is increased 
particles spend a lot of time in performing horizontal 
jumps which are a waste of time in the run towards the bottom 
exit.

In this section we show that in the regime in which 
the bouncing back phenomenon is present a small not zero 
horizontal probability displacement can favour the 
exit of the particles. 

We have performed the following simulations:
$\varrho_\rr{u}=1$ (as always), 
$L_1=100$,
$L_2=200$,
$\delta=0$, 
\begin{displaymath}
\varrho_\rr{d}=0,0.05,\dots,0.5,0.6,0.7,0.8,\;\;
\;\;\;\textrm{ and }\;\;\;
h=0,0.005,0.010,\dots,0.125\;\;.
\end{displaymath}
As before, 
in all these cases, the residence time has been evaluated 
by averaging over all the particles entered in the system through 
the top boundary after the time $t_\rr{term}$, namely, at 
stationarity, and exited through the bottom one at a time
smaller than $t_\rr{max}$. 
The simulations have been performed with 
$t_\rr{term}=1\times10^6$
and 
$t_\rr{max}=1\times10^7$.


\begin{figure}[htbp]
\begin{picture}(200,450)(0,10)
\put(0,450)
{
\resizebox{9cm}{!}{\rotatebox{0}{\includegraphics{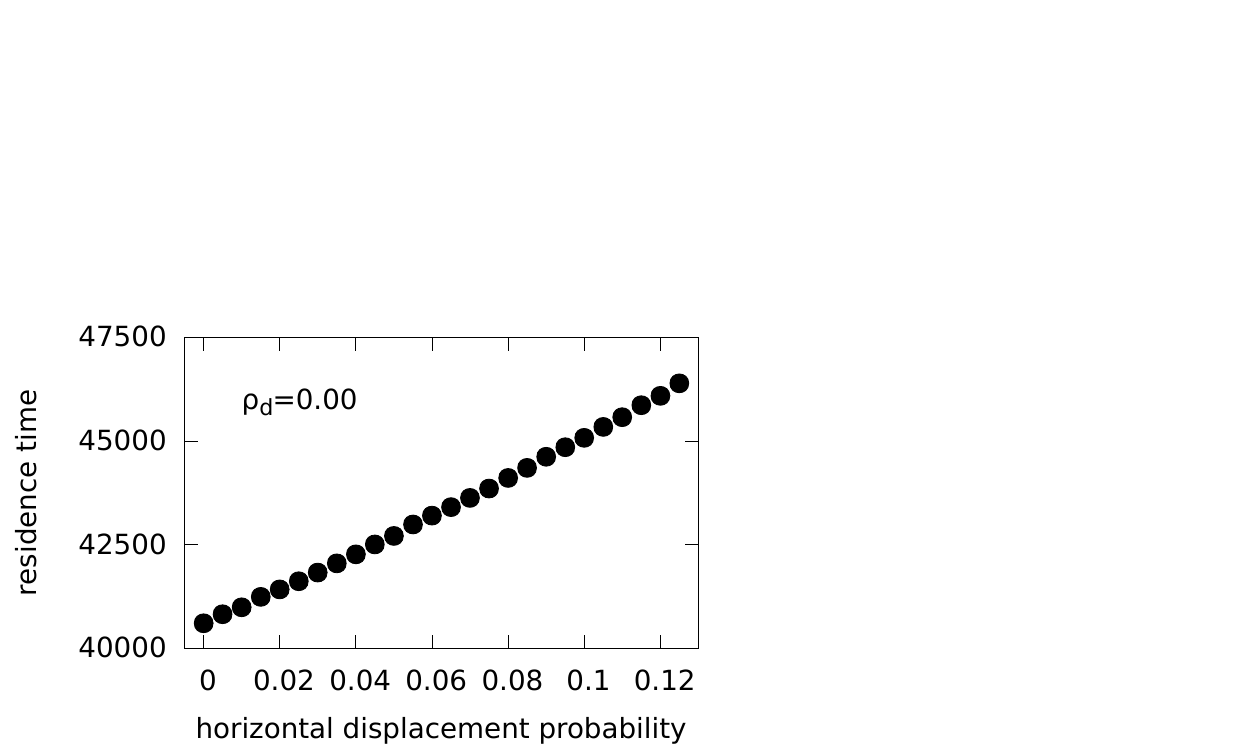}}} 
}
\put(160,450)
{
\resizebox{9cm}{!}{\rotatebox{0}{\includegraphics{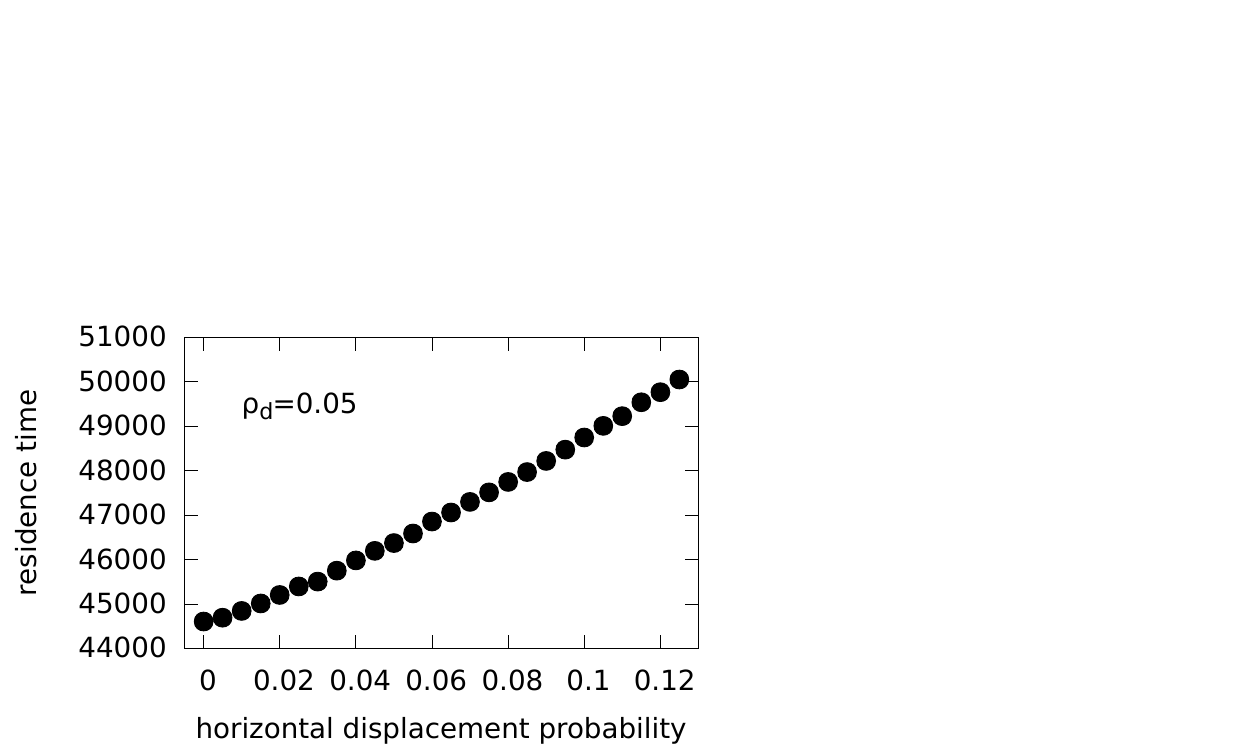}}} 
}
\put(320,450)
{
\resizebox{9cm}{!}{\rotatebox{0}{\includegraphics{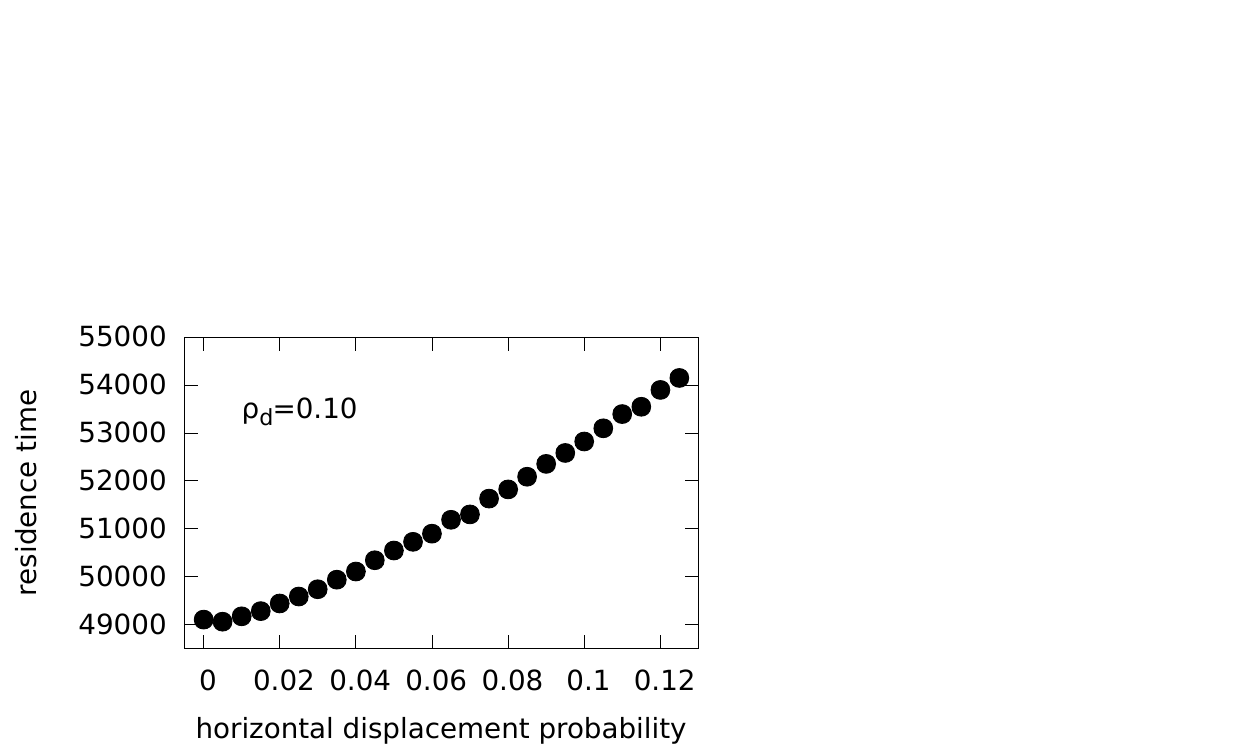}}} 
}
\put(0,350)
{
\resizebox{9cm}{!}{\rotatebox{0}{\includegraphics{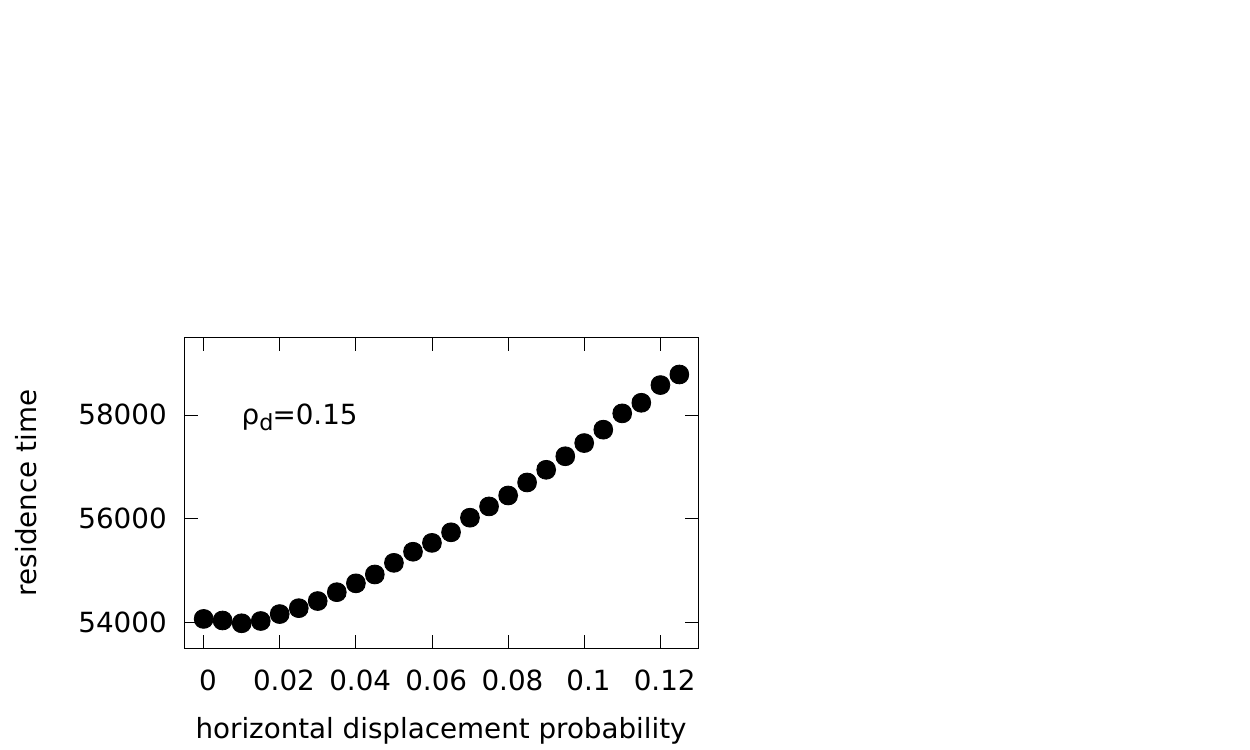}}} 
}
\put(160,350)
{
\resizebox{9cm}{!}{\rotatebox{0}{\includegraphics{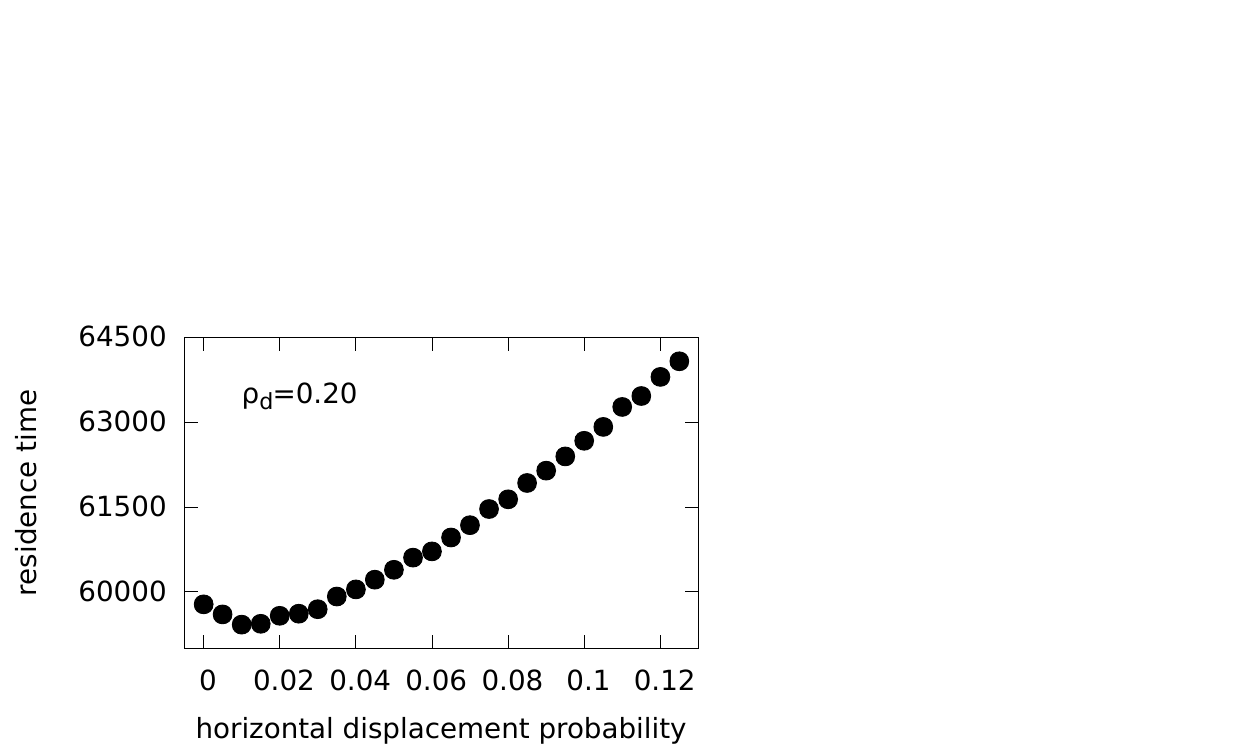}}} 
}
\put(320,350)
{
\resizebox{9cm}{!}{\rotatebox{0}{\includegraphics{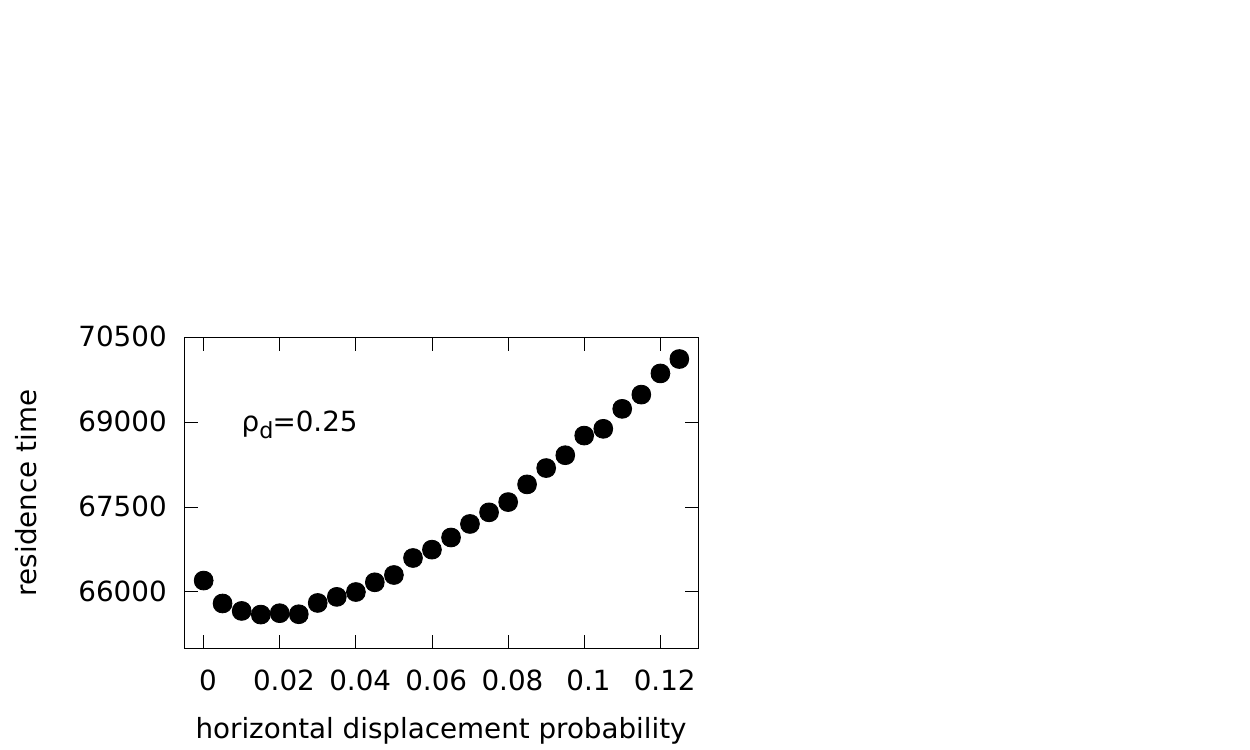}}} 
}
\put(0,250)
{
\resizebox{9cm}{!}{\rotatebox{0}{\includegraphics{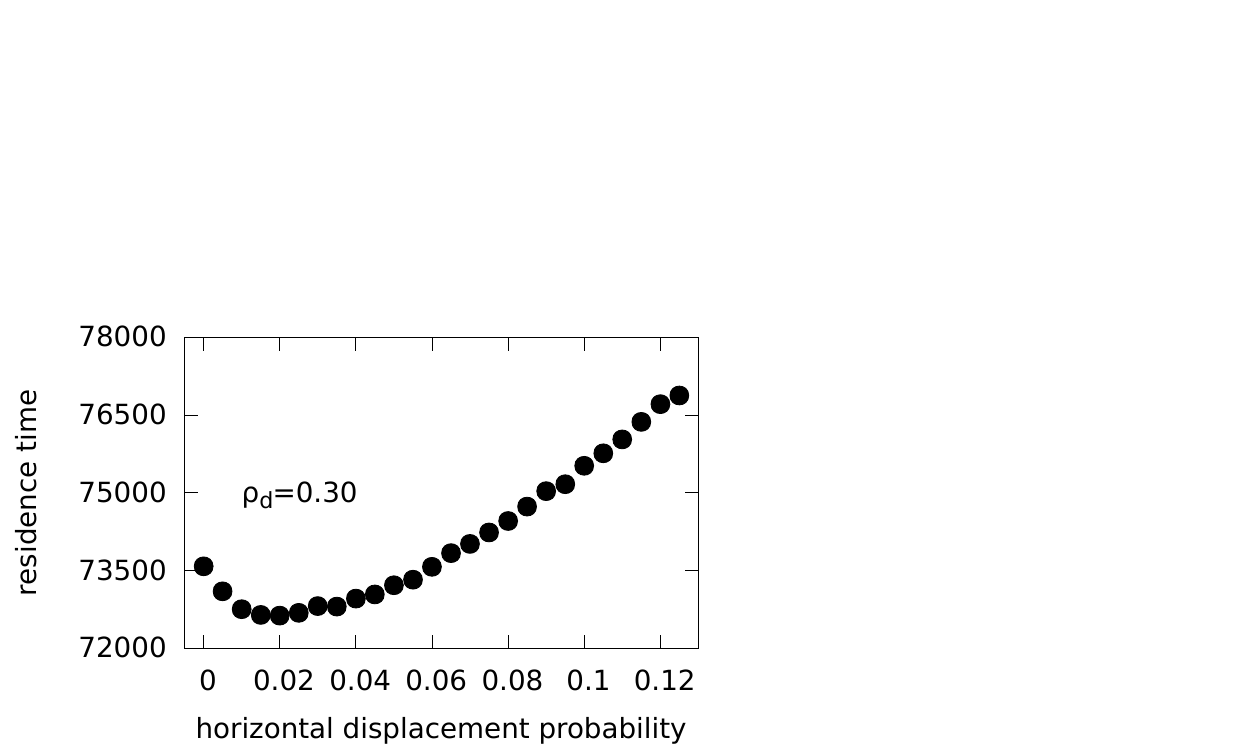}}} 
}
\put(160,250)
{
\resizebox{9cm}{!}{\rotatebox{0}{\includegraphics{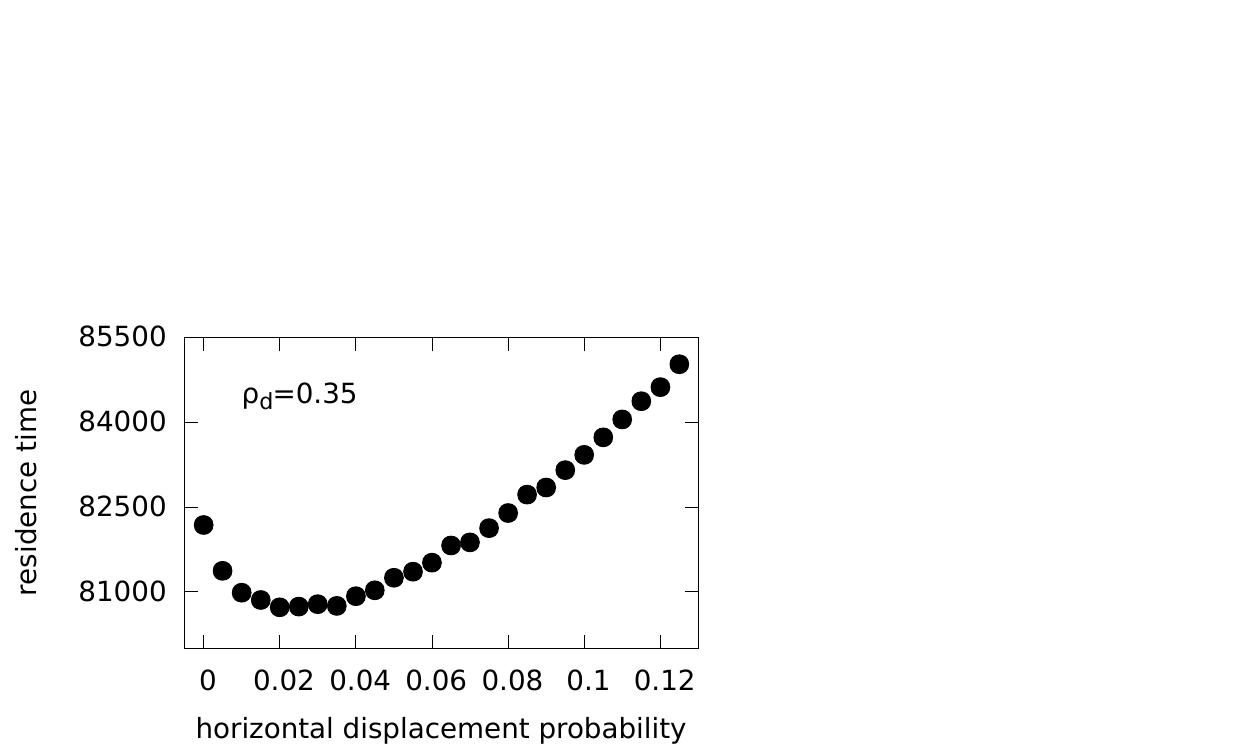}}} 
}
\put(320,250)
{
\resizebox{9cm}{!}{\rotatebox{0}{\includegraphics{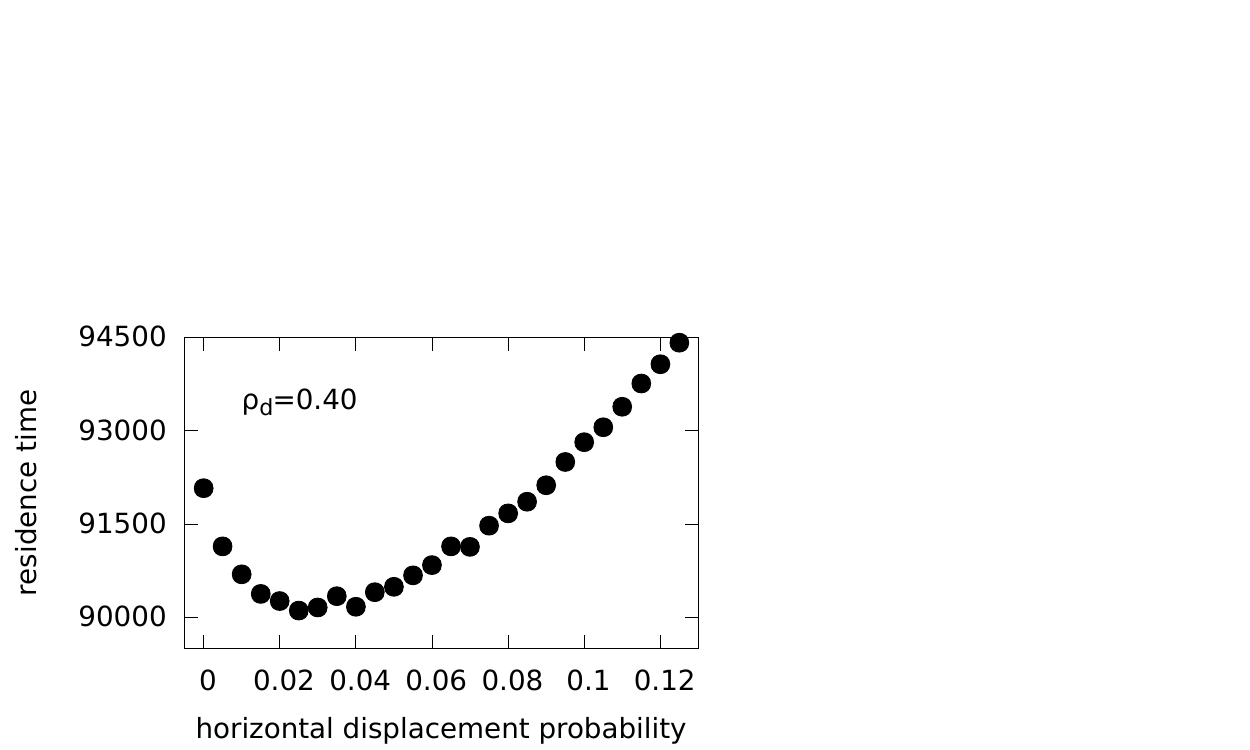}}} 
}
\put(0,150)
{
\resizebox{9cm}{!}{\rotatebox{0}{\includegraphics{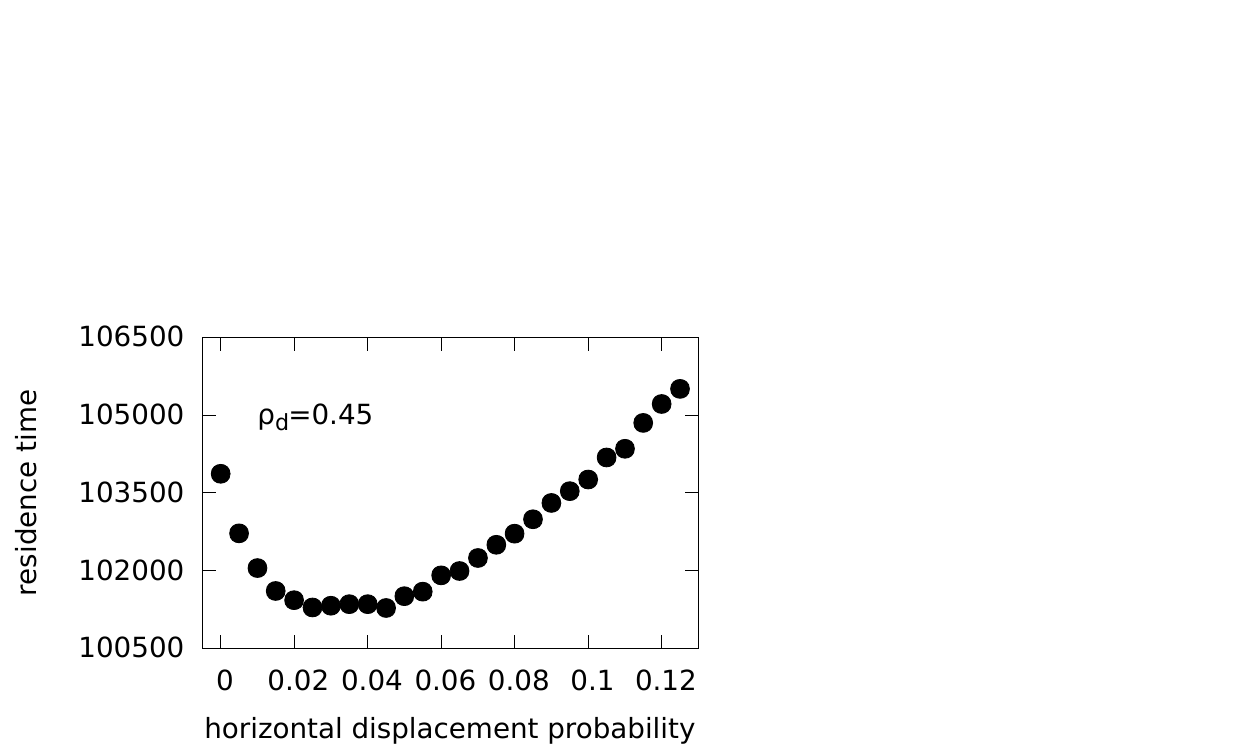}}} 
}
\put(160,150)
{
\resizebox{9cm}{!}{\rotatebox{0}{\includegraphics{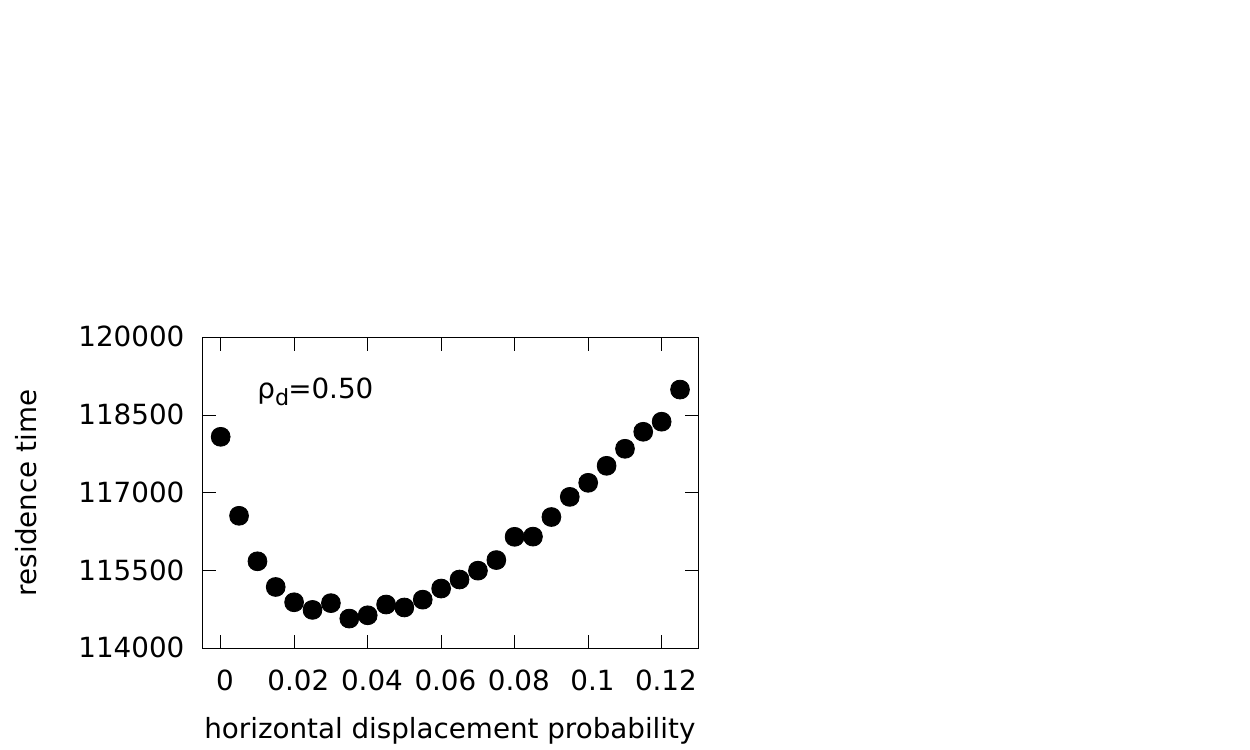}}} 
}
\put(320,150)
{
\resizebox{9cm}{!}{\rotatebox{0}{\includegraphics{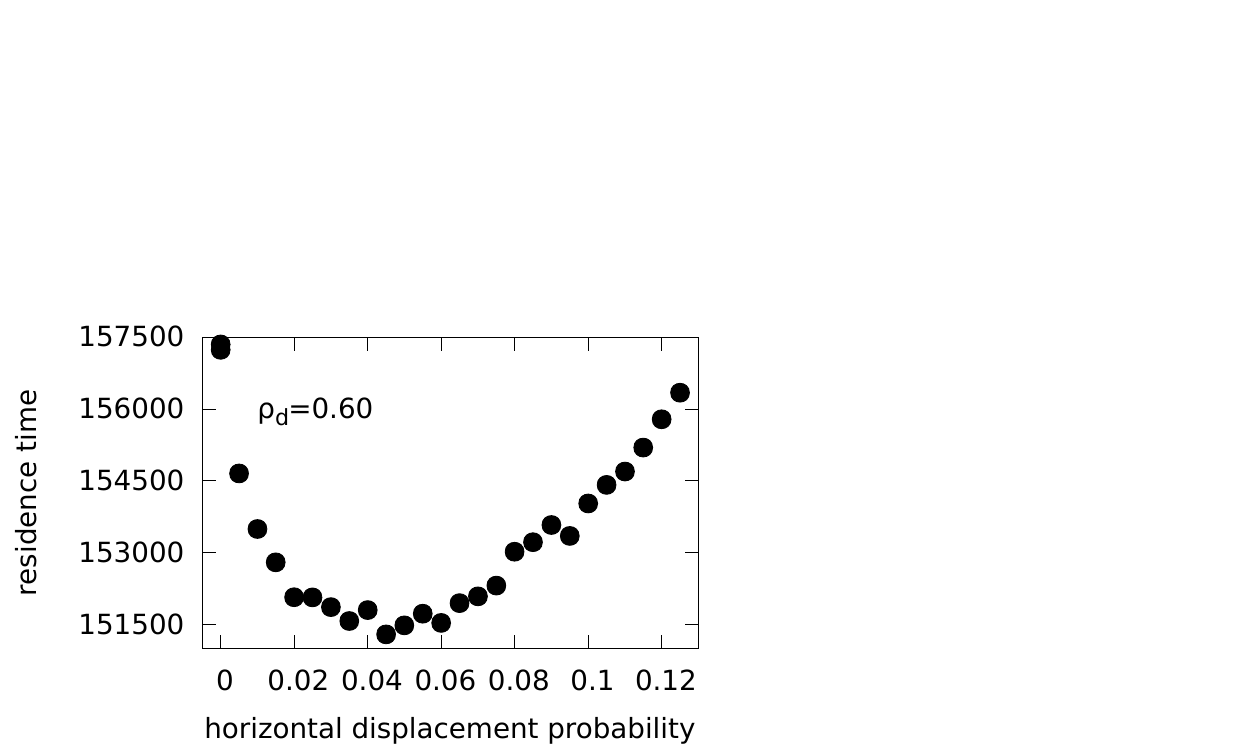}}} 
}
\put(0,50)
{
\resizebox{9cm}{!}{\rotatebox{0}{\includegraphics{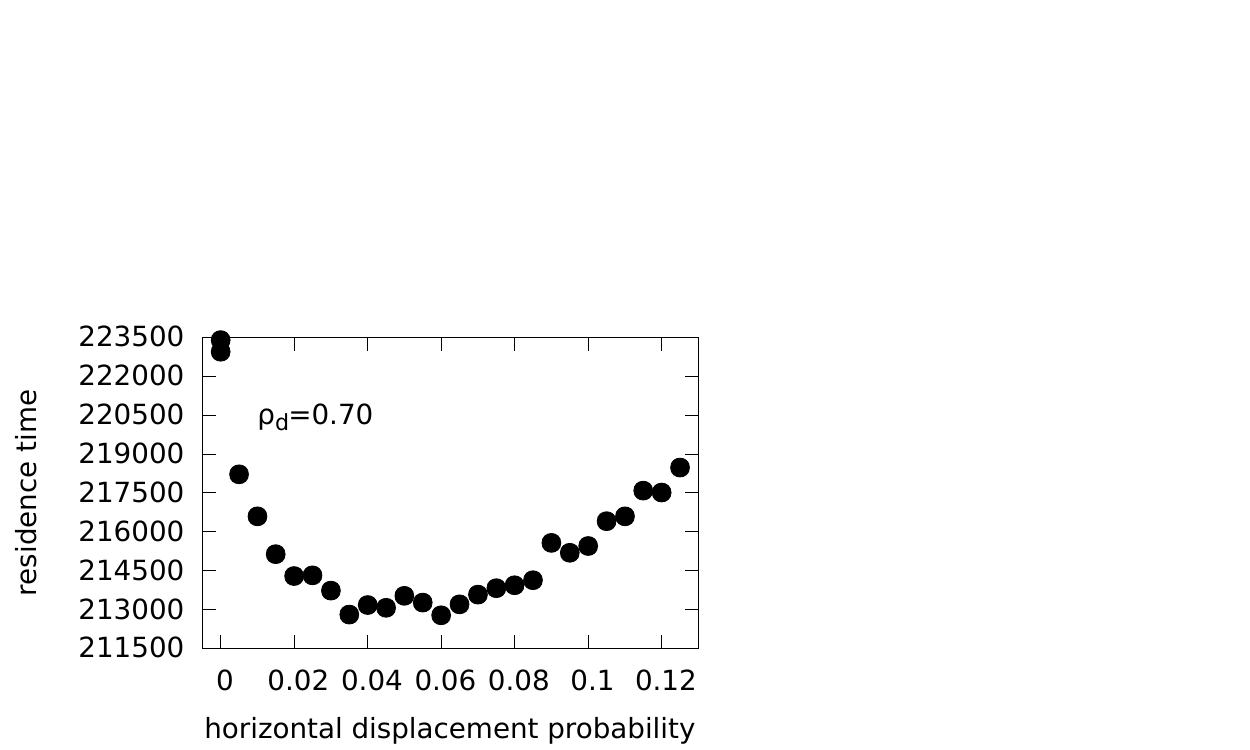}}} 
}
\put(160,50)
{
\resizebox{9cm}{!}{\rotatebox{0}{\includegraphics{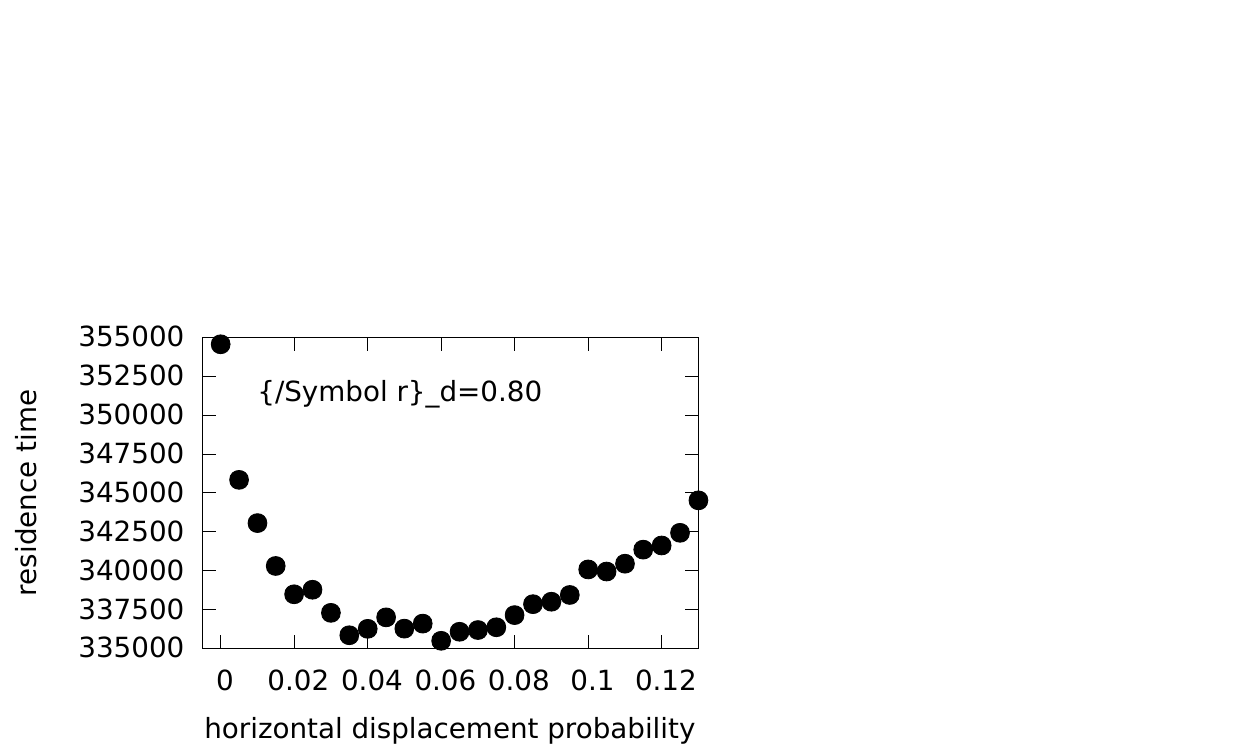}}} 
}
\end{picture}  
\vskip -0.5 cm 
\caption{Residence time versus the horizontal displacement 
probability $h$ in the case $\delta=0$ 
and $L_2=200$.
From the left to the right and from the top to the bottom 
the cases 
$\varrho_\rr{d}=0,0.05,0.01,\dots,0.045,0.5,0.6,0.7,0.8$
are depicted.
}
\label{f:res-notmono} 
\end{figure} 


In Figure~\ref{f:res-notmono}, the residence time is plotted 
as function of the horizontal displacement probability. It is 
remarkable the presence of a minimum at small values of $h$. 
This fact can be interpreted as follows: in the bouncing 
back regime, namely, when $\varrho_\rr{d}$ is high and $\delta$ low,
a particle
can find a blocking cluster of particles
in its way out through the bottom boundary. 
In such a situation, then, 
having a larger horizontal displacement probability can help 
the particle to bypass the obstacle. Obviously, if $h$ is increased 
too much this effect disappears due to the time wasted by the 
particles in horizontal movements. 


\section{Conclusions}
\label{s:conclusioni}
\par\noindent
We focused our attention on the study of the simplest 
2D model that mimics the flow of particles 
in a straight pipe under the effect of different driving boundary 
conditions and external fields. 
We studied the residence time, i.e., the typical time a particle
entering the strip at stationarity from the
top boundary needs to exit through the bottom one, 
under the assumption that
particles in the strip interact only via hard--core exclusion and 
vertical boundaries are reflecting. 

We explored the
dependence of the residence time on the
external driving force, 
length of the strip,
horizontal diffusion, and 
boundary conditions.
We have shown that, in almost all the considered regimes, 
the mean residence time is equal to the average time needed to cross the 
strip by a particle performing a random walk in a background 
prescribed by the stationary density profile of the original model. 
In this way, in particular, 
we recover the structure of the fluxes as well as the residence times  
proven mathematically in 1D in \cite{DEHP}.

A completely macroscopic point of view has also been alternatively adopted, 
i.e. a Mean Field computation of the residence time has been performed 
by connecting such a quantity to the stationary current in the 
system. It has been shown that also this point of view 
porvides very satisfactory results. 

This picture fails to be correct in the case in which there is 
a particle accumulation close to the bottom boundary (exit).
In this regime, we  discover new effects 
that are consequence of the two--dimensionality of the system.
The most relevant is the non--monotonic 
behavior in changes in the horizontal displacement 
probability in the bouncing back regime.

A second very interesting result is the fact that in this 
bouncing back regime the Mean Field approach gives a very good prediction 
of the residence time in the single file (zero horizontal 
displacement probability) case. On the other hand, 
if the horizontal displacement probability is large (close to one) 
the Mean Field prediction is poor and the one based on the analogy 
with the birth and death model on the background of the stationary 
density profile yield a very good prediction.


This particle accumulation 
situation can be realized artificially by inserting 
obstacles in the strip. We then expect interesting non--linear 
phenomena to show up. We shall study this regime in a 
follow--up paper.


\end{document}